\documentclass[longauth]{aa}

\usepackage{graphicx}
\usepackage{txfonts}
\usepackage{pifont} 
\usepackage[switch]{lineno}
\usepackage[dvipsnames]{xcolor}
\usepackage[acronym, shortcuts, nonumberlist]{glossaries} 
\usepackage{hyperref}
\usepackage{multicol}
\usepackage{amsmath}
\usepackage{caption}
\usepackage{float}
\usepackage{subcaption}
\usepackage{makecell}
\usepackage{svg}
\usepackage{utfsym}
\usepackage{xcolor}
\usepackage{cellspace}

\newacronym{agn}{AGN}{active galactic nuclei}
\newacronym{iact}{IACT}{Imaging Atmospheric Cherenkov Telescope}
\newacronym{sed}{SED}{spectral energy distribution}
\newacronym{eed}{EED}{electron energy distribution}
\newacronym{lc}{LC}{light curve}
\newacronym{mwl}{MWL}{multi-wavelength}
\newacronym{magic}{MAGIC}{Major Atmospheric Gamma Imaging Cherenkov}
\newacronym{lat}{\textit{Fermi}-LAT}{\emph{Fermi} Large Area Telescope}
\newacronym{bat}{\textit{Swift}-BAT}{\emph{Swift} Burst Alert Telescope}
\newacronym{xrt}{\textit{Swift}-XRT}{\emph{Swift} X-ray Telescope}
\newacronym{uvot}{\textit{Swift}-UVOT}{\emph{Swift} UV-Optical Telescope}
\newacronym{ctao}{CTAO}{Cherenkov Telescope Array Observatory}
\newacronym{421}{Mrk~421}{Markarian~421}
\newacronym{hsp}{HSP}{high synchrotron peaked}
\newacronym{ssc}{SSC}{synchrotron self-Compton}
\newacronym{ssa}{SSA}{synchrotron self-absorption}
\newacronym{ic}{IC}{inverse Compton}
\newacronym{ec}{EC}{external Compton}
\newacronym{ts}{\textit{TS}}{test statistic}
\newacronym{orm}{ORM}{Observatorio del Roque de los Muchachos}
\newacronym{swift}{\textit{Swift}}{\textit{Neil Gehrels Swift Observatory}}
\newacronym{webt}{WEBT}{Whole Earth Blazar Telescope}
\newacronym{gasp}{GASP}{\textit{GLAST-Agile} Support Program}
\newacronym{fsrq}{FSRQ}{flat spectrum radio quasar}
\newacronym{bllac}{BL~Lac}{BL Lacertae object}
\newacronym{smbh}{SMBH}{supermassive black hole}
\newacronym{vhe}{VHE}{very-high energy}
\newacronym{he}{HE}{high energy}
\newacronym{le}{LE}{low energy}
\newacronym{uv}{UV}{ultraviolet}
\newacronym{ebl}{EBL}{extra-galactic background light}
\newacronym{lppl}{LPPL}{log-parabola with a low energy power-law branch}
\newacronym{plc}{PLC}{power-law with an exponential cut-off}
\newacronym{mcmc}{MCMC}{Markov chain Monte Carlo}
\newacronym{nH}{$n_H$}{neutral Hydrogen}
\newacronym{kn}{\textit{KN}}{Klein-Nishina}
\newacronym{ixpe}{\textit{IXPE}}{Imaging X-ray Polarimetry Explorer}
\newacronym{vlbi}{VLBI}{Very Long Baseline Interferometry}

\makeglossaries

\newcommand{\ginj}{\gamma_\text{inj}}
\newcommand{\gammacut}{\gamma_\text{c}}

\newcommand{\ngamma}{n(\gamma)}
\newcommand{\gtp}{\gamma_\text{3p}}
\newcommand{\rtp}{r_\text{3p}}
\newcommand{\ngt}{\gamma^3 n(\gamma)}
\newcommand{\chisq}{\chi^{2}}
\newcommand{\jst}{\texttt{JetSeT}\,}
\newcommand{\nusync}{\nu^*_{\text{sync}}}
\newcommand{\nussc}{\nu^*_{\text{SSC}}}
\newcommand{\nuFnusync}{\nu F_{\nu, _\text{sync}}^*}

\newcommand{\bexp}{\beta_\text{exp}}
\newcommand{\texp}{t_\text{exp}}
\newcommand{\bsync}{b_\text{sync}}
\newcommand{\fvar}{F_\text{var}}
\newcommand{\thetaview}{\theta_{\text{view}}}

\definecolor{c08}{rgb}{0.18995, 0.07176, 0.23217}
\definecolor{c11}{rgb}{0.25369, 0.26327, 0.65406}
\definecolor{c12}{rgb}{0.27691, 0.44145, 0.91328}
\definecolor{c13}{rgb}{0.24427, 0.60937, 0.99697}
\definecolor{c14}{rgb}{0.13278, 0.77165, 0.88580}
\definecolor{c15}{rgb}{0.10342, 0.89600, 0.71500}
\definecolor{c16}{rgb}{0.27597, 0.97092, 0.51653}
\definecolor{c18}{rgb}{0.53255, 0.99919, 0.30581}
\definecolor{c19}{rgb}{0.72596, 0.96470, 0.20640}
\definecolor{c20}{rgb}{0.88331, 0.86553, 0.21719}
\definecolor{c21}{rgb}{0.98000, 0.73000, 0.22161}
\definecolor{c22}{rgb}{0.99297, 0.55214, 0.15417}
\definecolor{c23}{rgb}{0.94084, 0.35566, 0.07031}
\definecolor{c24}{rgb}{0.83926, 0.20654, 0.02305}
\definecolor{c25}{rgb}{0.68602, 0.09536, 0.00481}
\definecolor{c26}{rgb}{0.47960, 0.01583, 0.01055}

\hypersetup{ colorlinks=true, linkcolor={Cerulean}, filecolor=magenta, urlcolor={ForestGreen}, citecolor={Cerulean}}

\begin{document} 

    \title{The January 2010 flare of Mrk 421: Insights from a stochastic acceleration model}

    \subtitle{}

    \author{ 
    K.~Abe\inst{1} \and
    S.~Abe\inst{2} \and
    J.~Abhir\inst{3}\thanks{contact.magic@mpp.mpg.de} \and
    A.~Abhishek\inst{4} \and
    V.~A.~Acciari\inst{5} \and
    A.~Aguasca-Cabot\inst{6} \and
    I.~Agudo\inst{7} \and
    I.~Albanese\inst{8} \and
    T.~Aniello\inst{9} \and
    L.~A.~Antonelli\inst{9}
% ...
    }

\institute { Japanese MAGIC Group: Department of Physics, Tokai University, Hiratsuka, 259-1292 Kanagawa, Japan
    \and Japanese MAGIC Group: Department of Physics, Kyoto University, 606-8502 Kyoto, Japan
    \and ETH Z\"urich, CH-8093 Z\"urich, Switzerland
    \and Universit\`a di Siena and INFN Pisa, I-53100 Siena, Italy
    \and Institut de F\'isica d'Altes Energies (IFAE), The Barcelona Institute of Science and Technology (BIST), E-08193 Bellaterra (Barcelona), Spain
    \and Universitat de Barcelona, ICCUB, IEEC-UB, E-08028 Barcelona, Spain
    \and Instituto de Astrof\'isica de Andaluc\'ia-CSIC, Glorieta de la Astronom\'ia s/n, 18008, Granada, Spain
    \and Universit\`a di Padova and INFN, I-35131 Padova, Italy
    \and National Institute for Astrophysics (INAF), I-00136 Rome, Italy
% ...
}

    \date{Received \today; accepted \today}


    \abstract 
    {\acrshort{421} displayed its highest flux state ever observed in February of 2010 with very high TeV fluxes and interesting cross-band correlations and a \gls{sed} evolution not entirely consistent with the standard single zone leptonic \acrlong{ssc} model. The source was already in a high state in January 2010 and displayed strong variability in the days preceding the highest state. We study the temporal evolution of the spectra in January to extract information about the particle dynamics and the physical properties of the emission region.}
    {We build up on the temporal variability and correlations studied in the previous work \citep{felix_mrk421} and attempt to improve the \gls{sed} model fits with a physics oriented approach.}
    {The \acrlong{mwl} data was processed and the \glspl{sed} were fit using \jst. The \gls{sed} evolution and cross band correlations were modelled using leptonic \gls{lppl} and pile-up distributions that are predicted in a stochastic acceleration scenario. A simplified temporal evolution model was developed and fit to the \glspl{sed} and the resulting trends and phenomenology were characterised in context of theoretical literature. An expanding emission region model was also tested. }
    {We find the spectral variability to be well in agreement with stochastic acceleration. Our analysis suggests that the standard \gls{lppl} distribution develops a Maxwellian pile-up component at the transition from acceleration to cooling dominated phase on 3 nights in the dataset, as also hinted by the \acrlong{vhe} and X-ray \acrlong{lc}s. The resulting phenomenology of our sequential snapshot evolution \gls{sed} model agrees well with theoretical and numerical simulation studies on temporal evolution using the diffusion equation approach. Curvature in the \gls{eed} anti-correlates with the synchrotron peak frequency, as expected from stochastic acceleration. We test an alternate model with expanding emission region and the resulting \gls{eed} spectral index, expansion velocity and magnetic field index agree with prior works using datasets collected in different flaring states of the source.}
    {}

\maketitle
\glsresetall
\section{Introduction}\label{sec:intro}
Blazars, a class of jetted \gls{agn}, are known to be highly variable in the X-ray and \gls{vhe} $\gamma$-ray ($E>100$\,GeV) bands.
The blazar jet is powered by the \acrlong{smbh} in the central regions of these \gls{agn} and emission is beamed towards the observer, enhancing the intrinsic variability by relativistic effects. 
The archetypal blazar \gls{421} \citep[$z=0.031$;][]{Vaucouleurs1991} belongs to the \gls{hsp} \acrshort{bllac} object category as per the unified blazar sequence \citep{Urry_1995,Fossati_1998MNRAS.299..433F} with the synchrotron peak around $10^{17}$Hz, although the validity of this classification scheme has been called into question by more recent large sample size studies \citep{updated_blazar_sequence, 2022Galax..10...35P}. 
The \gls{vhe} flux of \gls{421} often changes by an order of magnitude between the flaring and quiet states. 

\gls{421} has been known to show a variety of spectral characteristics during different flaring periods \citep{Balokovic_2016, mrk421_2017_magic,Acciari_2020, 2021MNRAS.505.2712D}. 
The general structure of its \gls{sed} agrees well with a single zone \gls{ssc} model \citep[e.g.][]{magic_veritas_sed_2015A&A...578A..22A} typically used for modelling \acrshort{bllac} objects owing to the lack of evidence for external photon fields (see \cite{2019Galax...7...20B} for an overview of blazar modelling). 
However, these single and even multi-zone models are a highly simplified description of the emission from a complex object spanning several kiloparsecs. 
\gls{421} in particular is well known for showing strong variations in the X-ray and VHE $\gamma$-ray flux over short timescales \citep{2021A&A...647A..88A, 2024MNRAS.529.1450G, 2025arXiv250908686M}, which makes it important to consider this temporal variability aspect in any spectral modelling attempts.

The `snapshot' approach has often been used in the past to study the time averaged emission of blazar jets. 
To build such a snapshot, quasi-simultaneous \gls{mwl} observations are combined depending on the brightness of the source in different bands, which are then usually modelled independent of each other. 
The dynamics of the particle population are abstracted out using parametrised steady-state solutions to the diffusion equation. 
This approach leads to the loss of some of the physics information contained in the evolution of the states.
In contrast, temporal evolution models can provide more information about the relativistic flows and emission environments in the jet. 
These models use a self-consistent diffusion equation approach to model the time evolution of the particle population that drives the photon emission. 

The dramatic activity of \gls{421} in 2010 was covered in an extensive \gls{mwl} campaign \citep{magic_veritas_sed_2015A&A...578A..22A, magic_veritas_lc_2020ApJ...890...97A} from radio to \gls{vhe} bands. 
While the highest flux in the X-ray and \gls{vhe} bands was seen in February 2010 \citep[see models in ][]{magic_veritas_sed_2015A&A...578A..22A, 2021MNRAS.505.2712D}, the cadence of spectral data in this period is sparse compared to January 2010 where it is possible to build daily timescale snapshots.
This rich dataset from January allows us to extract the physical information regarding the acceleration of particles and the emission environment in the jet. 
The daily binned \glspl{sed} (in observer reference frame) minimise contamination due to averaging of different spectral states.
Additionally, we treat the \gls{sed} dataset as a sequence of states instead of as completely independent states, bringing our method closer to a temporal evolution model compared to the typical snapshot model. 
We combine this approach with a physically motivated particle distribution under the stochastic acceleration framework. 
This allowed us to probe the jet state transitioning between acceleration and cooling domination. 
The results of this study will be used to guide a fully self-consistent temporal evolution model in a future work. 

In Section~\ref{sec:data} we describe the processing and data reduction into a series of \glspl{sed}, while the fitting procedure and the physical setup of the models is explained in Section~\ref{sec:modelling}. 
The results are reported in Section~\ref{sec:fit_results} with the associated phenomenology and physics interpretation in Section~\ref{sec:phenomenology}. 
We build an expanding emission region model based on these results in Section~\ref{sec:expanding_blob} and provide conclusions, caveats and future outlook in Section~\ref{sec:conclusion}. 
Additional information on data processing and some plots for individual models that are not directly relevant to the discussion can be found in the Appendix.

\section{Observations and Data reduction}\label{sec:data}
This section describes the instrument specific processing relevant to building the \glspl{sed} sequence and the observations that motivate the physics oriented approach to setting up the emission region in Section~\ref{sec:modelling}. 
The full \gls{mwl} \gls{lc} dataset can be found in \cite{felix_mrk421}, while a summary of the analysis done for the individual instruments can be found in the Appendix \ref{appendix:mwl_data}. 

\begin{figure}
    \centering
    \includegraphics[width=0.49\textwidth]{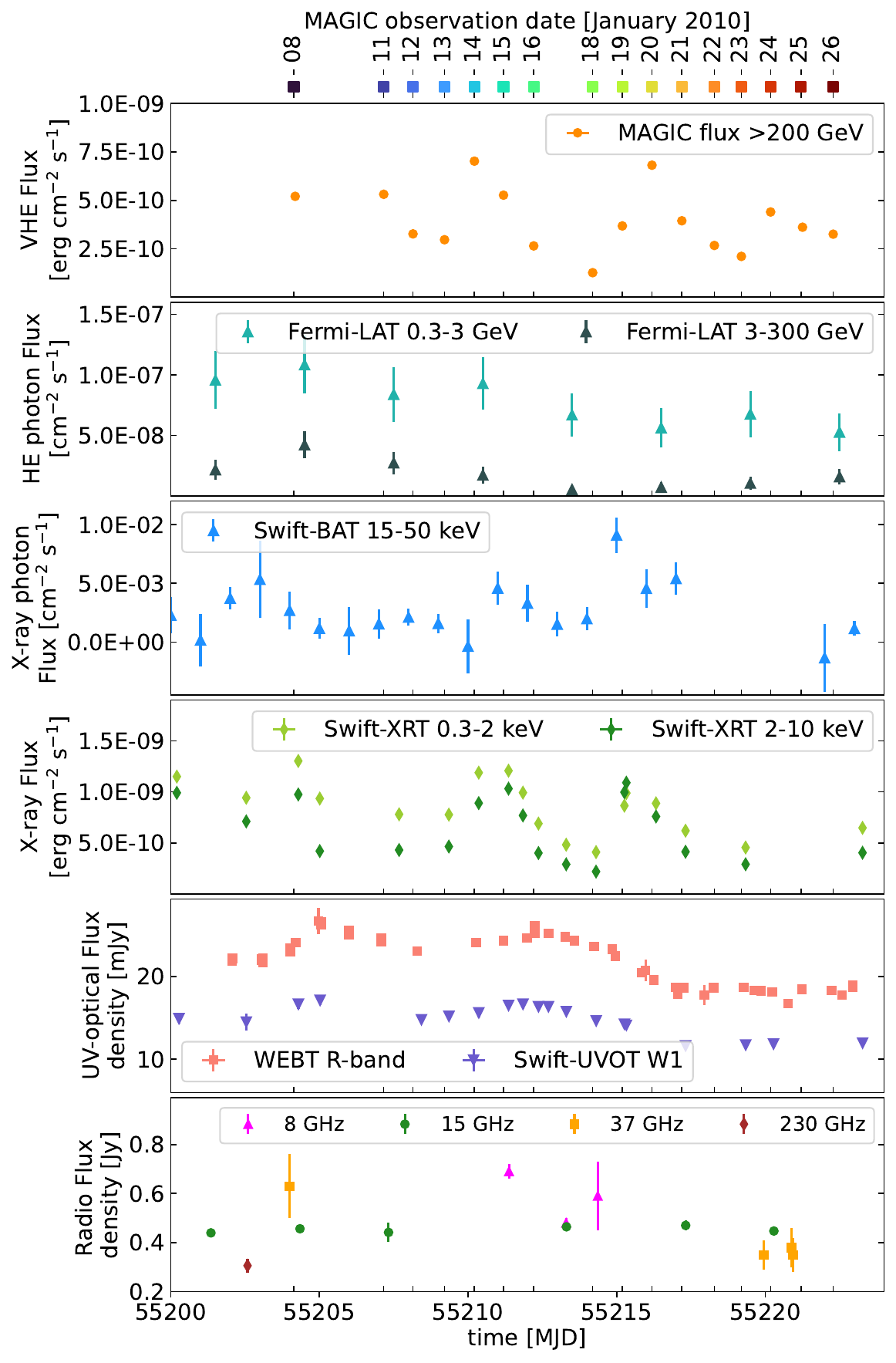}
    \caption{\gls{mwl} \gls{lc} during January 2010 from radio to VHE. The data are taken from \cite{felix_mrk421}. The top axis is marked with date colour key which is consistent across plots and tables in the paper. }
    \label{fig:lc_zoomed}
\end{figure}

Fig.~\ref{fig:long_term_lc} shows the \gls{vhe} and X-ray \glspl{lc} which were the bands with the strongest variability and historic high flux levels during the 2009-2010 observation campaign while Fig.~\ref{fig:lc_zoomed} shows the January 2010 dataset. 
The well sampled \gls{mwl} \gls{lc} and the multiple flaring incidents with the flux changing by a factor of $\sim 3$ in a matter of days motivated us to study January 2010 in detail.  
The January dataset was used to build and label the \glspl{sed} by \gls{magic} telescope observation start date as described in Section~\ref{sec:gamma_ray_obs}. 

\begin{figure}
    \centering
    \includegraphics[width=0.49\textwidth]{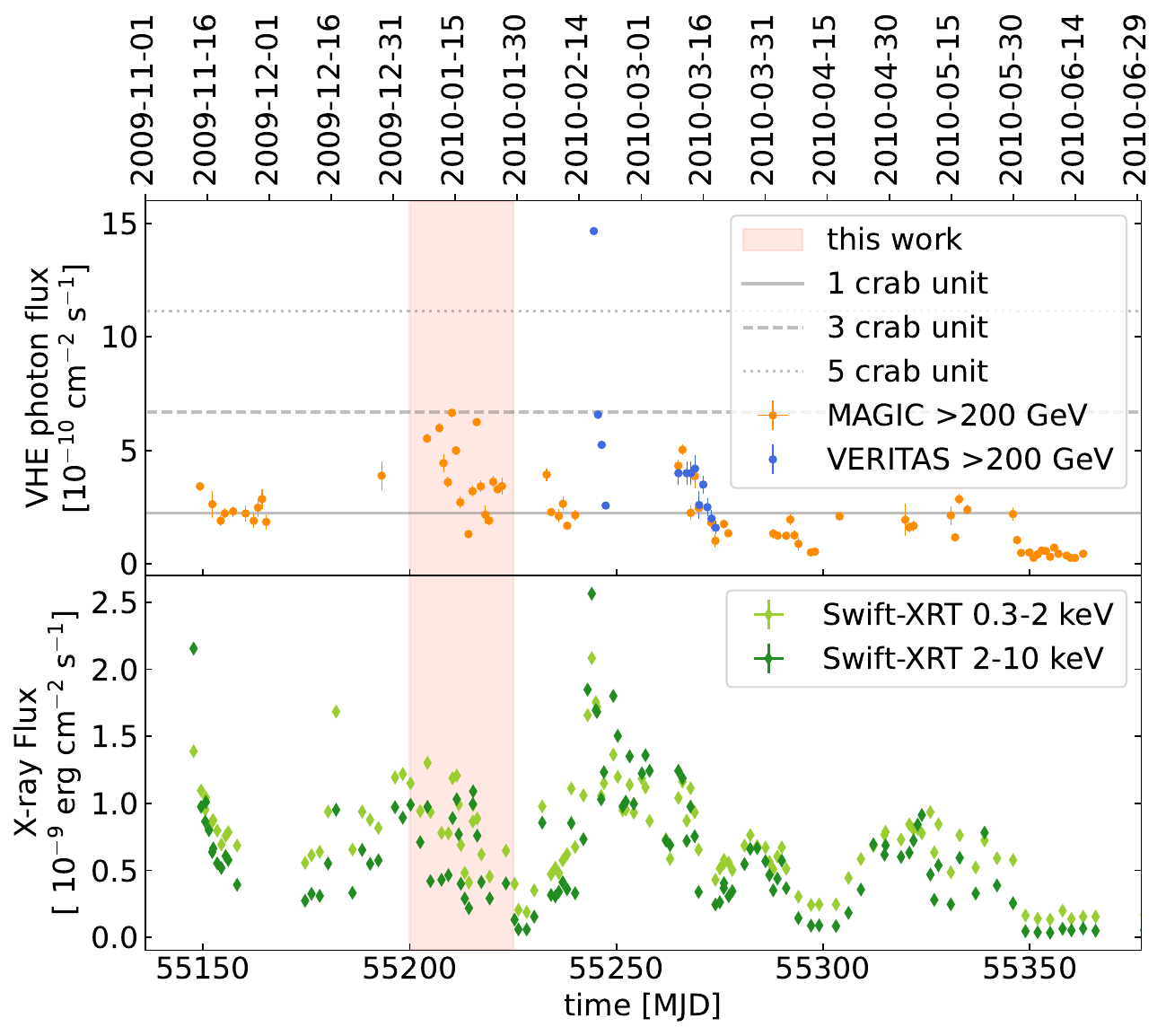}
    \caption{\gls{vhe} and X-ray \gls{lc} of \gls{421} during the 2009-2010 \gls{magic} observations campaign. The data are daily binned and taken from \cite{felix_mrk421}. Our broadband SED modelling is performed over the period highlighted with a vertical red band (January 2010). Fig.~\ref{fig:lc_zoomed} shows the \gls{mwl} data in the highlighted region in more detail.}
    \label{fig:long_term_lc}
\end{figure}

\subsection{\texorpdfstring{$\gamma$}{gamma}-ray bands}\label{sec:gamma_ray_obs}
We binned the spectral data according to the availability of the \gls{magic} telescope observations. 
There are 16 nights in our \gls{sed} dataset starting from 08 January 2010 until 26 January 2010 and the \gls{mwl} \glspl{sed} are labelled according to the date on which \gls{magic} observations started. 
Given the duty cycle and observation schedules of ground based \glspl{iact}, it is hard to obtain evenly sampled datasets so the \gls{mwl} observations were assigned to the temporally closest \gls{magic} observation. 
This assumes that a few short observations within a few hours of each other are contemporary and representative of the general behaviour of the source over multi-hour timescales. 
\cite{felix_mrk421} found significant intra-night variability on 15 January 2010 with flux changing by a factor of $\sim 2$ on sub-hour timescales at a $3.7 \sigma$ level.
However the rest of the dataset during January 2010 showed no such intra-night variability which supports the idea of the source being stable during individual \gls{vhe} observations. 

Given the sensitivity of \gls{lat} and the flux of \gls{421} in the \gls{he} ($\sim 10^{6-10}$ eV) band, it was not possible to obtain statistically significant energy bins for a large fraction of the dataset at daily timescales (see Appendix \ref{appendix:mwl_data} for details) for the \gls{lat} observations. 
Thus, a 3-day binning was used with the closest 3-day bin matched with each \gls{magic} observation night. 
Even with the 3-day bins, there are some nights in the dataset with a very large uncertainty in the spectrum. 
However, the variability in the \gls{he} band is low as quantified by the fractional variability in \cite{felix_mrk421} and we do not expect interesting dynamics in the \gls{he} band on the daily timescales for this particular source and the 2010 flare. 

To account for inter-instrument calibration effects, we added $20\%$ systematic uncertainty to the \gls{magic} data and $10\%$ to the \gls{lat} data (see Table~\ref{tab:systematics}) based on the published instrument performance \citep{2009PhRvL.103y1101A, 2016APh....72...76A}.

\subsection{X-ray bands} \label{sec:xray_data}
We converted the hard X-ray ($>10$ keV) excess count observed by \gls{bat} to Crab Units and subsequently to a spectral point using the method described in \cite{2013ApJS..209...14K}. 
As described by the authors, the systematic uncertainties in the spectral points thus obtained are large and we added a $30\%$ uncertainty to this data (Table~\ref{tab:systematics}) during the modelling. 
Additionally, for days with negative excess counts due to background fluctuations, no spectral points can be extracted (14 and 26 January). 
Despite this, the \gls{bat} data turned highly useful on 19 January when the peak of the synchrotron bump of the \gls{sed} was outside the energy range of the \gls{xrt} and hence it would be hard to constrain the peak position without the \gls{bat} spectral point. 

The data from the \gls{xrt} was processed as described in \cite{felix_mrk421}. 
The X-ray band spectral points at the edges of the ranges in the $0.3-2$ keV and $2-10$ keV bands of the \gls{xrt} were additionally checked for energy dependence of systematic uncertainties since the observed curvature in the data points directly affects the conclusions of our \gls{sed} modelling. 
Uncertainty in the \gls{nH} density along the line of sight to \gls{421} can mimic additional curvature  (see Appendix \ref{sec:mwl_xrt}). 
This systematic uncertainty was included in the \gls{sed} modelling and the conclusions regarding the curvature were unaffected.

\subsection{\texorpdfstring{\acrshort{uv}-optical and radio data}{UV-optical and radio data}}
We used the combined dataset from \cite{felix_mrk421} to generate our \glspl{sed} for \gls{uvot}, R-band and radio bands.
We combined the data to build as contemporaneous \glspl{sed} as possible, using the \gls{mwl} data which was temporally closest to the \gls{vhe} observation. 
We used a $10\%$ systematic uncertainty in the \gls{uv} and optical band data which is above the typical uncertainty assumed in these bands \citep{2016ApJ...819..156B, 2023ApJ...950..152A}.

The radio band emission is expected to originate from larger regions in the jet due to the \gls{ssa} process \citep[see for e.g.][]{tramacere2022} compared to the compact regions close to the central \gls{smbh} emitting higher energy photons. 
Hence, we restricted the minimiser to the spectral points with $\geq 10^{11}$ Hz to exclude the lowest energy radio data that is strongly affected by \gls{ssa}.
Typical systematic flux uncertainty for the individual telescopes and radio bands for our dataset range from $2-10\%$ \citep{2011ApJS..194...29R, 2016ApJ...821...61P} but we use a $20\%$ value in our modelling to account for the multiple different telescopes and the uneven sampling. 
A more detailed description of the data processing can be found in the Appendix \ref{appendix:mwl_data}.

\begin{table}
    \centering
    \begin{tabular}{cc}
        \hline
        Instrument & systematic uncertainty \\
        \hline
        \gls{magic}  & $20\%$     \\
        \gls{lat}    & $10\%$     \\
        \gls{bat}    & $30\%$     \\
        \gls{xrt}    & $10-16\%$$^*$ \\
        \gls{uvot}   & $10\%$     \\
        R-band       & $10\%$     \\
        radio        & $20\%$     \\
        \hline
    \end{tabular}
    \caption{Additional systematic uncertainty added to the spectral data for fitting models in \jst. The instrument performance references can be found in Section~\ref{sec:data} and Appendix~\ref{appendix:mwl_data}. $^*$: See discussion on energy dependent systematic uncertainty for \gls{xrt} data in Section~\ref{sec:xray_data}}
    \label{tab:systematics}
\end{table}

\subsection{Data phenomenology}\label{sec:data_phonomenology}
The X-ray \gls{lc} has peaks around 15 and 19 January and the \gls{vhe} \gls{lc} follows a similar trend with peaks on 14 and 20 January. 
These peaks hint at multiple re-acceleration and/or injection episodes. 
Combined with the variability and curvature information described in the subsequent paragraphs, this could be seen as a sign of a change from acceleration to cooling dominated evolution, plausibly from weakening injection or reduced acceleration efficiency of the particle population driving the emission.
We exploit this information in Section~\ref{sec:lepton_dist} to motivate a particle distribution function to capture the physics of this transition. 
This also differentiates the January flare from the February and March flares which show a single peak followed by a consistently decreasing flux \citep{2016APh....72...76A} indicative of a cooling dominated phase especially noticeable in the denser sampling of the X-ray band in Fig.~\ref{fig:long_term_lc}. 

\cite{felix_mrk421} carried out a detailed study of temporal variability and \gls{mwl} cross-correlation and we report the results that are relevant to our analysis here. 
The dataset shows the typical two peak structure in the fractional variability  \citep[$\fvar$;][]{2003MNRAS.345.1271V}, with the X-ray and \gls{vhe} bands showing the highest $\fvar > 0.8$.
Moreover, there was no delay between the X-ray and \gls{vhe} bands in the full $7$-month long period suggesting a co-spatial origin of the emission. 
Significant correlation and similar value of fractional variability were seen between the \gls{uv} and \gls{he} $\gamma$-ray bands, making the overall \gls{mwl} picture consistent with the standard \gls{ssc} scenario used to model the broadband emission of \gls{421}. 
There was no correlation between the \gls{uv} and X-ray data and a weak correlation between \gls{uv} and the \gls{vhe} band, which was interpreted by the authors as a multi-zone scenario with a compact \gls{vhe} emission zone embedded in a larger zone emitting at lower energies. 
However, given the longer synchrotron cooling times of electrons emitting at \gls{uv} energies versus the ones emitting at X-rays, the correlation between these bands can be weak even in a single zone scenario. 
We will use a single zone \gls{ssc} scenario in this work to minimise the degeneracies in the model as the parameter space is not fully constrained, with the goal of studying the acceleration-cooling domination transition hinted by the \gls{lc} using a physically motivated \gls{eed}. 

\begin{figure}
    \centering
    \includegraphics[width=0.48\textwidth]{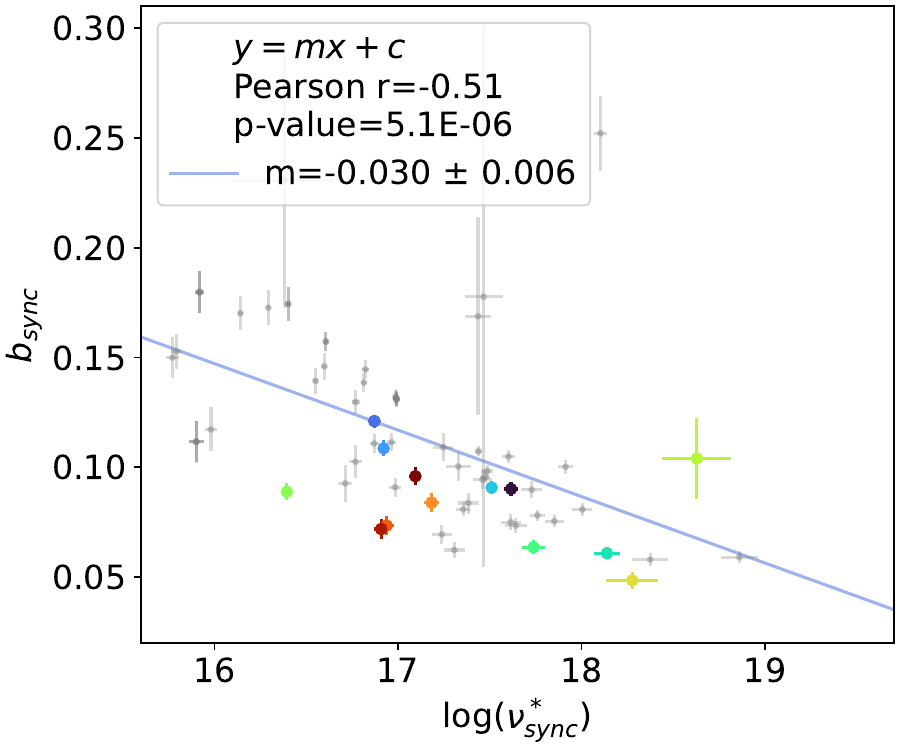}
    \caption{The anti-correlation between the peak curvature $\bsync$ vs the $\log_{10}$ of the peak frequency ($\nusync$) in Hz for the full 2009-2010 campaign data. The curvature was measured via a log-parabolic fit to the spectral data in the \gls{uv}--X-ray bands. The coloured data-points are from January 2010, with the same colour key as Fig.~\ref{fig:lc_zoomed} and Table~\ref{tab:mcmc_fit_par_list}.}
    \label{fig:data_Ep_bsync}
\end{figure}

Since the shape of the underlying \gls{eed} and stochastic acceleration can affect the curvature in the synchrotron peak \citep{2004A&A...413..489M, 2006A&A...448..861M, tramacere2007_mrk421_xray}, we tested for a log-parabolic curvature at the peak of the synchrotron bump.
The peak of the synchrotron emission lies in the \gls{xrt} energy window for \gls{421}, the \texttt{XSPEC} analysis of the \gls{xrt} data found that the log-parabola was a better fit than a power-law with a very high significance ($>>5\sigma$) for all days in January 2010. 
This analysis however gave a distorted picture of the phenomenology as will be seen later in Section~\ref{sec:fit_results}, since on 19 January, the peak position clearly lies outside the \gls{xrt} energy and the log-parabolic fit reports a very small curvature contrary to the true picture with a very large curvature shown by the \gls{mwl} dataset. 
Thus, we performed a log-parabolic fit to a larger frequency range, from \gls{uv} to X-ray bands (\gls{uvot}, \gls{xrt} and \gls{bat} data). 
The results of the analysis can be seen in Fig.~\ref{fig:data_Ep_bsync} with an anti-correlation trend between the peak frequency ($\nusync$) and the log-parabolic curvature measured in the \gls{uv}--X-ray range ($\bsync$). 
This behaviour is expected in a stochastic acceleration scenario which was then used for choosing the parametrised form of the \gls{eed} as described in Section~\ref{sec:lepton_dist}.

\section{SED modelling}\label{sec:modelling}
The \gls{sed} modelling was carried out using \jst \footnote{\url{https://github.com/andreatramacere/jetset}, \url{https://jetset.readthedocs.io/en/latest/index.html}} \texttt{ v1.3.1} \citep{tramacere2020Jetset}. 
The input relativistic particle distribution radiates via synchrotron and \gls{ic} processes, and all relevant internal absorption processes are also included in the code (\gls{ssa} and $\gamma$-$\gamma$ photo-absorption). 
The software is capable of evolving the particle distribution in a self-consistent time-dependent approach considering acceleration and radiative cooling. 
The parameter space of such fully self-consistent time-dependent models is much larger than snapshot models. 
This highly complicates the implementation of a single time evolving model to account for all of the complex time variability and phenomenology shown by \gls{421}.
Hence, we start with a snapshot model using phenomenological \glspl{eed} sensitive to cooling and acceleration, driven by hints of stochastic acceleration seen in Figs.~\ref{fig:lc_zoomed} and \ref{fig:data_Ep_bsync}. 
The current analysis will build the baseline for future temporal evolution studies of this flare. 

\subsection{Particle population distribution} \label{sec:lepton_dist}
The observations provide a rich phenomenology, with the source going through a complex pattern of acceleration and cooling dominated episodes. 
The acceleration and cooling timescales are a function of electrons energy and the physics of the acceleration and cooling processes. 
In all subsequent mentions, we will refer to these timescales for electrons emitting in the X-ray and VHE bands, where the largest spectral variability occurs.
Modelling this phenomenology is a challenging task since the competition between cooling and acceleration timescales adds further degeneracies to the ones intrinsic to the SSC scenario, plus the possible coexistence of multiple emitting regions. 
Hence, we focus on a one-zone SSC leptonic scenario, based on an \gls{eed} with parameters sensitive to both the cooling and the acceleration processes during flaring sequences. 

The physics of these processes can be captured in a self-consistent way using differential kinetic equations \citep[see][]{ramaty1979AIPC...56..135R, becker2006ApJ...647..539B, tramacere2011}. 
In particular, as a consequence of stochastic Fermi acceleration in a turbulent magnetic field, analytical and numerical solutions of the kinetic equations naturally lead to the formation of a \gls{lppl} \gls{eed} \citep{stawarz2008, tramacere2009, tramacere2011} through the momentum-diffusion process. 
The \gls{lppl} \gls{eed} consists of a power-law low energy branch characterised by spectral index $s$ and a log-parabolic high energy branch with curvature $r$. 
This provides a good description of the \gls{eed}, as long as the process is in the acceleration dominated regime, that is, the electron acceleration times are shorter than the cooling times.
The low-energy branch power-law index, $s$ can be driven by either the first order or the second order Fermi processes, depending on the efficiency of confinement of the particles within the acceleration region. 
The log-parabolic curvature $r$ is driven by the momentum-diffusion process, and its value can be used to track the evolution of the system from acceleration to cooling-dominated regime \citep{tramacere2011}.
In this regard, and based on the spectral patterns described in \ref{sec:data_phonomenology}, specifically the log-parabolic shape observed in the X-ray window, we model the January 2010 period with a \gls{lppl} distribution.
Mono-energetic relativistic particles are assumed to be injected into the acceleration region at Lorentz factor $\gamma = \ginj$, which can be of the order of $\sim10^2 - 10^3$ as a consequence of electron preheating within (mildly relativistic) electron-ion plasma shocks \citep{2021A&A...654A..96Z, 2025A&A...702A.255A}.
An additional simple power-law component develops for $\gamma \leq \ginj$ in the \gls{eed} at equilibrium because of the cooling of particles at higher energies \citep{rybicki_lightman}. 
The acceleration process is also limited at the highest energies by an exponential cut-off. 
The overall particle number density distribution can be parametrised in the following way \citep{tramacere2011}:
\begin{align}
    n(\gamma) &\propto \begin{cases}
        (\gamma / \gamma_0) ^ {{(2s+1)}/{2}} & \text{for } \gamma \leq \, \ginj, \\
        (\gamma / \gamma_0) ^ {-s} & \text{for } \ginj < \gamma < \gamma_0, \\
        (\gamma / \gamma_0) ^ {-s - r \cdot log(\gamma / \gamma_0)} & \text{for } \gamma > \gamma_0\\
    \end{cases} \\
    F_\text{cut-off}(\gamma) &= \exp \left( -\frac{1}{a} \left( \frac{\gamma}{\gammacut} \right)^a \right) \label{eq:cut_off_define} \\
    n_\text{LPPL}(\gamma) &= n(\gamma) \times F_\text{cut-off}(\gamma) \label{eq:lppl_define}
\end{align}
where $\gamma_0$ is the onset of the spectral curvature $r$, $\gammacut$ is the limit of the acceleration process and $a$ is the cut-off index. 
For our \gls{lppl} model, $\gammacut = 1 \times 10^7$, capping the maximum energies the particles can be accelerated to, and $a=1$ which yields the standard exponential cut-off functional form.
This choice of $\gammacut$ is consistent with the lack of an observation of an exponential cut-off in the X-ray data and avoids a non-physical sharp cut-off at the high energy edge of the numerical grid of $\gamma$.
Hence we have a total of 4 free parameters for the \gls{lppl} distribution used in the modelling ($\ginj$, $\gamma_0$, $s$, $r$). 

We found that this \gls{eed} parametrization provides a good description for most of the \glspl{sed} in our dataset, but failed to capture the curvature seen in the synchrotron peak of the \gls{sed} on some days, especially 19 January 2010. 
This suggested that some of these states, which were all close to the peak of the flares could be the result of the transition of the system towards a cooling-dominated regime.

When the system moves from an acceleration-dominated to a cooling-dominated regime, i.e. when the acceleration timescales become longer than the cooling timescales, a fraction of the population can `thermalise' close to the equilibrium energy where the acceleration and cooling effects are balanced. 
As a result, the formation of a pile-up component \citep{1984A&A...136..227S, 1985A&A...143..431S, stawarz2008} described by a relativistic Maxwellian distribution is expected at the high energy tail of the \gls{eed}.
Such a transition might be triggered by changes in the acceleration efficiency, variations in the injection luminosity of the leptons or sudden changes in the jet environment, however the approach in this work cannot distinguish between these scenarios and we defer testing these cases to a future work. 
Instead, we use a parametric distribution to capture the signature of this transition in the spectra.
The radiative signature of such a pile-up appears close to the \gls{sed} peak frequencies, i.e. in the X-ray and \gls{vhe} bands in the case of \gls{hsp} blazars.
The general form of the Maxwellian component is as follows \citep{stawarz2008}:
\begin{align} 
    n (\gamma) &\propto \gamma^{2} \cdot F_\text{cut-off}(\gamma) 
    \label{eq:pileup_form}
\end{align}
which can be combined with a power-law to get the final \gls{eed} with pile-up:
\begin{align}
    n(\gamma) &\propto \begin{cases}
        (\gamma / \gamma_0) ^ {{(2s+1)}/{2}} & \text{for } \gamma \leq \, \ginj, \\
        (\gamma / \gamma_0) ^ {-s} & \text{for } \gamma > \ginj \\
    \end{cases} \\
    n_{\text{pile-up}} &= (n(\gamma) + f \; \gamma^2 ) \times F_\text{cut-off}(\gamma)
    \label{eq:pileup_define}
\end{align}
where the same cut-off function as the \gls{lppl} case applies (Eq.~\ref{eq:cut_off_define}). 
Here, $\gammacut$ is the acceleration-cooling equilibrium Lorentz factor and $a$ depends on the dominant cooling process and the magnetic turbulence index \citep[parameter $q$ in][]{stawarz2008}.
$f$ is a scale factor for the relative strength of the pile-up with respect to the standard power-law component. 
This functional form is an analytical simplification of the equilibrium approximations of simulated mono-energetic particle injection and evolution under stochastic acceleration and radiative cooling using the diffusion equation approach, as seen in \cite{stawarz2008} and \cite{tramacere2011}. 

The pile-up \gls{eed} defined above has a total of 5 free parameters ($\ginj$, $\gammacut$, $s$, $a$, $f$). 
We fit the \gls{lppl} and pile-up \glspl{eed} to all the \glspl{sed} and compared the fits via a likelihood ratio test described in Section~\ref{sec:fit_results}. 
For plots visualising the \gls{eed} in the subsequent text, we will use the $\gamma^2 n(\gamma)$ distribution instead of just $n(\gamma)$ to emphasize and compare the curvature in the \gls{eed}. 
However, for simplicity, we will keep referring to the $\gamma^2 n(\gamma)$ distribution as the \gls{eed}.

\subsection{Physical setup of the emitting zone}
The emission region is assumed to be spherical blob with radius $R$ that is permeated by a tangled magnetic field of strength $B$. The particle density in the emitting region is given by $N$. 
The emitting plasma moves relativistically towards the observer with a bulk Lorentz factor $\Gamma$. This leads to relativistic beaming of the radiation with a Doppler factor $\delta \sim \Gamma$. Finally, the ratio of relativistic electrons to cold protons, which influences the jet power computation, is fixed to 1 in our modelling.

\subsection{Model fitting procedure - `parallel' and `sequential' fits}\label{sec:model_fitting}
We fit our model to the daily binned \glspl{sed} of the January 2010 epoch. 
In particular, we focus on the evolution of the \gls{eed} whose shape is physically motivated assuming stochastic acceleration (see Section~\ref{sec:lepton_dist}) to obtain insights on the cooling and acceleration processes at play.

We performed an initial fit of each \gls{sed} independently using a frequentist approach via \texttt{iMinuit} to minimise the $\chi^2$.
Since the parameters for each day were independent of each other, we refer to this as the `parallel fit'. 
The parallel fitting process was repeated a few times iteratively, which allowed us to restrict the fit ranges and to identify parameters that can be reasonably kept constant without significantly degrading the goodness of fit. 
The Doppler beaming factor $\delta$ was fixed to $45$ in agreement with the average value derived from preliminary fits which did not reveal significant temporal variations.
For a small angle of the jet axis with the observer line of sight as is the case for blazars ($< 5^\circ$), $\delta \sim \Gamma$ which makes our choice of Doppler factor consistent with the typical bulk Lorentz factor used in the literature \citep{magic_veritas_sed_2015A&A...578A..22A, mrk421_2017_magic, 2019MNRAS.487..845B}. 

In order to better capture the temporal evolution of the emitting region, we developed the `sequential fit' procedure using the parameter values and ranges of the parallel fit to setup the model as described:
\begin{itemize}
    \item The parameter fit ranges in the sequential fit are set based on the maximum and minimum range of the parallel fit with additional padding of $50 \%$ to avoid convergence on the boundary.
    \item The parallel fit of the first \gls{sed} (8 January 2010) is used as the starting point for the sequential fit, subsequent \glspl{sed} use the best fit values obtained for the preceding day as initial guess, leading to a temporally evolving snapshot model. 
    \item We compute the sequential fit with the \gls{lppl} \gls{eed} ($n_\text{LPPL}$, Eq.~\ref{eq:lppl_define}) as the baseline or reference, and then similarly with pile-up\gls{eed} ($n_\text{pile-up}$, Eq.~\ref{eq:pileup_define}) for each day.
    \item We pass the best fit model for both \glspl{eed} to a \gls{mcmc} sampler (\texttt{emcee}\footnote{\cite{emcee}} interface in \jst ) to get the Bayesian error bars and SED model ranges. 
\end{itemize}

\begin{table}[h!]
    \centering
    \begin{tabular}{ScSrSrSrSc} 
        \hline
        date & \makecell{ \gls{lppl} \\ $\chisq_\text{red}$} & \makecell{ pile-up \\ $\chisq_\text{red}$} & \multicolumn{1}{c}{TS$_\text{pile-up}$} & \makecell{ pile-up scale \\ $f [\times10^{-3}]$ } \\ 
        \hline
        \textcolor{c08}{$\blacksquare$} 08 & $17.41/26$ & $14.01/25$ &  \textcolor{gray}{$ 3.40 $} & $ 29.32^{+3.51}_{-3.96} $  \\
        \textcolor{c11}{$\blacksquare$} 11 & $10.84/26$ & $10.45/25$ &  \textcolor{gray}{$ 0.39 $} & $ 10.57^{+0.97}_{-1.09} $  \\
        \textcolor{c12}{$\blacksquare$} 12 & $ 8.81/23$ & $10.51/22$ &  \textcolor{gray}{$-1.70 $} & $  2.44^{+0.49}_{-0.45} $  \\
        \textcolor{c13}{$\blacksquare$} 13 & $ 8.24/25$ & $ 7.37/24$ &  \textcolor{gray}{$ 0.87 $} & $  7.45^{+0.92}_{-0.98} $  \\
        \textcolor{c14}{$\blacksquare$} 14 & $ 6.87/26$ & $ 5.55/25$ &  \textcolor{gray}{$ 1.32 $} & $  7.64^{+0.85}_{-0.94} $  \\
        \textcolor{c15}{$\blacksquare$} 15 & $13.71/27$ & $ 8.49/26$ & \textcolor{black}{$ 5.22 $} & $  7.70^{+0.62}_{-0.68} $  \\
        \textcolor{c16}{$\blacksquare$} 16 & $13.16/23$ & $10.53/22$ &  \textcolor{gray}{$ 2.63 $} & $ 11.26^{+1.42}_{-1.40} $  \\
        \textcolor{c18}{$\blacksquare$} 18 & $12.39/24$ & $11.83/23$ &  \textcolor{gray}{$ 0.56 $} & $  0.48^{+0.07}_{-0.07} $  \\
        \textcolor{c19}{$\blacksquare$} 19 & $22.61/25$ & $16.03/24$ & \textcolor{black}{$ 6.58 $} & $  5.39^{+0.88}_{-0.92} $  \\
        \textcolor{c20}{$\blacksquare$} 20 & $17.04/29$ & $12.21/28$ & \textcolor{black}{$ 4.83 $} & $  2.43^{+0.36}_{-0.43} $  \\
        \textcolor{c21}{$\blacksquare$} 21 & $ 7.60/25$ & $ 8.44/24$ &  \textcolor{gray}{$-0.84 $} & $  1.66^{+0.26}_{-0.28} $  \\
        \textcolor{c22}{$\blacksquare$} 22 & $10.05/21$ & $ 9.37/20$ &  \textcolor{gray}{$ 0.68 $} & $  0.33^{+0.05}_{-0.06} $  \\
        \textcolor{c23}{$\blacksquare$} 23 & $14.03/23$ & $13.92/22$ &  \textcolor{gray}{$ 0.11 $} & $  0.27^{+0.04}_{-0.04} $  \\
        \textcolor{c24}{$\blacksquare$} 24 & $11.89/24$ & $ 8.90/23$ &  \textcolor{gray}{$ 2.99 $} & $  2.98^{+0.40}_{-0.45} $  \\
        \textcolor{c25}{$\blacksquare$} 25 & $16.74/25$ & $13.92/24$ &  \textcolor{gray}{$ 2.83 $} & $  3.51^{+0.39}_{-0.49} $  \\
        \textcolor{c26}{$\blacksquare$} 26 & $ 7.22/21$ & $ 6.93/20$ &  \textcolor{gray}{$ 0.29 $} & $ 10.15^{+1.25}_{-1.19} $  \\
        \hline
    \end{tabular}
    \caption{Significance of the pile-up vs the \gls{lppl} \gls{eed} model, with the days which have \gls{ts}$>4$ (p-value $< 0.05$) highlighted in darker text. The \gls{ts} is the likelihood ratio of the models. $\chisq_\text{red}$ is the reduced-$\chisq$, where the numerator is the $\chisq$ and the denominator is the number of degrees-of-freedom of the model. The 3 days which satisfy the \gls{ts} threshold for pile-up are marked with a star marker `\ding{88}' on the phenomenology plots in Sections~\ref{sec:phenomenology}, \ref{sec:expanding_blob} and Appendix~\ref{appendix:phenomenology}. }
    \label{tab:significance}
\end{table}

The comparison between the sequential fit results of the \gls{lppl} and pile-up model can be found in Section~\ref{sec:fit_results} and \ref{sec:phenomenology}. 
In the fitting process, we found that on average the emitting region size $R$ had the tendency to expand along with an anti-correlation with $B$ and $N$ (Section~\ref{sec:par_evo}). 
This  expansion is expected if the region travels downstream a jet with conical (or parabolic) shape. 
Motivated by these findings, we performed additional investigation attempting to capture the physics of this expansion. 
This model will be discussed in Section~\ref{sec:expanding_blob}.

\section{Best fit model and results}\label{sec:fit_results}
\subsection{Comparing the LPPL and pile-up models}
\begin{figure}
    \includegraphics[width=0.48\textwidth]{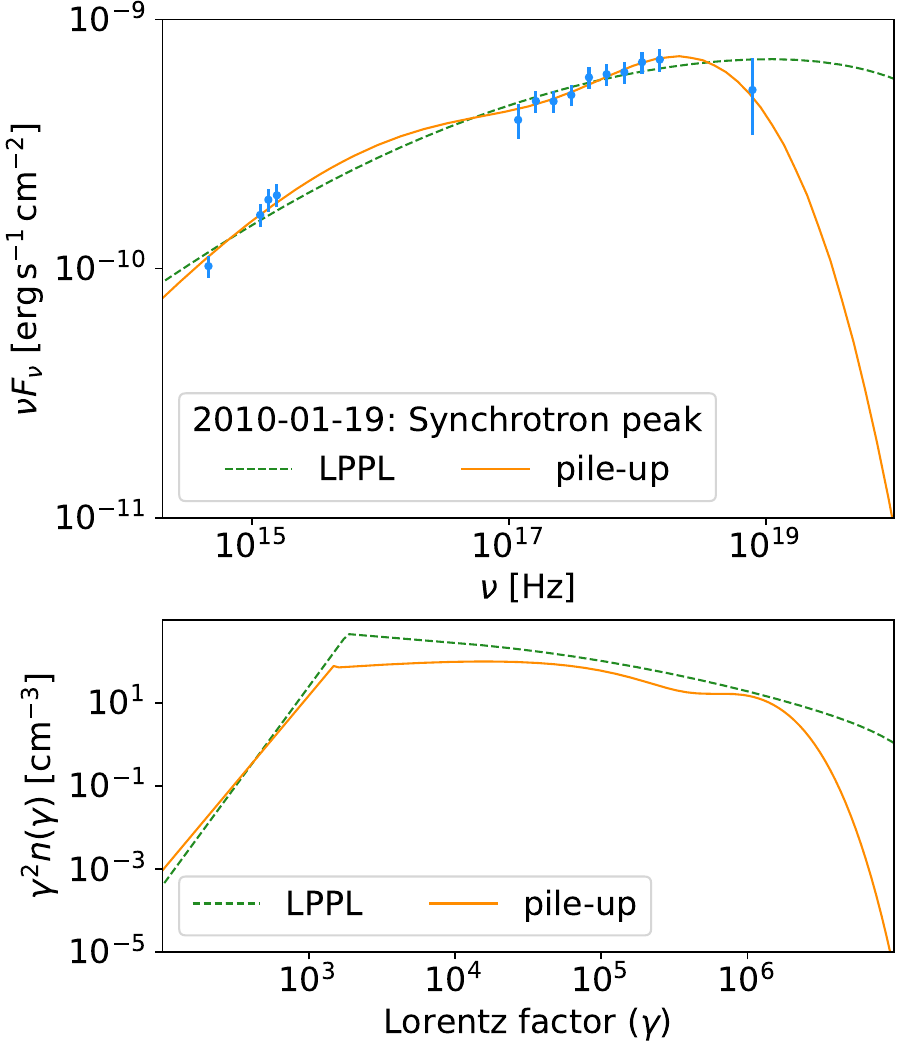}
    \caption{The top panel illustrates the optical-to-X-ray SED using the pile-up (orange line) and LPPL (green dashed line) \gls{eed} on 19 January 2010, the day with the strongest indication of a pile-up (see Section~\ref{sec:fit_results}). The bottom panel presents the corresponding \gls{eed}, where $\ngamma$ is the particle number density distribution. The full frequency range of this \gls{sed} model can be seen in Fig.~\ref{fig:sed_combined_1}(i). } 
    \label{fig:compare_peak_19_jan}
\end{figure}

Using the $\chisq$ of the models with the \gls{lppl} and pile-up \glspl{eed}, we obtained the likelihood $\mathcal{L}$ as $\mathcal{L} \approx \mathrm{e}^{-\frac{1}{2}\chisq}$ and subsequently the \gls{ts} of the pile-up model as $\gls{ts}_\text{pile-up} = -2 \, \ln (\mathcal{L}_\text{LPPL} / \mathcal{L}_\text{pile-up})$. 
Based on our sequential fitting strategy and a $\gls{ts}>4$ cut  (p-value $< 0.05$), we found that the pile-up model was significant on 15, 19 and 20 January 2010.
A comparison of the \gls{ts} of the models can be seen in Table~\ref{tab:significance} including the reduced $\chisq$ and the pile-up scale parameter $f$. 
A major difference appears in the synchrotron peak shape and position, sharply evident on 19 January in Fig.~\ref{fig:compare_peak_19_jan} despite both models having a good fit to the data as indicated by the reduced $\chisq$ (see Fig.~\ref{fig:compare_peak_15_20_jan} for the same comparison for 15 and 20 January).
The \gls{ts} of the pile-up model is positive except on 12 and 21 January, although not always statistically significant. 
Since $n_{\text{pile-up}}$ in Eq.~\ref{eq:pileup_define} has a power-law with an exponential cut-off which has a similar effect as the curvature in $n_\text{LPPL}$ (albeit an energy dependent curvature compared to the constant curvature in the \gls{lppl} case), it can mimic the general shape of the \gls{lppl} distribution and produce similar $\chisq$. 
However, the \gls{lppl} and pile-up distributions appear at different stages of the temporal evolution of the \gls{eed} and on 12 and 21 January, the evolution could be in a strictly acceleration dominated phase better described by the \gls{lppl} distribution.

\begin{figure}[h!]
    \resizebox{0.49\textwidth}{!}{
    \includegraphics[width=\textwidth]{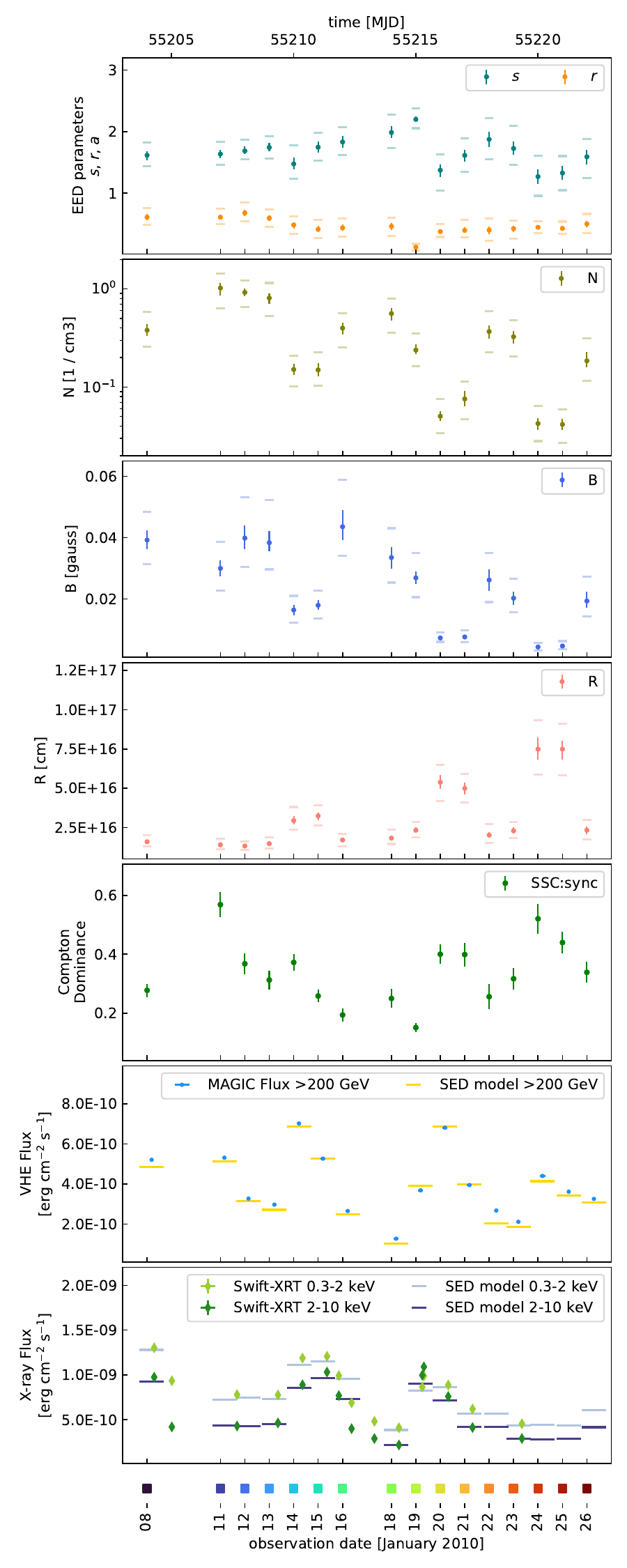}}
    \caption{\gls{lppl} model results. The marker with errorbar represents $[0.15, 0.50, 0.85]$ quantiles of $10^4$ \gls{mcmc} realisations and the short horizontal dashes represent $[0.01, 0.99]$ quantile range. }
    \label{fig:mcmc_par_evolution_lppl}
\end{figure}

The best fit parameter values are reported in Table~\ref{tab:mcmc_fit_par_list}. 
The temporal evolution of the parameters can be seen in Fig.~\ref{fig:mcmc_par_evolution_lppl} and \ref{fig:mcmc_par_evolution_pileup} for the \gls{lppl} and pile-up models respectively. 

\begin{table*} 
   \begin{center}
        \begin{tabular}{ScSrSrSrSrScScScScScSc}
            \hline
            \multicolumn{1}{l}{date} & \multicolumn{1}{c}{$B$}  & \multicolumn{1}{c}{$R$} & \multicolumn{1}{c}{$N$} & \multicolumn{1}{c}{$s$} &  \multicolumn{1}{c}{$r$} & $a$ & $f$ & $\ginj$ & $\gamma_{0}$ & $\gammacut$ \\
            & \multicolumn{1}{c}{$mG$} & \multicolumn{1}{c}{$\times 10^{16} cm$} & \multicolumn{1}{c}{$cm^{-3}$} & & & & \multicolumn{1}{c}{$\times 10^{-3}$} & \multicolumn{1}{c}{$\times 10^3$} & \multicolumn{1}{c}{$\times 10^3$} & \multicolumn{1}{c}{$\times 10^3$} \\
            \hline
            \multicolumn{11}{l}{\gls{lppl}} \\
            \hline
            \textcolor{c08}{$\blacksquare$} 08 Jan & $ 39.4^{+3.0}_{-3.0}$ & $ 1.60^{+0.11}_{-0.12} $ & $ 0.38^{+0.05}_{-0.05} $ & $ 1.62^{+0.07}_{-0.08} $ & $ 0.61^{+0.05}_{-0.05} $ & $1^*$         & $-$        & $ 0.7^{+0.1}_{-0.1} $ & $13.1^{+1.5}_{-1.6} $ & $10^{4*}$  \\
            \textcolor{c11}{$\blacksquare$} 11 Jan & $ 30.0^{+2.6}_{-2.6}$ & $ 1.40^{+0.11}_{-0.11} $ & $ 1.01^{+0.14}_{-0.16} $ & $ 1.64^{+0.07}_{-0.07} $ & $ 0.61^{+0.04}_{-0.04} $ & $1^*$         & $-$        & $ 0.6^{+0.1}_{-0.1} $ & $ 9.7^{+1.4}_{-1.4} $ & $10^{4*}$  \\
            \textcolor{c12}{$\blacksquare$} 12 Jan & $ 40.2^{+3.7}_{-3.8}$ & $ 1.32^{+0.10}_{-0.10} $ & $ 0.92^{+0.09}_{-0.08} $ & $ 1.69^{+0.06}_{-0.06} $ & $ 0.68^{+0.05}_{-0.05} $ & $1^*$         & $-$        & $ 0.5^{+0.0}_{-0.0} $ & $11.7^{+1.2}_{-1.2} $ & $10^{4*}$  \\
            \textcolor{c13}{$\blacksquare$} 13 Jan & $ 38.8^{+3.4}_{-3.3}$ & $ 1.48^{+0.11}_{-0.11} $ & $ 0.81^{+0.10}_{-0.11} $ & $ 1.75^{+0.07}_{-0.06} $ & $ 0.59^{+0.04}_{-0.04} $ & $1^*$         & $-$        & $ 0.5^{+0.0}_{-0.0} $ & $10.3^{+1.2}_{-1.2} $ & $10^{4*}$  \\
            \textcolor{c14}{$\blacksquare$} 14 Jan & $ 16.5^{+1.7}_{-1.7}$ & $ 2.96^{+0.25}_{-0.26} $ & $ 0.15^{+0.02}_{-0.02} $ & $ 1.48^{+0.10}_{-0.10} $ & $ 0.48^{+0.05}_{-0.05} $ & $1^*$         & $-$        & $ 0.9^{+0.1}_{-0.2} $ & $ 8.5^{+1.0}_{-1.0} $ & $10^{4*}$  \\
            \textcolor{c15}{$\blacksquare$} 15 Jan & $ 18.0^{+1.6}_{-1.6}$ & $ 3.23^{+0.24}_{-0.26} $ & $ 0.15^{+0.02}_{-0.02} $ & $ 1.75^{+0.09}_{-0.09} $ & $ 0.42^{+0.05}_{-0.05} $ & $1^*$         & $-$        & $ 1.0^{+0.2}_{-0.2} $ & $12.2^{+1.9}_{-2.1} $ & $10^{4*}$  \\
            \textcolor{c16}{$\blacksquare$} 16 Jan & $ 44.0^{+4.8}_{-4.7}$ & $ 1.70^{+0.15}_{-0.14} $ & $ 0.40^{+0.05}_{-0.06} $ & $ 1.83^{+0.09}_{-0.09} $ & $ 0.44^{+0.05}_{-0.05} $ & $1^*$         & $-$        & $ 0.9^{+0.1}_{-0.1} $ & $ 8.3^{+1.5}_{-1.5} $ & $10^{4*}$  \\
            \textcolor{c18}{$\blacksquare$} 18 Jan & $ 33.5^{+3.5}_{-3.6}$ & $ 1.84^{+0.16}_{-0.16} $ & $ 0.56^{+0.08}_{-0.08} $ & $ 1.99^{+0.09}_{-0.09} $ & $ 0.46^{+0.05}_{-0.05} $ & $1^*$         & $-$        & $ 1.0^{+0.2}_{-0.2} $ & $ 6.5^{+1.2}_{-1.2} $ & $10^{4*}$  \\
            \textcolor{c19}{$\blacksquare$} 19 Jan & $ 26.9^{+1.9}_{-2.1}$ & $ 2.34^{+0.17}_{-0.14} $ & $ 0.24^{+0.03}_{-0.03} $ & $ 2.20^{+0.04}_{-0.04} $ & $ 0.12^{+0.01}_{-0.01} $ & $1^*$         & $-$        & $ 1.9^{+0.1}_{-0.1} $ & $ 3.3^{+0.2}_{-0.2} $ & $10^{4*}$  \\
            \textcolor{c20}{$\blacksquare$} 20 Jan & $  7.4^{+0.6}_{-0.6}$ & $ 5.38^{+0.45}_{-0.41} $ & $ 0.05^{+0.01}_{-0.01} $ & $ 1.36^{+0.10}_{-0.10} $ & $ 0.38^{+0.03}_{-0.03} $ & $1^*$         & $-$        & $ 1.7^{+0.2}_{-0.2} $ & $ 4.3^{+0.6}_{-0.7} $ & $10^{4*}$  \\
            \textcolor{c21}{$\blacksquare$} 21 Jan & $  7.7^{+0.7}_{-0.7}$ & $ 4.98^{+0.36}_{-0.38} $ & $ 0.08^{+0.01}_{-0.01} $ & $ 1.61^{+0.09}_{-0.10} $ & $ 0.40^{+0.04}_{-0.05} $ & $1^*$         & $-$        & $ 1.4^{+0.3}_{-0.3} $ & $ 7.3^{+1.1}_{-1.1} $ & $10^{4*}$  \\
            \textcolor{c22}{$\blacksquare$} 22 Jan & $ 26.2^{+3.3}_{-3.5}$ & $ 2.03^{+0.18}_{-0.18} $ & $ 0.37^{+0.05}_{-0.06} $ & $ 1.88^{+0.13}_{-0.12} $ & $ 0.40^{+0.06}_{-0.07} $ & $1^*$         & $-$        & $ 1.0^{+0.2}_{-0.2} $ & $ 8.2^{+1.5}_{-1.5} $ & $10^{4*}$  \\
            \textcolor{c23}{$\blacksquare$} 23 Jan & $ 20.3^{+2.1}_{-2.2}$ & $ 2.30^{+0.20}_{-0.19} $ & $ 0.33^{+0.04}_{-0.05} $ & $ 1.73^{+0.11}_{-0.11} $ & $ 0.42^{+0.05}_{-0.05} $ & $1^*$         & $-$        & $ 1.0^{+0.2}_{-0.2} $ & $ 5.1^{+0.7}_{-0.8} $ & $10^{4*}$  \\
            \textcolor{c24}{$\blacksquare$} 24 Jan & $  4.4^{+0.5}_{-0.5}$ & $ 7.52^{+0.71}_{-0.71} $ & $ 0.04^{+0.01}_{-0.01} $ & $ 1.27^{+0.12}_{-0.12} $ & $ 0.44^{+0.03}_{-0.04} $ & $1^*$         & $-$        & $ 1.2^{+0.2}_{-0.2} $ & $ 4.2^{+0.6}_{-0.6} $ & $10^{4*}$  \\
            \textcolor{c25}{$\blacksquare$} 25 Jan & $  4.7^{+0.5}_{-0.5}$ & $ 7.47^{+0.56}_{-0.64} $ & $ 0.04^{+0.01}_{-0.01} $ & $ 1.33^{+0.11}_{-0.10} $ & $ 0.43^{+0.04}_{-0.03} $ & $1^*$         & $-$        & $ 1.2^{+0.2}_{-0.2} $ & $ 4.1^{+0.6}_{-0.7} $ & $10^{4*}$  \\
            \textcolor{c26}{$\blacksquare$} 26 Jan & $ 19.8^{+2.5}_{-2.5}$ & $ 2.32^{+0.23}_{-0.25} $ & $ 0.19^{+0.03}_{-0.03} $ & $ 1.59^{+0.12}_{-0.12} $ & $ 0.50^{+0.05}_{-0.05} $ & $1^*$         & $-$        & $ 1.3^{+0.2}_{-0.2} $ & $ 7.9^{+1.0}_{-1.1} $ & $10^{4*}$  \\
            \\
            \hline 
            \multicolumn{11}{l}{Pile-up} \\
            \hline
              \textcolor{c15}{$\blacksquare$} 15 Jan & $ 15.6^{+1.6}_{-1.6}$ & $ 3.56^{+0.32}_{-0.33} $ & $ 0.11^{+0.01}_{-0.01} $ & $ 1.46^{+0.06}_{-0.06} $ & $-$         & $ 0.71^{+0.02}_{-0.02} $ & $ 7.70^{+0.62}_{-0.68} $ & $ 1.2^{+0.1}_{-0.1} $ & $-$        & $ 63.9^{+5.9}_{-6.1}   $ \\
              \textcolor{c19}{$\blacksquare$} 19 Jan & $ 17.5^{+1.3}_{-1.4}$ & $ 3.45^{+0.20}_{-0.20} $ & $ 0.11^{+0.01}_{-0.01} $ & $ 1.69^{+0.06}_{-0.06} $ & $-$         & $ 0.66^{+0.03}_{-0.03} $ & $ 5.39^{+0.88}_{-0.92} $ & $ 1.4^{+0.2}_{-0.2} $ & $-$        & $ 86.9^{+7.5}_{-7.8}   $ \\
              \textcolor{c20}{$\blacksquare$} 20 Jan & $  6.8^{+0.4}_{-0.5}$ & $ 5.81^{+0.39}_{-0.35} $ & $ 0.06^{+0.01}_{-0.01} $ & $ 1.32^{+0.07}_{-0.07} $ & $-$         & $ 0.61^{+0.02}_{-0.02} $ & $ 2.43^{+0.36}_{-0.43} $ & $ 0.9^{+0.1}_{-0.1} $ & $-$        & $ 87.0^{+13.0}_{-12.4} $ \\
            \hline
    \end{tabular} 
    \end{center}
    \caption{\gls{mcmc} sampler results for the models. Pile-up model is only shown for the dates when it was significant (see Section~\ref{sec:fit_results} and Table~\ref{tab:significance})
    The first column represents the colour key for the plots and the observation dates are from January 2010.
    $B$: bulk magnetic field | $R $: size of emission region | $N$: lepton number density | $s$: power-law index | $r$: log-parabolic curvature | $a$: \gls{he} cut-off index | $f$: pile-up scale factor | $\ginj$: lepton injection Lorentz factor | $\gamma_0$: log-parabolic curvature onset | $\gammacut$: \gls{he} cut-off onset. 
    $z=0.0308$ | $\delta=45$ | $^*$: Parameter set to constant for given \gls{eed}. 
    }
    \label{tab:mcmc_fit_par_list}
\end{table*}

\subsection{Combining the LPPL and pile-up models}\label{sec:combined_model}
Based on the results reported in the previous sections, we combined the \gls{lppl} and pile-up models by picking the results from the pile-up model on 15, 19 and 20 January when \gls{ts}$_\text{pile-up} >4$.
This however, requires an additional step, since simply mixing the model results breaks the sequential fit approach. 
The resulting fits in each sequential fit step depends on the prior step's best fit which would not be the case if we just mix the results.
Thus, to properly combine the models, we run our sequential fitting pipeline again with the \gls{eed} changed from \gls{lppl} ($n_\text{LPPL}$, Eq.~\ref{eq:lppl_define}) to pile-up ($n_{\text{pile-up}}$, Eq.~\ref{eq:pileup_define}) for the days when pile-up was significant, carrying over the compatible parameters ($s$ and $\ginj$).
This model will be referred to as the `combined model' henceforth. 

\subsection{SED fits for the combined model}
\begin{figure*}[h!]
    \resizebox{\textwidth}{!}{
        \begin{tabular}{ccc}
            \subfloat[\textcolor{c08}{$\blacksquare$} 08 January 2010]{\includegraphics[width=0.3\textwidth]{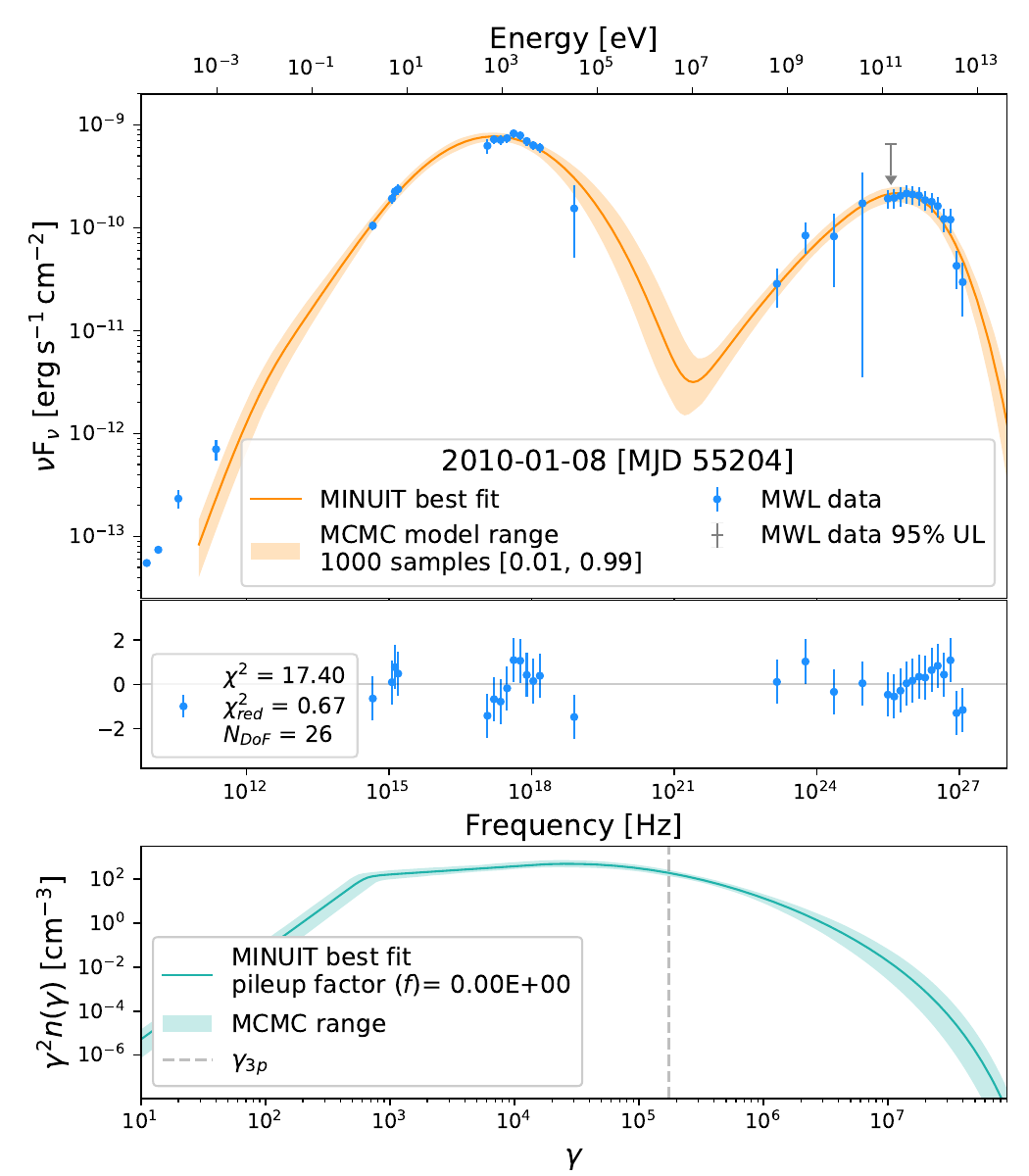}} &
            \subfloat[\textcolor{c11}{$\blacksquare$} 11 January 2010]{\includegraphics[width=0.3\textwidth]{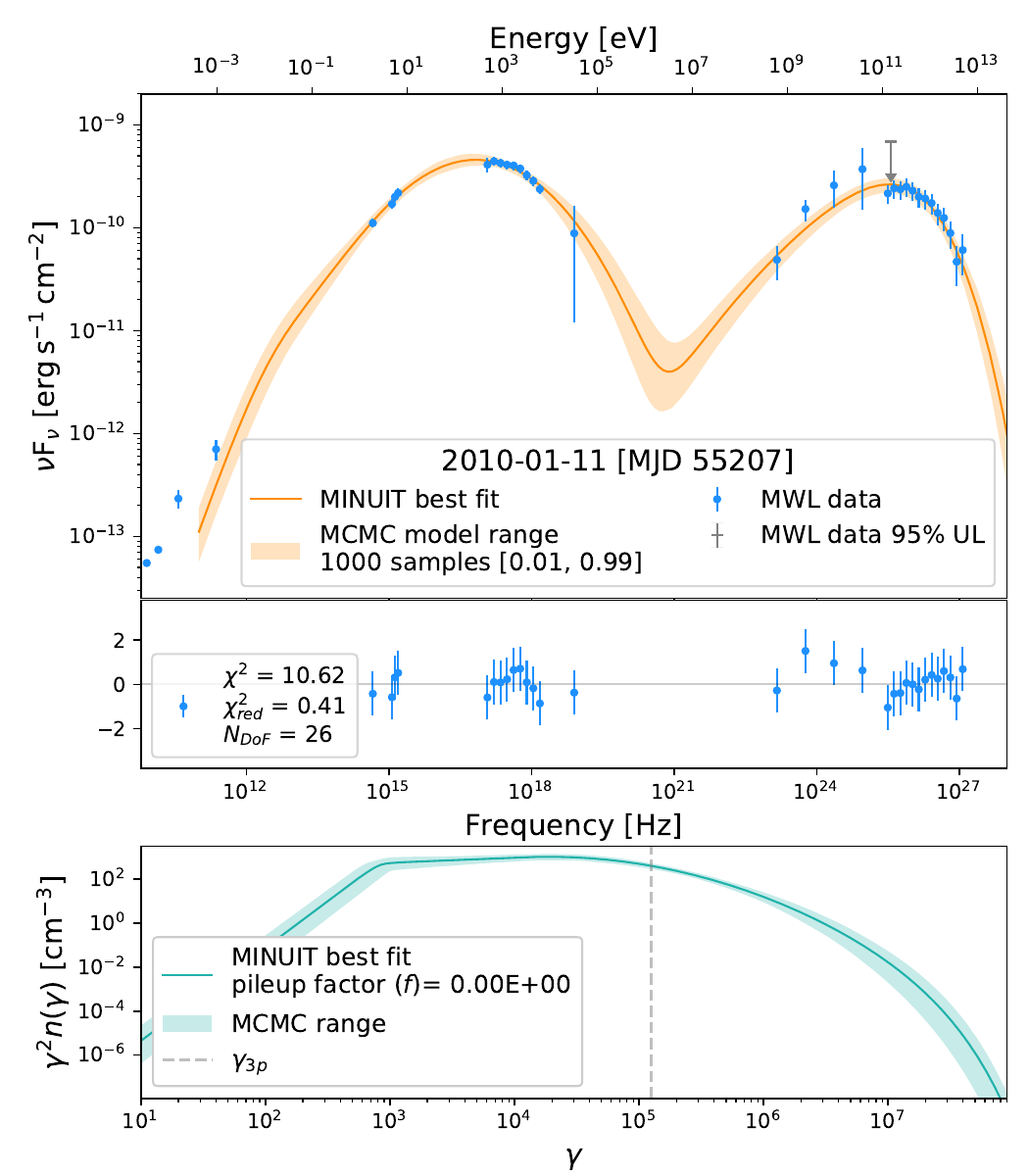}} &
            \subfloat[\textcolor{c12}{$\blacksquare$} 12 January 2010]{\includegraphics[width=0.3\textwidth]{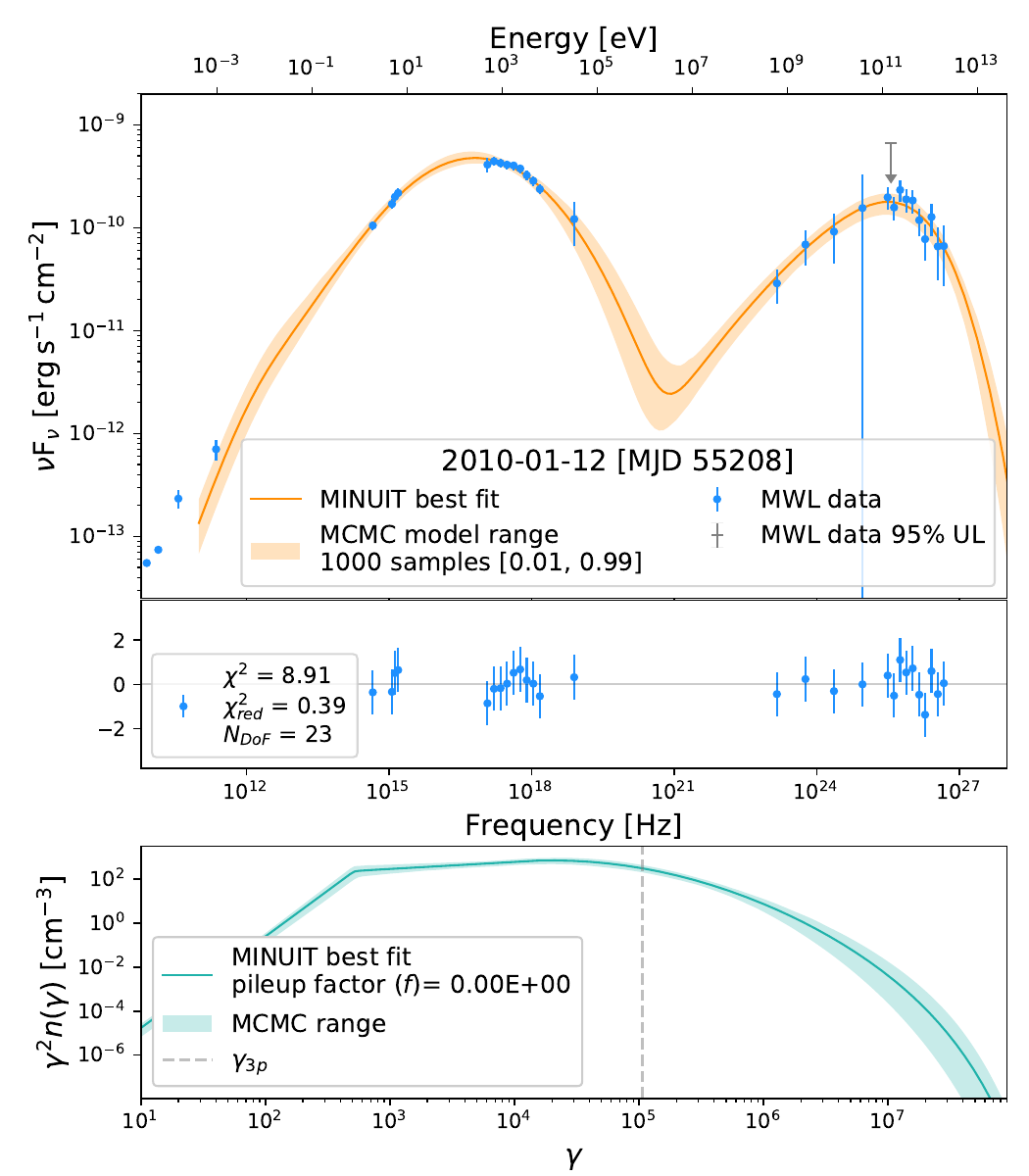}} \\
            \subfloat[\textcolor{c13}{$\blacksquare$} 13 January 2010]{\includegraphics[width=0.3\textwidth]{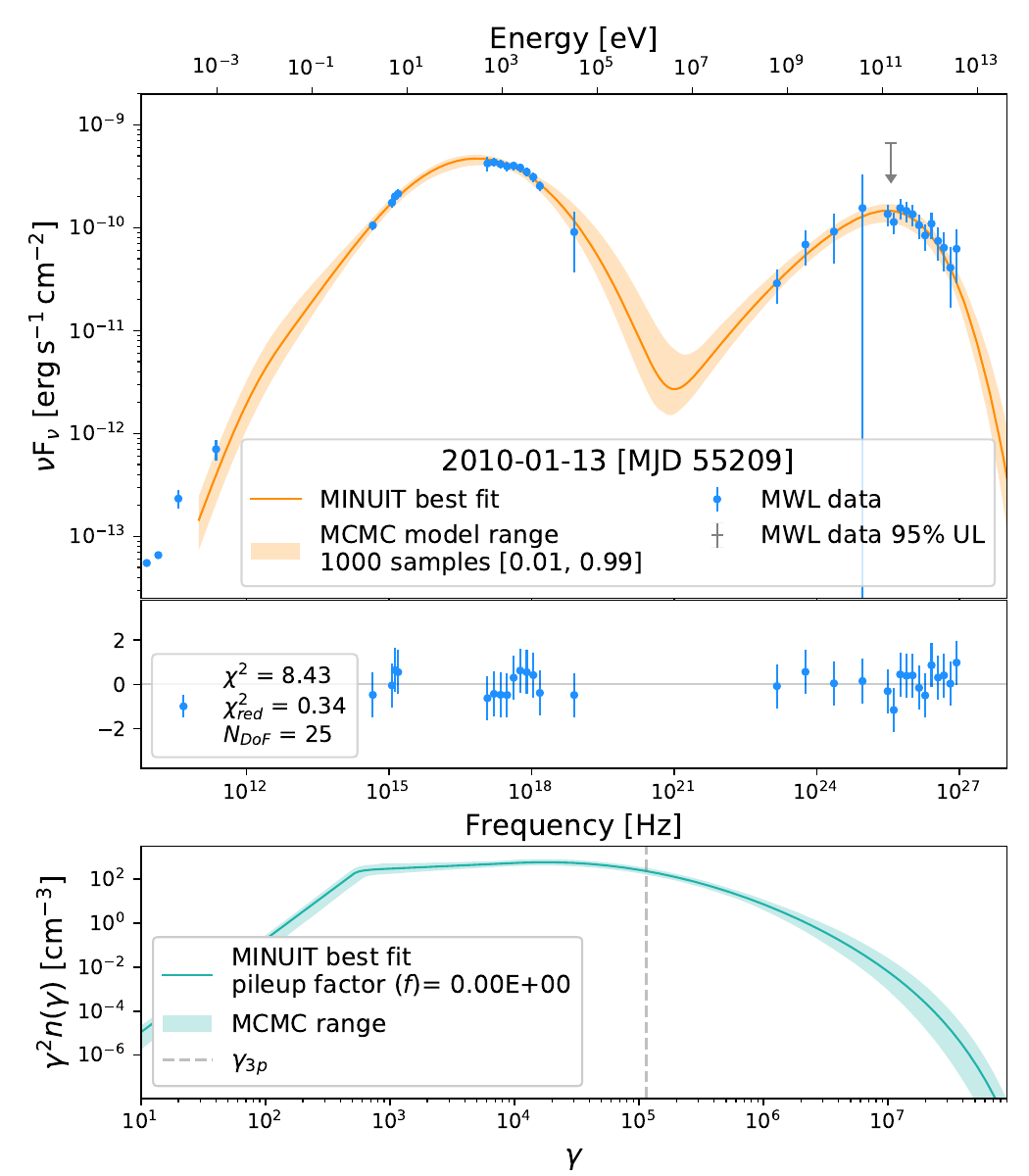}} &
            \subfloat[\textcolor{c14}{$\blacksquare$} 14 January 2010]{\includegraphics[width=0.3\textwidth]{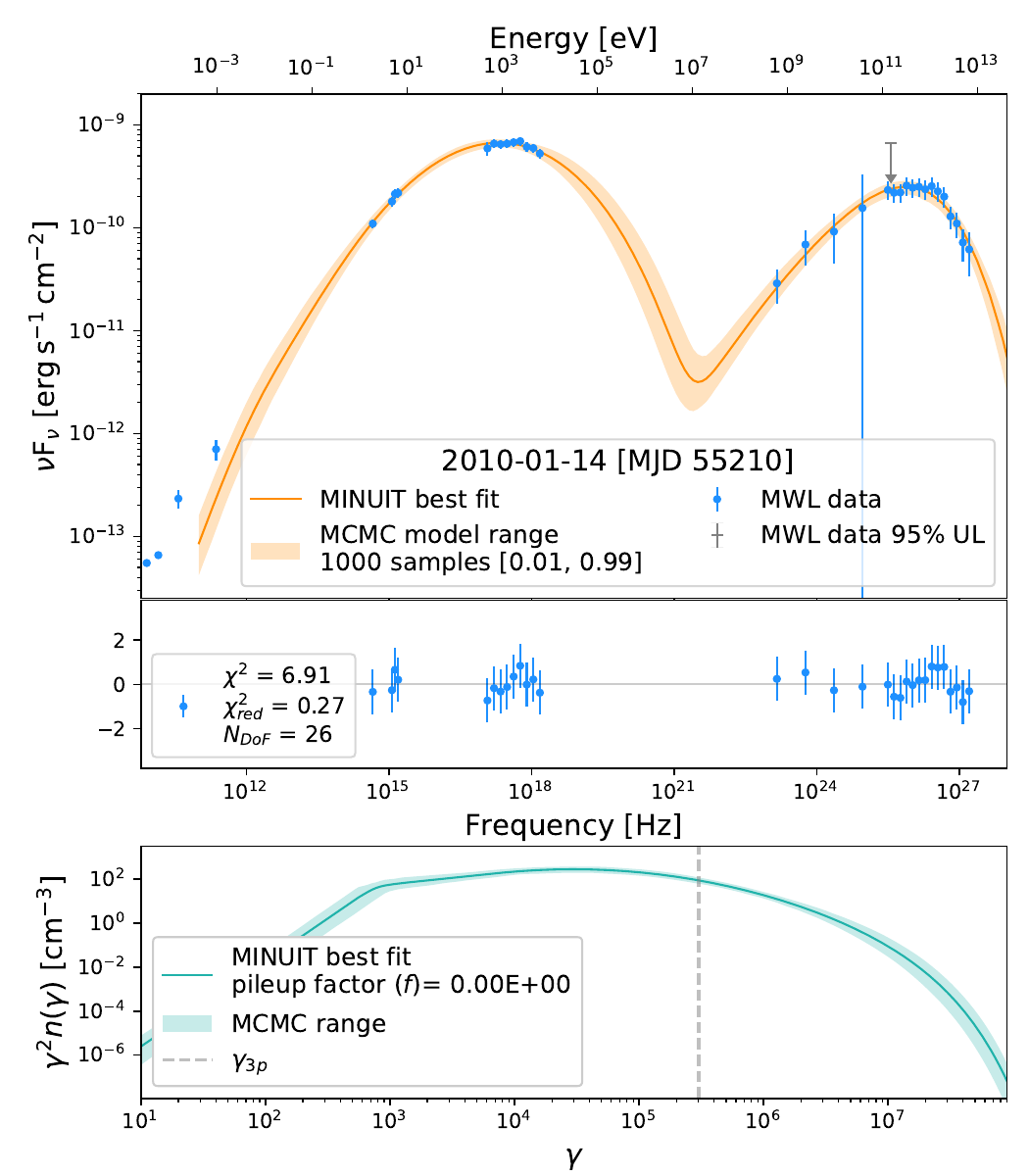}} &
            \subfloat[\textcolor{c15}{$\blacksquare$} 15 January 2010]{\includegraphics[width=0.3\textwidth]{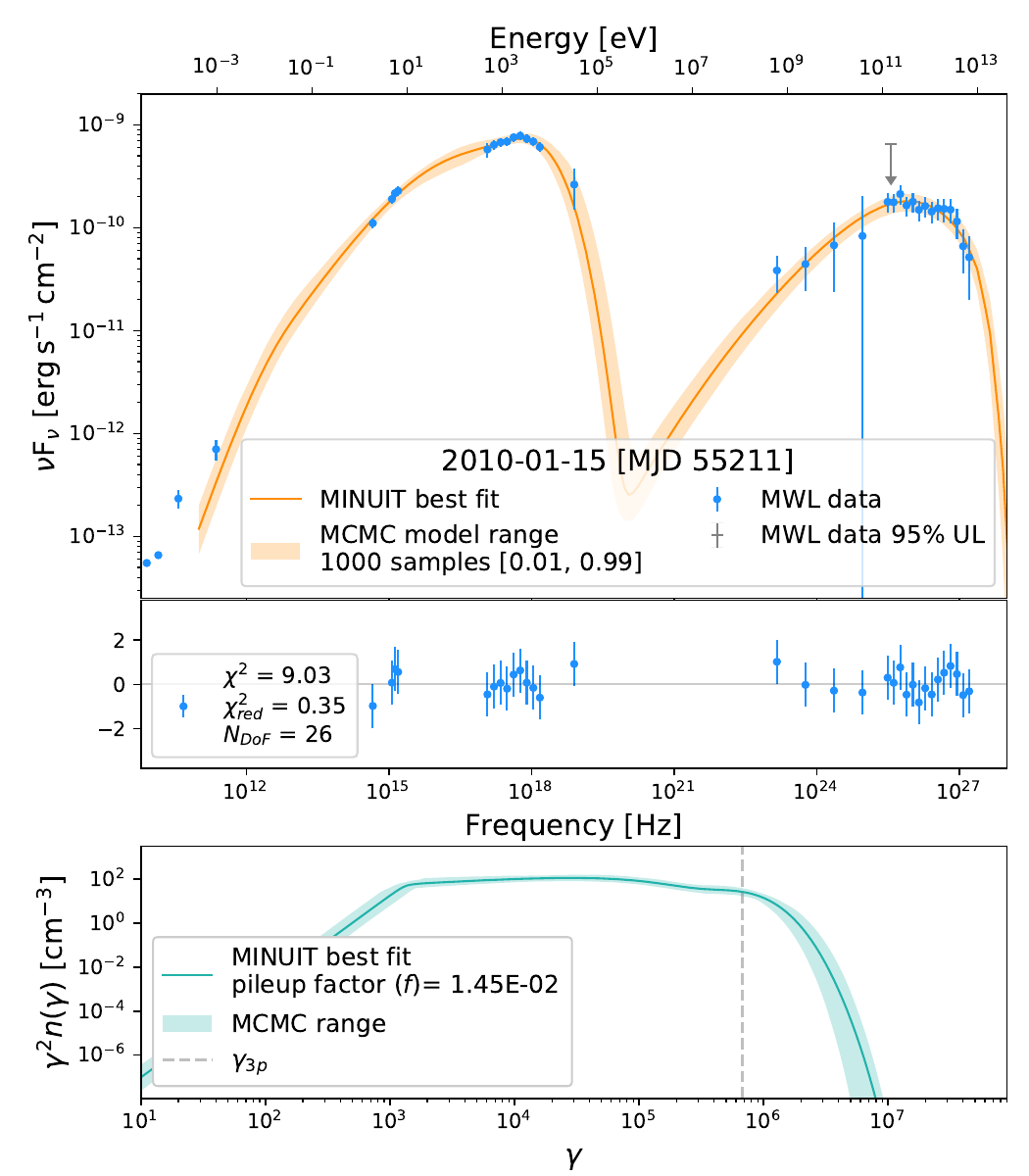}} \\ 
            \subfloat[\textcolor{c16}{$\blacksquare$} 16 January 2010]{\includegraphics[width=0.3\textwidth]{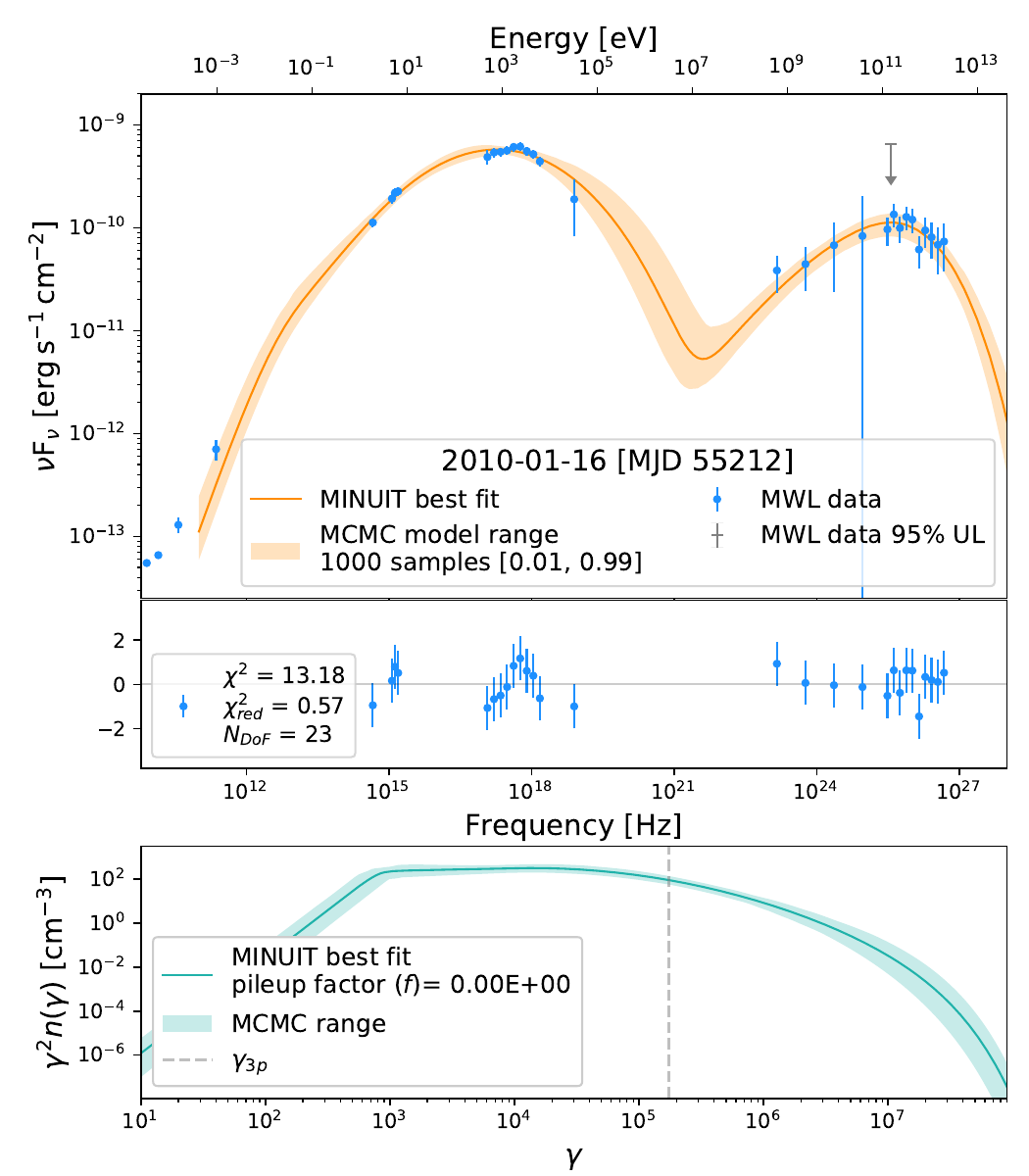}} &
            \subfloat[\textcolor{c18}{$\blacksquare$} 18 January 2010]{\includegraphics[width=0.3\textwidth]{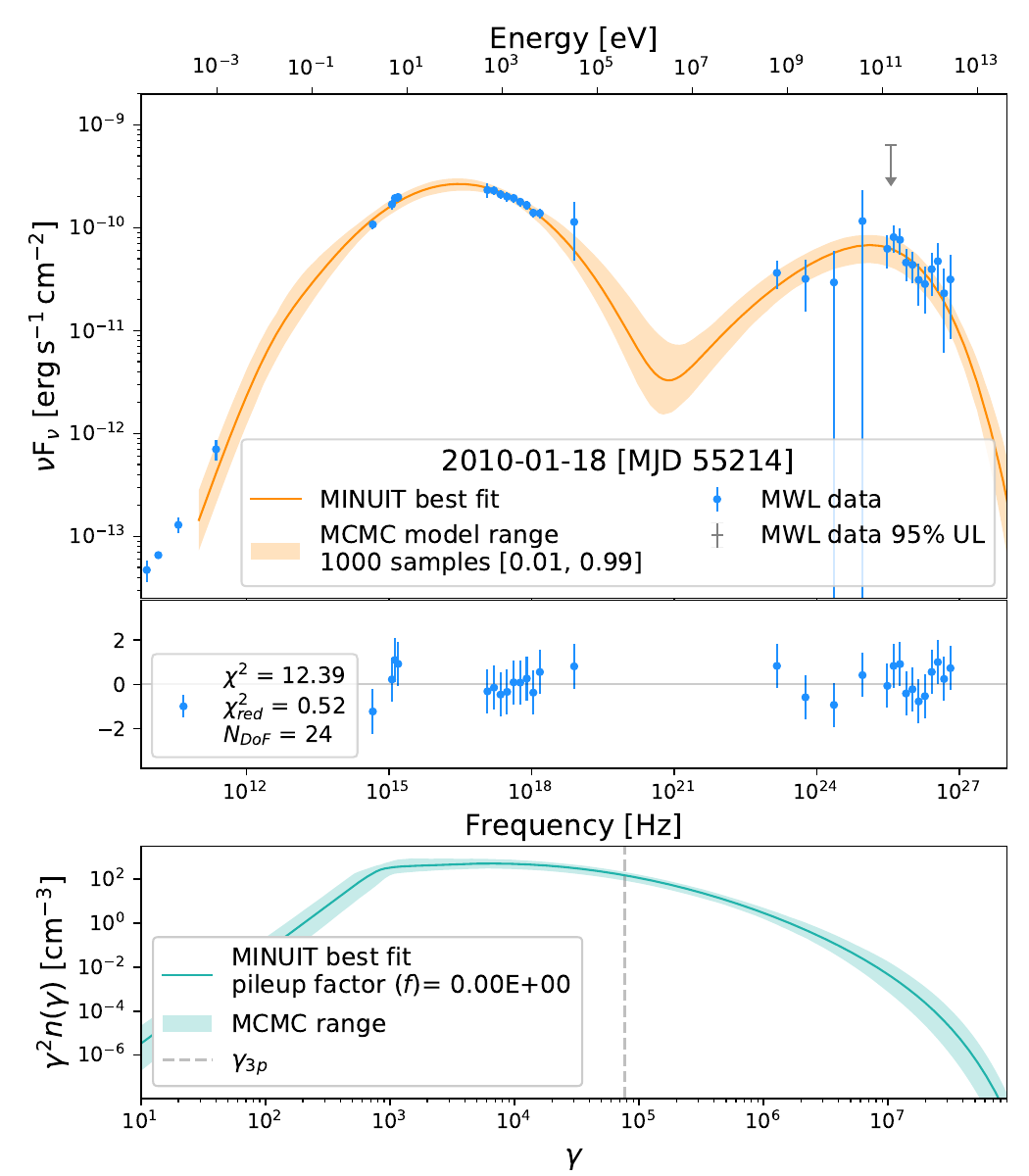}} &
            \subfloat[\textcolor{c19}{$\blacksquare$} 19 January 2010]{\includegraphics[width=0.3\textwidth]{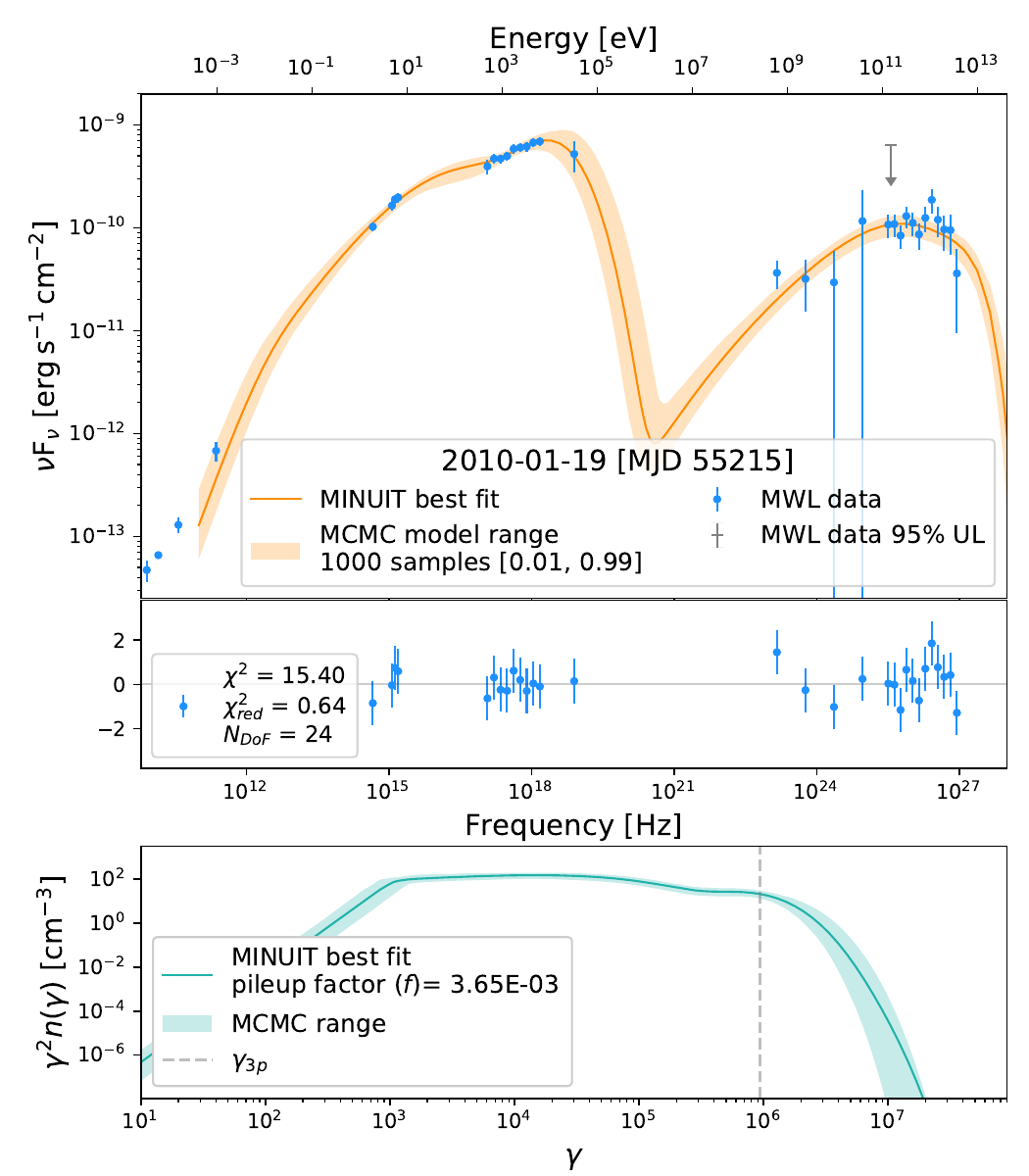}} \\
        \end{tabular}}
    \caption{\gls{sed} fits: combined model. The best fit and \gls{mcmc} range for the \gls{sed} (top subplot) and the \gls{eed} (bottom subplot)  for each observation date in January 2010 are shown. The residuals of the model and the $\chisq$, reduced $\chisq$ ($\chisq_\text{red}$) and degrees-of-freedom (N$_\text{DoF}$) of the fit are mentioned in the subplot underneath the \gls{sed}. 15, 19 and 20 January have the pile-up \gls{eed} $(n_{\text{pile-up}}$, Eq.~\ref{eq:pileup_define}) and the rest are \gls{lppl} $(n_\text{LPPL}$, Eq.~\ref{eq:lppl_define}) as explained in Section~\ref{sec:combined_model}. $\gtp$ is the Lorentz factor of the leptons emitting at the peak of the synchrotron bump (see Section~\ref{sec:Ep_g3p_r3p}). }
    \label{fig:sed_combined_1}
\end{figure*}

\begin{figure*}[h!]
    \resizebox{\textwidth}{!}{
        \begin{tabular}{cccc}
            \subfloat[\textcolor{c20}{$\blacksquare$} 20 January 2010]{\includegraphics[width=0.3\textwidth]{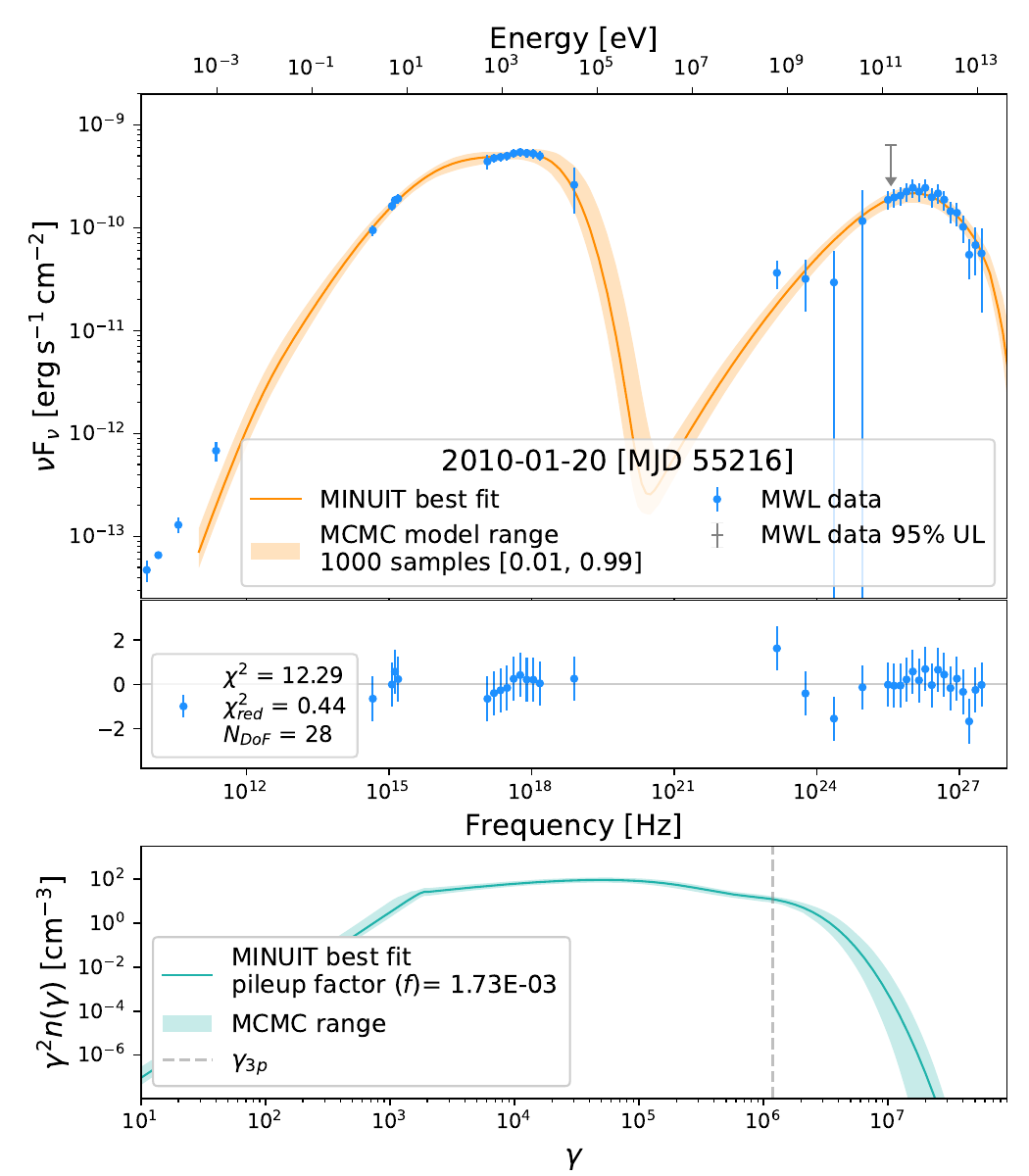}} &
            \subfloat[\textcolor{c21}{$\blacksquare$} 21 January 2010]{\includegraphics[width=0.3\textwidth]{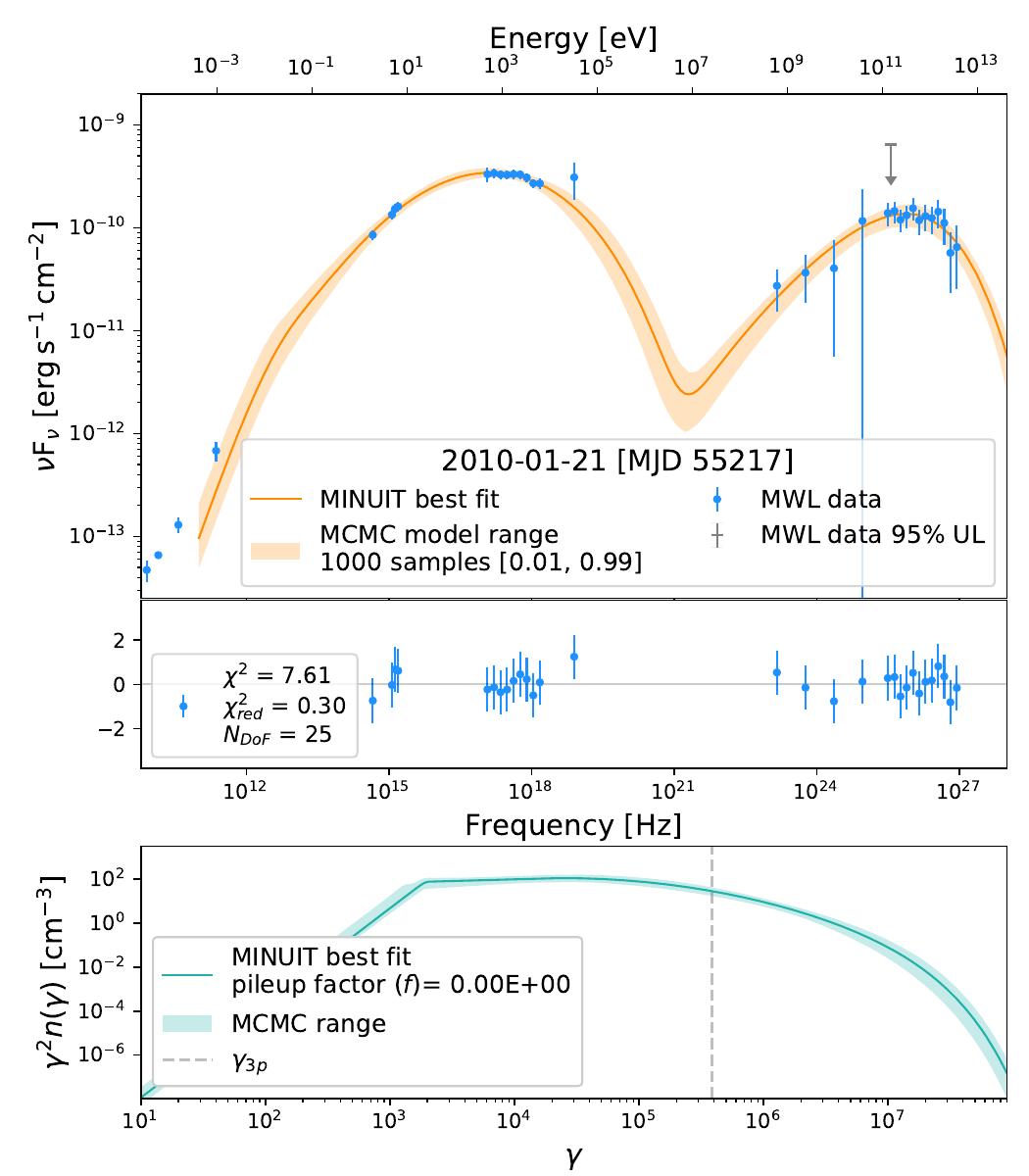}} &
            \subfloat[\textcolor{c22}{$\blacksquare$} 22 January 2010]{\includegraphics[width=0.3\textwidth]{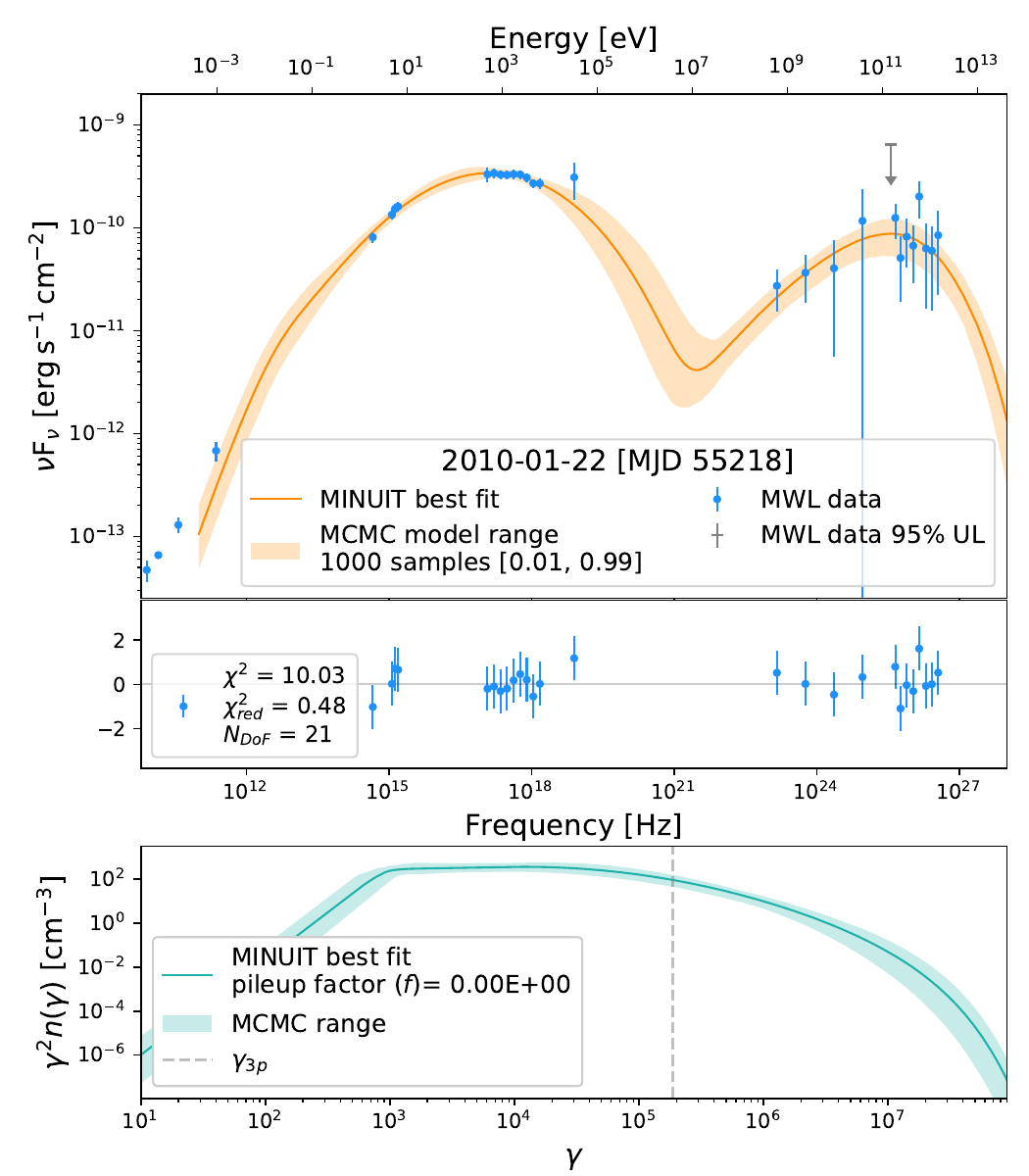}} \\
            \subfloat[\textcolor{c23}{$\blacksquare$} 23 January 2010]{\includegraphics[width=0.3\textwidth]{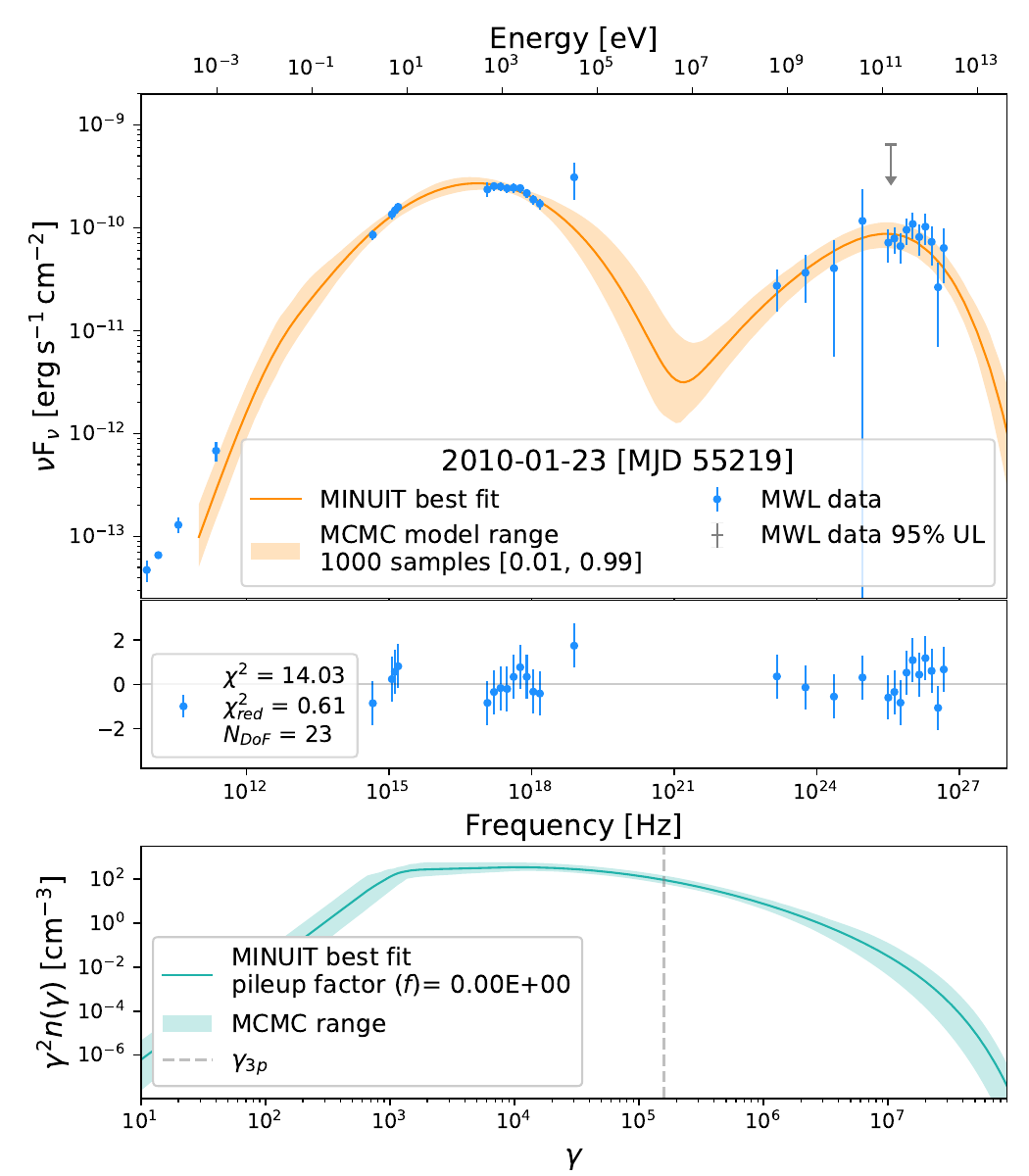}} &
            \subfloat[\textcolor{c24}{$\blacksquare$} 24 January 2010]{\includegraphics[width=0.3\textwidth]{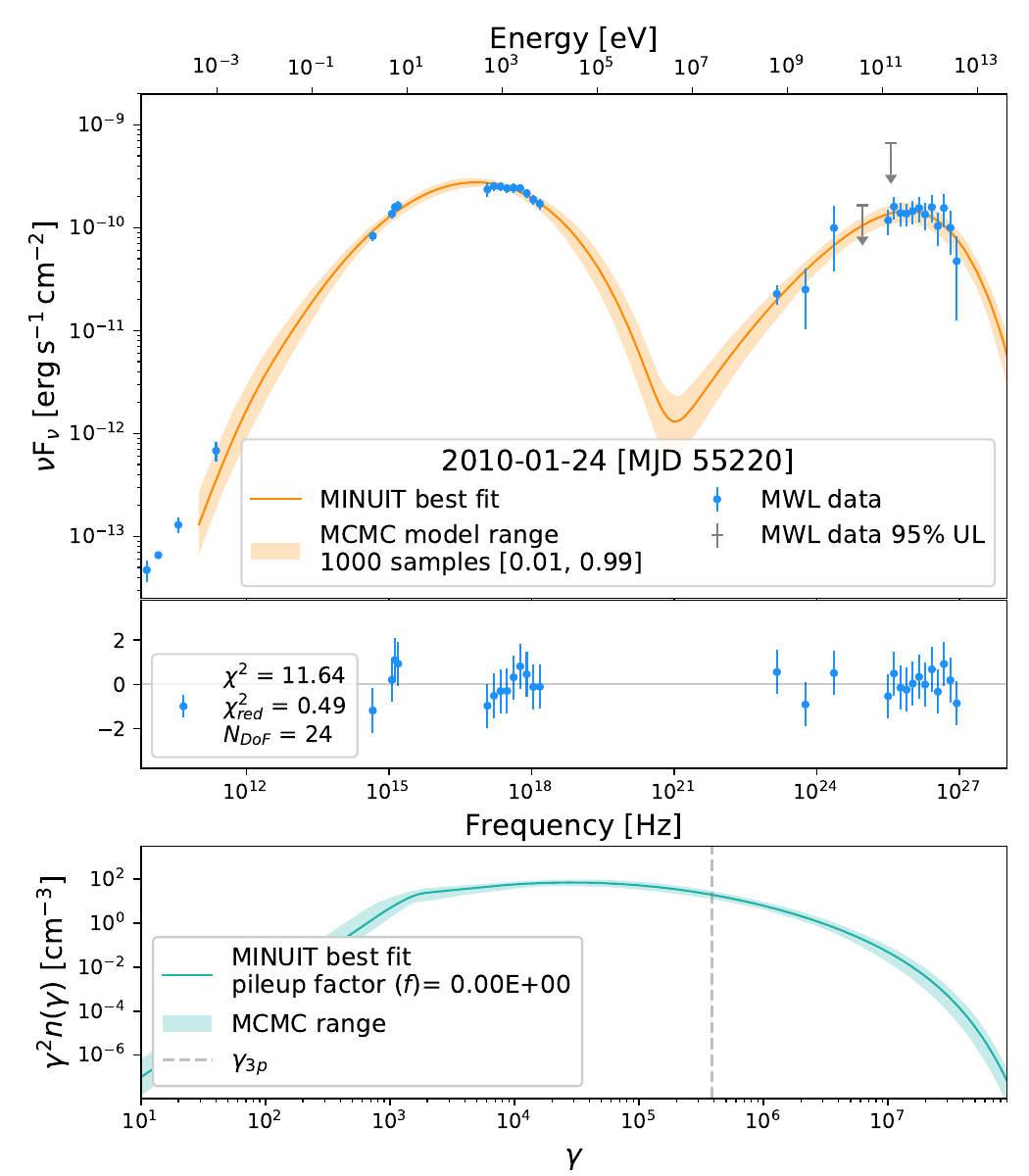}} &
            \subfloat[\textcolor{c25}{$\blacksquare$} 25 January 2010]{\includegraphics[width=0.3\textwidth]{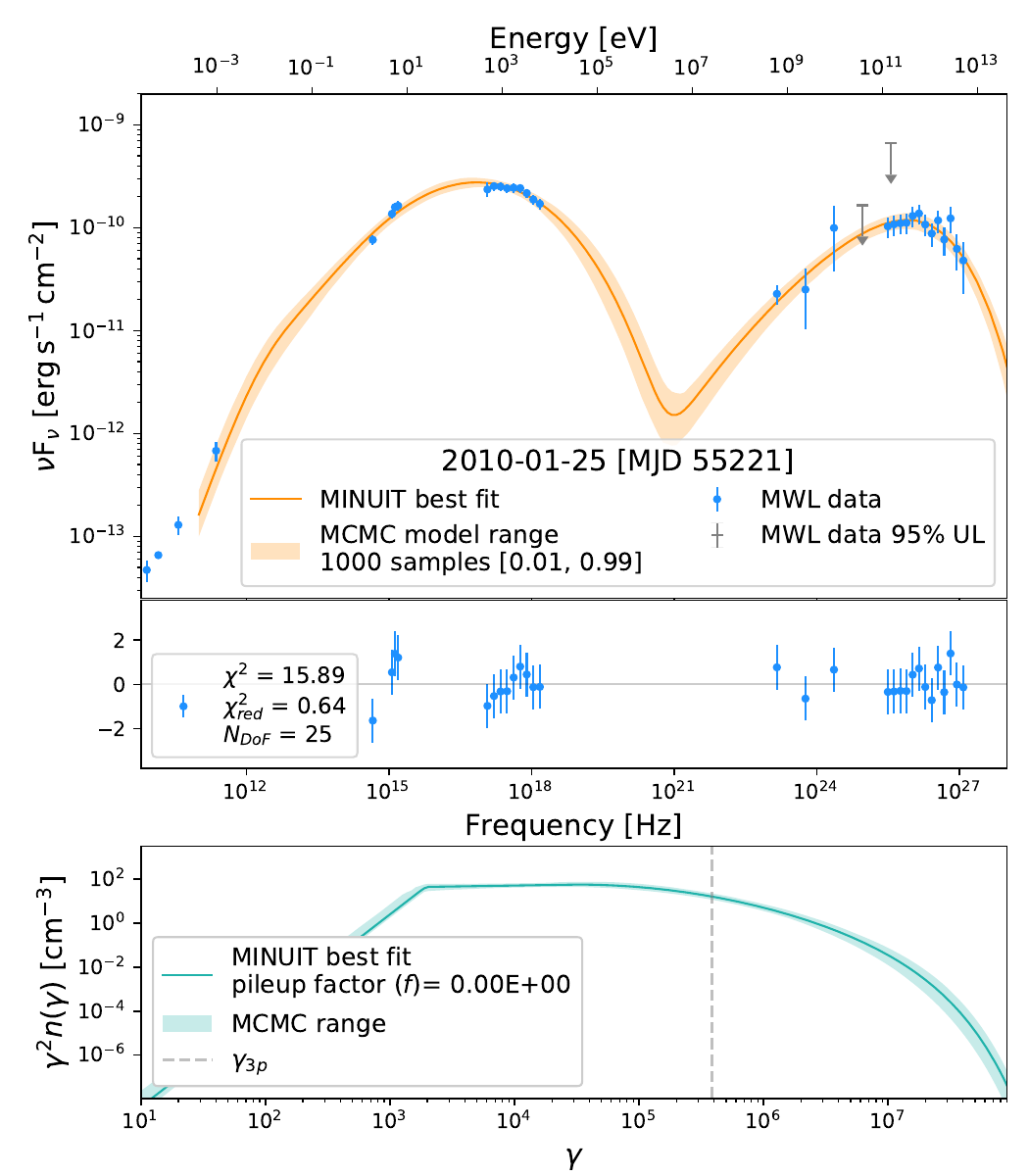}} \\
            \subfloat[\textcolor{c26}{$\blacksquare$} 26 January 2010]{\includegraphics[width=0.3\textwidth]{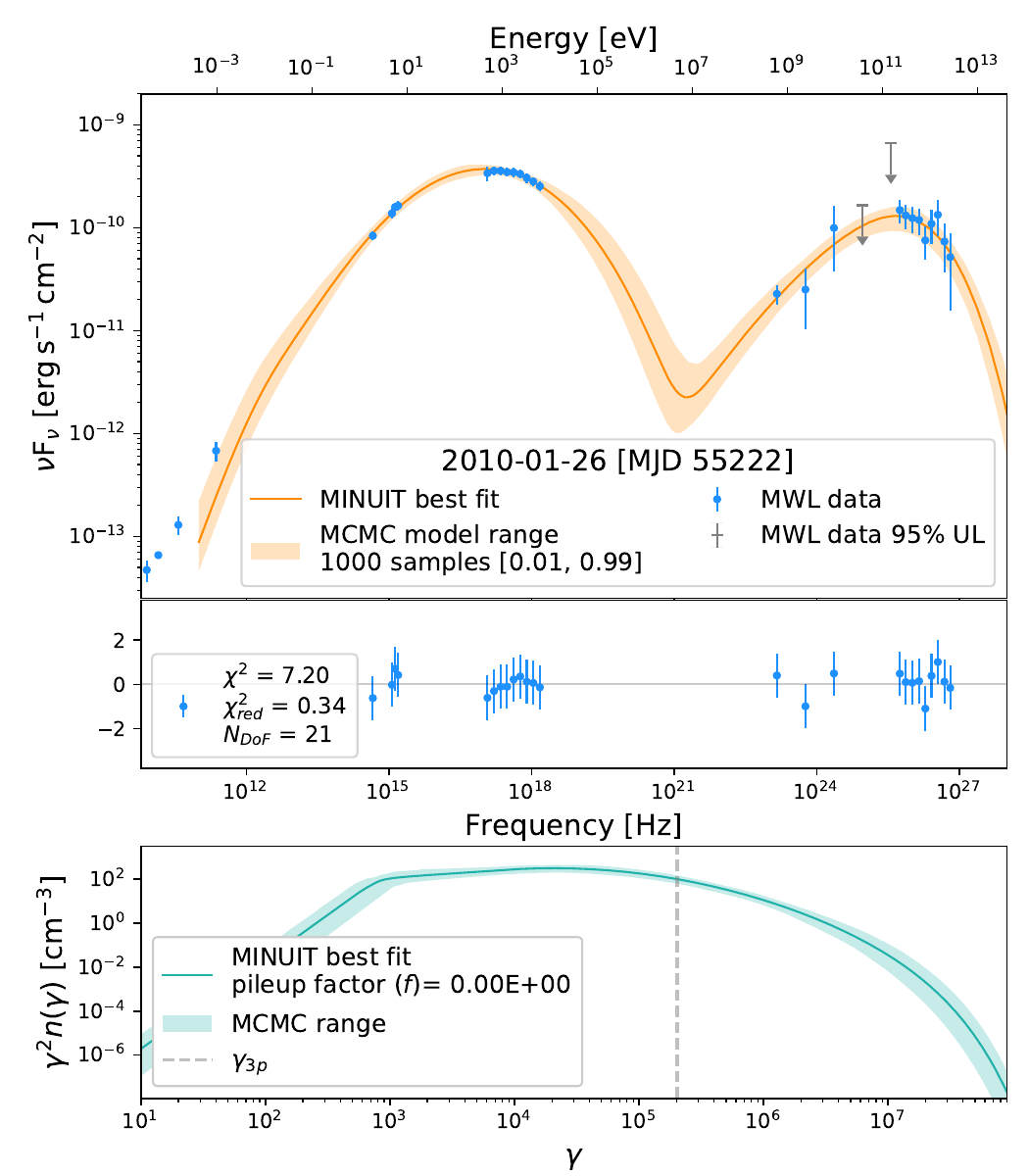}}
    \end{tabular}}
    \caption{\gls{sed} fits: combined model. Continued from Fig.~\ref{fig:sed_combined_1}.}
    \label{fig:sed_combined_2}
\end{figure*}

The Bayesian $98$\% quantile model ranges of the \gls{sed} fits of the combined model can be seen in Fig.~\ref{fig:sed_combined_1} and \ref{fig:sed_combined_2}. 
While the biggest difference in the \gls{lppl} and pile-up models was in the synchrotron peak (Fig.~\ref{fig:compare_peak_19_jan}), pile-up can also appear as a narrow feature in the \gls{vhe} spectrum as hinted by the residuals of 15 and 19 January in panels (f) and (i) of Fig.~\ref{fig:sed_combined_1}.
Narrow features in the \gls{vhe} spectrum have been reported by \cite{mrk501_pileup} for Mrk~501 and hinted for \gls{421} in \cite{magic_veritas_sed_2015A&A...578A..22A} in March 2010, however in our case this effect was small compared to the systematic uncertainty of the \gls{magic} telescopes, requiring further specialised analysis of the data. 
In our analysis, the hint of pile-up is mostly suggested by the X-ray spectra (Section~\ref{sec:xray_data}). 
Although the synchrotron peak frequency of Mrk~501 lies completely outside the \gls{xrt} energy range and the peak shape is not fully sampled by the data, a hint of pile-up was also present in the \gls{bat} spectral point in \cite{mrk501_pileup}. 

\section{Connection between SED modelling, phenomenology and temporal evolution}\label{sec:phenomenology}
In this section, we discuss the associated phenomenology and implications of the time evolving values of observable as well as model parameters, attempting to link the modelling with the theory. 
The time evolution trends and correlations of the model parameters and the expectations based on a stochastic acceleration framework are compared. 
The combined model described in Section~\ref{sec:combined_model} will be used for the plots, with the pile-up states highlighted with a star marker `\ding{88}'. 
Errorbars for all the plots are drawn from $10^4$ \gls{mcmc} realisations. 

\subsection{\texorpdfstring{Defining the parameters: Synchrotron peak frequency, $\gtp$ and curvature $\rtp$}{Defining the parameters: Synchrotron peak frequency and curvature}}
\label{sec:Ep_g3p_r3p}
The motivation to use a stochastic acceleration model was the curvature trend vs the peak frequency seen in Fig.~\ref{fig:data_Ep_bsync}. 
The synchrotron peak frequency depends on the bulk magnetic field ($B$) and beaming factor ($\delta$) as \citep{rybicki_lightman}: 
\begin{align}
    \nusync \propto \gamma_{3p}^2 \cdot \delta \cdot B
    \label{eq:Ep}
\end{align}
where $\gtp$ is the Lorentz factor of the leptons emitting at the peak of the synchrotron bump in the \gls{sed}. 
Mathematically, $\gtp$ is calculated using $\ngt$ distribution, which itself is the distribution of synchrotron photons under the $\delta$-approximation of synchrotron emission \citep{rybicki_lightman} for a given \gls{eed} ($\ngamma$).
This makes $\gtp$ a proxy for $\nusync$, as can be seen in the linear correspondence seen in our model fit in Fig.~\ref{fig:gamma3p_vs_nu_sync} and characterised in detail in \cite{tramacere2009,tramacere2011}. 
Thus, to disentangle the driver of the phenomenology as being changes in the \gls{eed} ($\ngamma$, indirectly $\gtp$) or changes in the physical environment, i.e. $\delta$ and/or $B$, we can study the trend of the curvature vs $\gtp$ instead of $\nusync$. 

Similar to our approach for $\nusync$, we use derived quantities for studying effect of changes in the \gls{eed} on the curvature. 
The quantity $\rtp$ is the synchrotron peak curvature under $\delta$-approximation, allowing us to directly compare our data with the simulations presented in \cite{tramacere2011}. 

\subsection{EED curvature trend - signature of stochastic acceleration}\label{sec:curvature}
\cite{tramacere2011} found that the anti-correlation of spectral curvature and peak energy is a robust prediction of stochastic acceleration. 
In the acceleration dominated phase of temporal evolution of the \gls{eed} from mono-energetic injection to \gls{lppl} to finally pile-up, the curvature $\rtp$ tends to decrease monotonically until it approaches $ \sim 0.5$, and then increases significantly to $ \sim 4$ once pile-up begins \citep{tramacere2011}. 

\begin{figure}[h]
    \centering
    \includegraphics[width=0.48\textwidth]{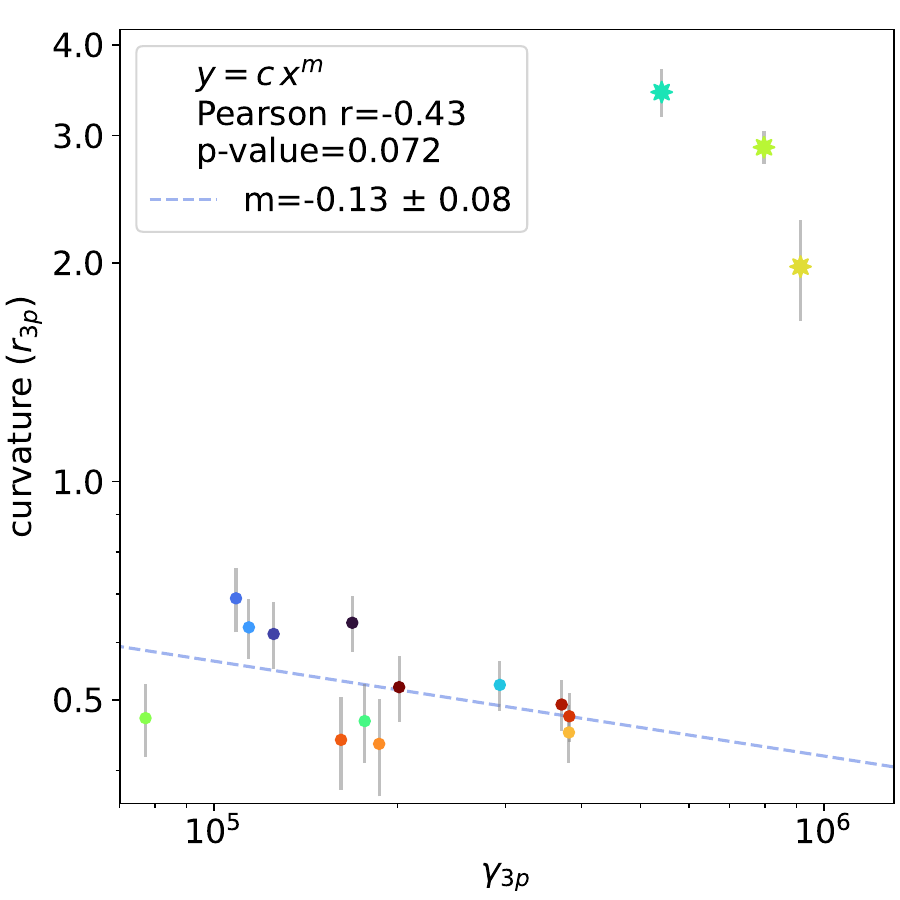}
    \caption{Lepton distribution curvature ($\rtp$) at the Lorentz factor of leptons emitting at the peak of the synchrotron bump ($\gtp$). The fit highlights the trend between spectral curvature and peak energy for the sub-sample of days with \gls{lppl} \gls{eed} in the combined model, with the pile-up states in a separate cluster at much higher curvature as predicted by the theory. The colour key is the same as Table~\ref{tab:mcmc_fit_par_list} and the star markers `\ding{88}' are the pile-up states.}
    \label{fig:r3p_vs_g3p}
\end{figure}

In our dataset, \gls{421} shows the typical phenomenology associated with an emitting population undergoing stochastic acceleration and the \gls{vhe} radiation produced via \gls{ic} process in the \gls{kn} regime during the January 2010 flare. 
We see a cluster of data-points in Fig.~\ref{fig:r3p_vs_g3p} with $\rtp$ around $0.4$ to $0.7$ and an anti-correlation with $\gtp$ as expected in a standard stochastic acceleration scenario. 
The p-value of the Pearson correlation is not statistically robust, however, the sub-sample of data with \gls{lppl} \gls{eed} shows the trend one would expect from the acceleration-dominated stage of stochastic acceleration, that is $\rtp$ decreasing as  $\gtp$ increases, whilst for the pile-up states occurring in cooling-dominated stages, an abrupt increase in curvature is seen. 
As previously explained, when emitters around $\gtp$ are at an energy such that acceleration timescales dominate over cooling timescales, we refer to the evolution as acceleration-dominated and vice-versa. 
We notice that values of $\rtp \approx [2-4]$ for the pile-up states, and the values of $\rtp \lesssim 1$ match those observed in numerical solutions of the momentum diffusion equation, as reported in \cite{tramacere2011}. 
In particular, the pile-up states on 15, 19 and 20 January have $\rtp \sim 3$ which is expected from a well developed pile-up component approaching its asymptotic equilibrium shape and curvature, particularly for the Kraichnan turbulence spectrum scenario described in \cite{tramacere2011}. 
In the simulations presented by the authors in the same paper, the magnetic field strength was a constant which is not the case for our models, hence the slight mismatch in the pile-up curvature, however the general trends still agree well with the numerical predictions. 

It is to be noted that this trend is robust, we obtain the same anti-correlation in the \gls{lppl} and pile-up models, as is expected given that the corresponding physical quantities, $\nusync$ and $\bsync$ are observables and not model dependent (synchrotron bump peak frequency and the corresponding log-parabolic spectral curvature at this frequency respectively).

\subsection{Emergence of the pile-up at LC peaks}\label{sec:pileup}
The X-ray and \gls{vhe} bands sample the peak of the synchrotron and \gls{ic} peaks respectively for \gls{hsp} blazars, thus providing a window into the leptons emitting and \gls{ic} scattering photons to very high energies. 
These highest energy leptons also suffer from the strongest cooling, so any changes in the acceleration-cooling equilibrium are promptly observable without a time lag if the sampling is short enough. 
Some conclusions can already be drawn by just looking at the \glspl{lc} of these bands. 
The \gls{xrt} X-ray \gls{lc} has local peaks on 15 and 19 January (See Fig.~\ref{fig:long_term_lc}, \ref{fig:lc_zoomed}), after which the flux decays, suggesting a change in the acceleration-cooling equilibrium. 

We see significant pile-up on 15, 19 and 20 January which is exactly where one would expect the pile-up to exist given the X-ray and \gls{vhe} \glspl{lc} hinting at a shift in the acceleration-cooling equilibrium. 
There is a small offset in the peaks of the \gls{vhe} \gls{lc} compared to the X-ray dataset, which can be attributed to the additional steps involved in the \gls{ic} process, leading to a time delay for changes in the \gls{eed} to appear in the \gls{sed}. 
\cite{felix_mrk421} did not find significant X-ray - \gls{vhe} band delays in the overall campaign of 2009-2010 but this result is still compatible with the day scale delay we see in January 2010 which is a subset of the larger dataset in that paper (similarly reported in \cite{2015A&A...576A.126A} for the same period and in \cite{2019hepr.confE..32S} for a longer and higher sampling rate \gls{vhe} dataset). 
Additionally, the increased curvature in the synchrotron peak, as discussed in more detail in Section~\ref{sec:curvature} is also a sign of pile-up. 

There is a hint of pile-up on 16 January, similar to the case of 20 January with the pile-up evolving from the previous day but the fit improvement was not above our \gls{ts} threshold. 
However, we still see signs of increased curvature as a signature of pile-up on 16 January visible in the residuals of the model, despite the pile-up model not exceeding our \gls{ts} requirements. 
This could be a result of averaging of states between the cooling dominated pile-up and a fresh acceleration dominated flare. 
The same can be said about 24 and 25 January with weaker improvement (\gls{ts}$\sim 3$) which unfortunately also aligns with a lack of coverage by \gls{xrt} preventing a clear conclusion about the acceleration-cooling transition, but the factor $\sim2$ higher flux peak in the \gls{vhe} band suggests an increased X-ray band flux which could be the reason for the appearance of (weak) pile-up.

\subsection{Peak flux and frequency trends}
Under the $\delta$-approximation of synchrotron emission, the synchrotron peak flux $\nuFnusync$ is related to the synchrotron peak frequency $\nusync$ as follows \citep{rybicki_lightman}:
\begin{equation}
    \nuFnusync \propto n(\gtp) \cdot \gtp^3 \cdot B^2 \cdot \delta^4
    \label{eq:Sp}
\end{equation}
which when combined with Eq.~\ref{eq:Ep} gives the following dependence between the peak flux and peak frequency:
\begin{equation}
    \nuFnusync \propto (\nusync)^m 
    \label{eq:Ep_vs_Sp}
\end{equation}

\begin{figure}
    \centering
    \includegraphics[width=0.48\textwidth]{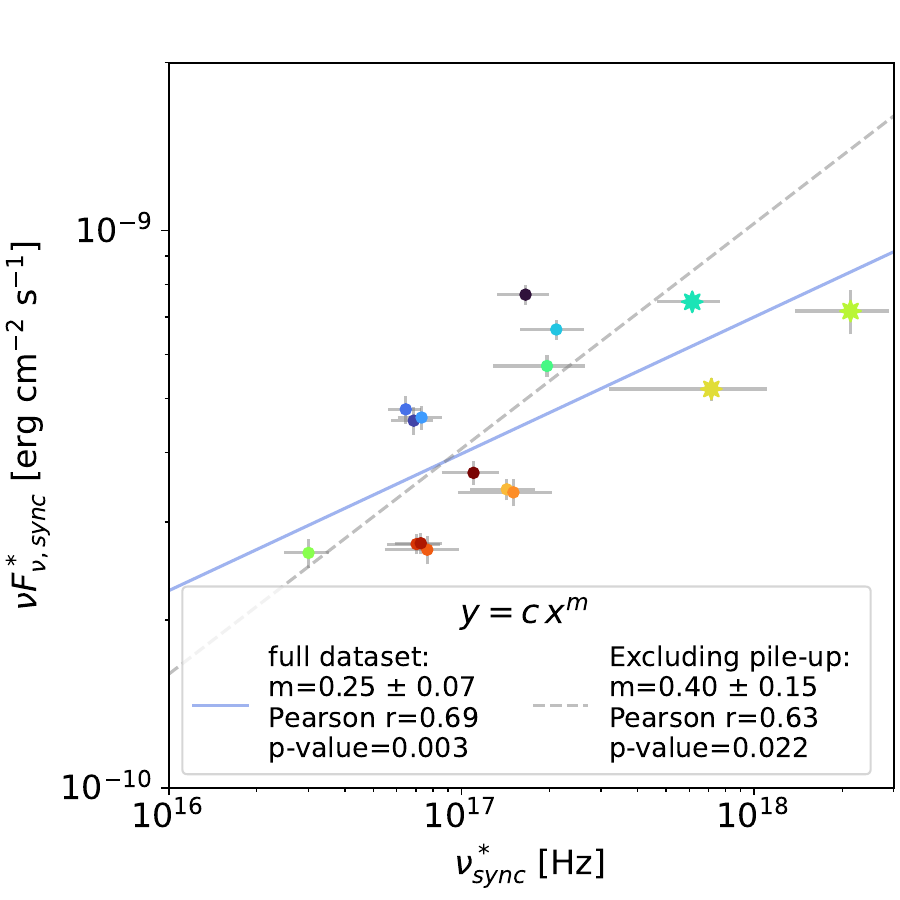}
    \caption{Trend for the synchrotron peak frequency vs peak flux, with the fit done for the overall dataset as well as excluding the pile-up states. Colour key and markers same as Fig.~\ref{fig:r3p_vs_g3p}. }
    \label{fig:sync_peak_Ep_Sp}
\end{figure}

This relationship is shown in Fig.~\ref{fig:sync_peak_Ep_Sp} where we found that $\nuFnusync$ and $\nusync$ show positive correlation with the fits including and excluding the pile-up days producing slightly different power-law indices ($m = 0.25 \pm 0.07$, $0.40 \pm 0.15$). 
This trend is robust and model independent. 
Using log-log polynomial fits to the observational data to estimate the peak flux and frequency also resulted in similar values of index $m$.
This rules out $B$ or $\delta$ as the main drivers of the emission, since then $m$ would have to be $2$ or $4$ respectively \citep{2004A&A...413..489M,tramacere2011}. 
One must note the caveats of this assessment and the snapshot method itself, $\gamma_{3p}$ and $B$ are not independent parameters. 
In a self-consistent temporal evolution approach, the synchrotron cooling timescales ($\propto 1/B^{2}$) and the acceleration timescales depend on the magnetic field \citep[via the momentum diffusion coefficient, see e.g.][]{1989ApJ...336..243S, stawarz2008}, thus affecting the \gls{eed} and $\gtp$. 
These dependencies have been characterised in detail in \cite{tramacere2011} for stochastic acceleration. 
The authors conclude that the index $m=2$ expected from the $\delta$-approximation assuming $B$ to be an independent driver hold in simulations of self-consistent temporal evolution for values of $B<0.2$G which is the case for our dataset. 
A statistical analysis of the long-term historic data of \gls{421} taking the covariance of the parameters determining $\nuFnusync$ and $\nusync$ can be found in \cite{tramacere2007_mrk421_xray} where the authors found that covariance of the parameters has a relatively minor effect on the trend.

The slope of peak flux ($\nuFnusync$) vs peak frequency ($\nusync$) trend depends on the spectral state of the source and our results are consistent with previous studies done using completely different datasets. 
Over a 6\,month long period in 2017, \cite{mrk421_2017_magic} found an index of $0.25 \pm 0.02$ with average flux levels typical of \gls{421} and short duration outbursts during this epoch. 
This is compatible with our result of $m=0.25$ despite 2010 being a particularly exceptional high state of the source. 
\cite{tramacere2007_mrk421_xray} found $m=0.54$ for a much larger X-ray dataset collected prior to 2006. 

\subsection{Emission region parameter evolution}\label{sec:par_evo}
Given the degeneracy between the parameters of the emission zone ($N$, $B$ and $R$ in our simple one zone scenario), it is possible to fit the \gls{sed} with different values of these parameters. 
From the temporal evolution of the emission zone parameters of the combined model (Fig.~\ref{fig:mcmc_par_evolution_lppl} and \ref{fig:mcmc_par_evolution_pileup}), we see a tendency for anti-correlation between the radius of the region ($R$) and the Bulk magnetic field strength ($B$) and the particle density ($N$) with a correlation coefficient of $-0.86$ and $-0.78$ respectively, considering the temporal evolution of the parameters as independent time-series. 
This anti-correlation was independent of the choice of the \gls{eed} with only minor differences in the values of the co-efficients for the \gls{lppl} ($-0.84$ and $-0.77$) and pile-up ($-0.94$ and $-0.78$) models. 

We computed the jet power assuming an equal number of relativistic electrons and cold protons in the emitting region, using the prescription in \cite{2010MNRAS.402..497G}. 
We obtain results consistent with the expected range for \acrshort{bllac} objects and with the results for \gls{421} in \cite{2010MNRAS.402..497G}. 
The jet power throughout January is in the order of $10^{44}$ erg $s^{-1}$ (radiative power $\sim 10^{42}$ erg $s^{-1}$) which is much smaller compared to the Eddington luminosity of $2\times10^{46}$ erg $s^{-1}$ for \gls{421} using a \gls{smbh} mass of $\sim 2\times10^8 \text{M}_\odot$ \citep{2003ApJ...583..134B}.
The total Jet power fluctuates by a factor of 5 in the combined model, however, given the constraints imposed by fixing the beaming factor $\delta$ on the Jet Power, we cannot make strong conclusions. 
Jet power will be investigated in more detail in a future work.

\cite{tramacere2022} found that the Compton dominance of singular flaring events decreases as time progresses for an adiabatically expanding emission region and this trend can be seen in evolution of Compton dominance in Fig.~\ref{fig:mcmc_par_evolution_lppl} between the decaying edges of flares in the \gls{vhe} \gls{lc}.

In isolation, the sequence of expansion and contraction of $R$  seen in Fig.~\ref{fig:mcmc_par_evolution_lppl} could be interpreted as a re-confinement of the jet \citep{2021A&A...649A.153C, 2015ApJ...809...38M}. On the other hand, the degeneracy between $B$, $N$ and $R$ also allows for a scenario where the emitting region undergoes a steady expansion overtime. Expansion is expected if the region travels downstream a jet with conical (or parabolic) shape, as predicted by simulations of jet geometry close to the \gls{smbh} with  $B$ and $N$ decreasing as the emission region propagates along the jet. 
To test the idea of emission region expansion, we modified our sequential fitting pipeline into what we would refer to as the `expanding blob' model henceforth, discussed in Section~\ref{sec:expanding_blob}, assuming a simplified scenario where we only have a constant expansion velocity.

\section{Expanding blob model}\label{sec:expanding_blob}
As described in Section~\ref{sec:par_evo}, the trends for the emission region parameters and the Compton dominance hint at an expansion over time. 
This tendency towards expansion was also seen in the preliminary models with unconstrained beaming factor using the parallel fitting scheme.
To test the idea of expansion, we modified our sequential fitting pipeline to build the `expanding blob' model. 

\subsection{Model setup and fit results}
For the expanding blob model, we followed the sequential fit approach described in Section~\ref{sec:model_fitting} with the fitting range of $R$ for the $(i+1)^{th}$ \gls{sed} additionally constrained by the best fit for the $(i)^{th}$ \gls{sed} in the sequence as follows: 
\begin{equation}
    R_{(i+1)} \in [R_{(i)}, \,\, f_\text{exp} \times R_{(i)}]
\end{equation}
with the factor $f_\text{exp} \sim 1.2$ obtained from a linear fit to the temporal evolution of $R$ in the combined model, while still ensuring that the minimiser doesn't converge on the fit boundaries.

\begin{figure}
    \resizebox{0.5\textwidth}{!}{
    \includegraphics[width=0.5\textwidth]{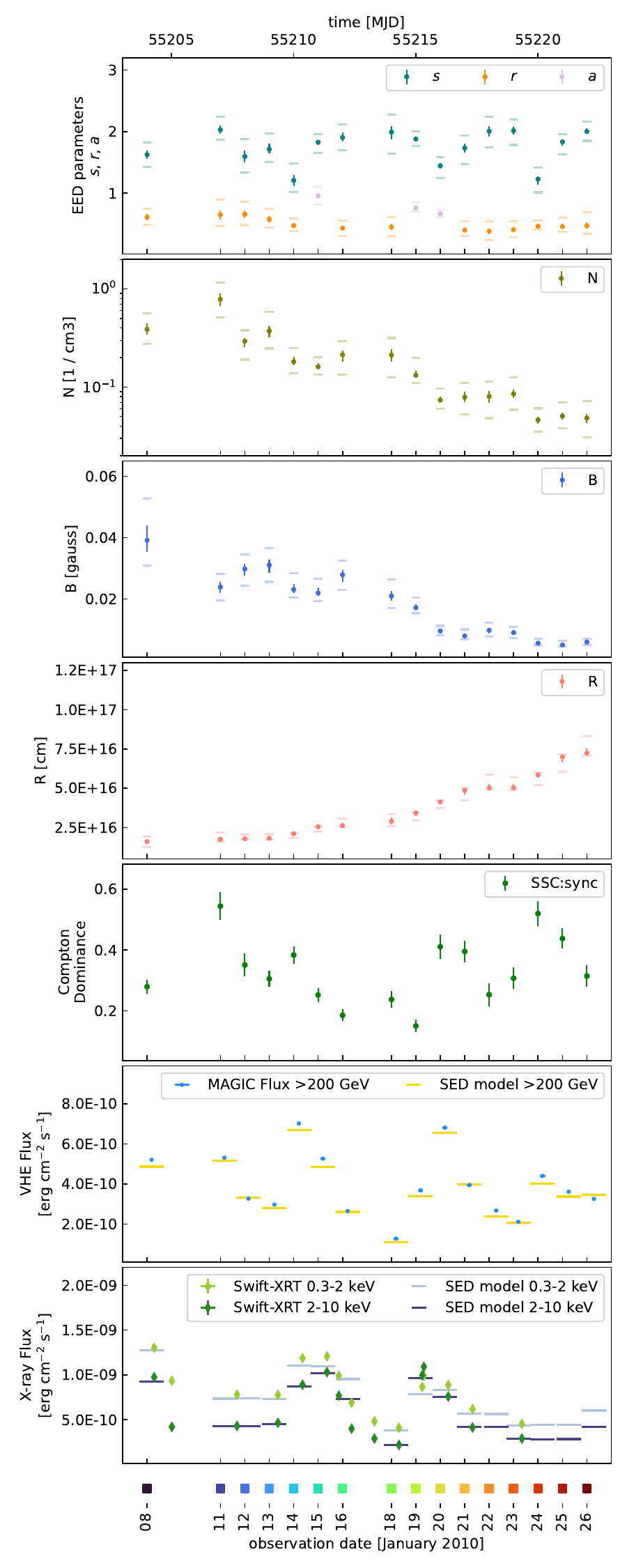}}
    \caption{Parameter evolution for the expanding blob model. Same as Fig.~\ref{fig:mcmc_par_evolution_lppl}.}
    \label{fig:mcmc_par_evolution_exp_blob}
\end{figure}

The resulting parameter values of the expanding blob model can be seen in Table~\ref{tab:mcmc_fit_par_list_exp_blob} and Fig.~\ref{fig:mcmc_par_evolution_exp_blob}. 
Since the constraints on the boundary of $R$ were applied to the combined model as base, 15, 19 and 20 January have a pile-up \gls{eed} while the rest have the \gls{lppl} \gls{eed}, as is apparent from the corresponding rows in Table~\ref{tab:mcmc_fit_par_list_exp_blob}. 
$B$ and $N$ show a consistent decrease as would be expected from falling density through the expansion while $s$, $r$ and $\ginj$ (the \gls{eed} parameters) remain in agreement with the \gls{lppl} and pile-up models which had no expansion constraints on $R$. 
We also studied the same phenomenological trends and correlations described for the combined model in Section~\ref{sec:phenomenology} for the expanding blob model, with the results consistent with the non-expanding models. 
The signature of stochastic acceleration for the expanding blob can be seen in Fig.~\ref{fig:r3p_vs_g3p_exp_blob}. 

\begin{table*} 
    \begin{center}
        \begin{tabular}{ScSrSrSrSrScScScScScSc}
            \hline
            \multicolumn{1}{l}{date} & \multicolumn{1}{c}{$B$}  & \multicolumn{1}{c}{$R$} & \multicolumn{1}{c}{$N$} & \multicolumn{1}{c}{$s$} &  \multicolumn{1}{c}{$r$} & $a$ & $f$ & $\ginj$ & $\gamma_{0}$ & $\gammacut$ \\
            & \multicolumn{1}{c}{$mG$} & \multicolumn{1}{c}{$\times 10^{16} cm$} & \multicolumn{1}{c}{$cm^{-3}$} & & & & \multicolumn{1}{c}{$\times 10^{-3}$} & \multicolumn{1}{c}{$\times 10^3$} & \multicolumn{1}{c}{$\times 10^3$} & \multicolumn{1}{c}{$\times 10^3$} \\
            \hline
            \textcolor{c08}{$\blacksquare$} 08 Jan & $ 39.7^{+4.2}_{-4.2}$ & $ 1.59^{+0.14}_{-0.14} $ & $ 0.39^{+0.05}_{-0.05} $ & $ 1.62^{+0.07}_{-0.07} $ & $ 0.61^{+0.05}_{-0.05} $ & $1^*                   $ & $-                     $ & $ 0.7^{+0.1}_{-0.1} $ & $ 13.3^{+1.7}_{-1.9} $ & $10^{4*}              $  \\
            \textcolor{c11}{$\blacksquare$} 11 Jan & $ 23.7^{+1.8}_{-1.9}$ & $ 1.78^{+0.11}_{-0.11} $ & $ 0.79^{+0.11}_{-0.12} $ & $ 2.03^{+0.07}_{-0.07} $ & $ 0.65^{+0.07}_{-0.08} $ & $1^*                   $ & $-                     $ & $ 1.2^{+0.2}_{-0.2} $ & $ 26.3^{+4.4}_{-4.1} $ & $10^{4*}              $  \\
            \textcolor{c12}{$\blacksquare$} 12 Jan & $ 29.6^{+1.8}_{-2.0}$ & $ 1.81^{+0.07}_{-0.06} $ & $ 0.29^{+0.03}_{-0.03} $ & $ 1.60^{+0.09}_{-0.09} $ & $ 0.66^{+0.06}_{-0.06} $ & $1^*                   $ & $-                     $ & $ 1.1^{+0.2}_{-0.2} $ & $ 10.6^{+1.2}_{-1.2} $ & $10^{4*}              $  \\
            \textcolor{c13}{$\blacksquare$} 13 Jan & $ 30.9^{+2.0}_{-2.3}$ & $ 1.84^{+0.09}_{-0.08} $ & $ 0.37^{+0.04}_{-0.05} $ & $ 1.72^{+0.08}_{-0.07} $ & $ 0.58^{+0.05}_{-0.05} $ & $1^*                   $ & $-                     $ & $ 0.9^{+0.1}_{-0.1} $ & $ 10.4^{+1.3}_{-1.4} $ & $10^{4*}              $  \\
            \textcolor{c14}{$\blacksquare$} 14 Jan & $ 23.5^{+1.5}_{-1.4}$ & $ 2.08^{+0.06}_{-0.06} $ & $ 0.18^{+0.02}_{-0.02} $ & $ 1.21^{+0.09}_{-0.09} $ & $ 0.48^{+0.04}_{-0.03} $ & $1^*                   $ & $-                     $ & $ 1.3^{+0.2}_{-0.2} $ & $  3.7^{+0.5}_{-0.4} $ & $10^{4*}              $  \\
            \textcolor{c15}{$\blacksquare$} 15 Jan & $ 22.3^{+1.5}_{-1.2}$ & $ 2.52^{+0.05}_{-0.04} $ & $ 0.16^{+0.01}_{-0.01} $ & $ 1.82^{+0.04}_{-0.04} $ & $ -                    $ & $ 0.96^{+0.04}_{-0.03} $ & $25.50^{+2.21}_{-2.32} $ & $ 1.9^{+0.1}_{-0.1} $ & $-                   $ & $ 110.6^{+6.6}_{-7.3} $  \\
            \textcolor{c16}{$\blacksquare$} 16 Jan & $ 27.7^{+1.8}_{-2.1}$ & $ 2.66^{+0.08}_{-0.08} $ & $ 0.21^{+0.03}_{-0.03} $ & $ 1.91^{+0.08}_{-0.07} $ & $ 0.43^{+0.04}_{-0.04} $ & $1^*                   $ & $-                     $ & $ 1.0^{+0.2}_{-0.2} $ & $ 12.4^{+1.8}_{-2.0} $ & $10^{4*}              $  \\
            \textcolor{c18}{$\blacksquare$} 18 Jan & $ 21.1^{+1.5}_{-1.6}$ & $ 2.91^{+0.21}_{-0.21} $ & $ 0.21^{+0.03}_{-0.03} $ & $ 1.99^{+0.10}_{-0.11} $ & $ 0.45^{+0.06}_{-0.05} $ & $1^*                   $ & $-                     $ & $ 1.4^{+0.2}_{-0.3} $ & $  7.8^{+1.1}_{-1.2} $ & $10^{4*}              $  \\
            \textcolor{c19}{$\blacksquare$} 19 Jan & $ 17.3^{+0.9}_{-0.8}$ & $ 3.37^{+0.11}_{-0.13} $ & $ 0.14^{+0.01}_{-0.01} $ & $ 1.88^{+0.04}_{-0.04} $ & $ -                    $ & $ 0.77^{+0.03}_{-0.02} $ & $12.74^{+0.81}_{-0.92} $ & $ 1.5^{+0.2}_{-0.1} $ & $-                   $ & $ 120.8^{+7.7}_{-7.9} $  \\
            \textcolor{c20}{$\blacksquare$} 20 Jan & $  9.6^{+0.4}_{-0.5}$ & $ 4.09^{+0.07}_{-0.08} $ & $ 0.07^{+0.01}_{-0.01} $ & $ 1.44^{+0.04}_{-0.04} $ & $ -                    $ & $ 0.67^{+0.02}_{-0.02} $ & $ 3.73^{+0.26}_{-0.27} $ & $ 1.9^{+0.1}_{-0.1} $ & $-                   $ & $  94.2^{+5.2}_{-6.5} $  \\
            \textcolor{c21}{$\blacksquare$} 21 Jan & $  8.1^{+0.6}_{-0.6}$ & $ 4.80^{+0.18}_{-0.21} $ & $ 0.08^{+0.01}_{-0.01} $ & $ 1.73^{+0.07}_{-0.07} $ & $ 0.41^{+0.04}_{-0.03} $ & $1^*                   $ & $-                     $ & $ 1.6^{+0.2}_{-0.1} $ & $ 10.3^{+0.8}_{-0.9} $ & $10^{4*}              $  \\
            \textcolor{c22}{$\blacksquare$} 22 Jan & $  9.8^{+0.8}_{-0.9}$ & $ 5.11^{+0.17}_{-0.15} $ & $ 0.08^{+0.01}_{-0.01} $ & $ 2.01^{+0.08}_{-0.09} $ & $ 0.38^{+0.05}_{-0.05} $ & $1^*                   $ & $-                     $ & $ 1.5^{+0.3}_{-0.2} $ & $ 18.2^{+2.0}_{-1.9} $ & $10^{4*}              $  \\
            \textcolor{c23}{$\blacksquare$} 23 Jan & $  9.0^{+0.6}_{-0.6}$ & $ 5.10^{+0.16}_{-0.14} $ & $ 0.09^{+0.01}_{-0.01} $ & $ 2.01^{+0.06}_{-0.06} $ & $ 0.41^{+0.04}_{-0.04} $ & $1^*                   $ & $-                     $ & $ 1.7^{+0.2}_{-0.2} $ & $ 15.7^{+2.1}_{-2.0} $ & $10^{4*}              $  \\
            \textcolor{c24}{$\blacksquare$} 24 Jan & $  5.7^{+0.3}_{-0.3}$ & $ 5.80^{+0.12}_{-0.13} $ & $ 0.05^{+0.00}_{-0.00} $ & $ 1.21^{+0.05}_{-0.07} $ & $ 0.47^{+0.03}_{-0.02} $ & $1^*                   $ & $-                     $ & $ 1.9^{+0.1}_{-0.1} $ & $  3.8^{+0.3}_{-0.3} $ & $10^{4*}              $  \\
            \textcolor{c25}{$\blacksquare$} 25 Jan & $  5.1^{+0.3}_{-0.3}$ & $ 6.88^{+0.21}_{-0.26} $ & $ 0.05^{+0.00}_{-0.00} $ & $ 1.82^{+0.05}_{-0.06} $ & $ 0.46^{+0.04}_{-0.04} $ & $1^*                   $ & $-                     $ & $ 1.7^{+0.2}_{-0.1} $ & $ 18.2^{+2.5}_{-2.6} $ & $10^{4*}              $  \\
            \textcolor{c26}{$\blacksquare$} 26 Jan & $  5.9^{+0.3}_{-0.4}$ & $ 7.34^{+0.24}_{-0.21} $ & $ 0.05^{+0.00}_{-0.01} $ & $ 2.01^{+0.05}_{-0.04} $ & $ 0.48^{+0.05}_{-0.05} $ & $1^*                   $ & $-                     $ & $ 1.7^{+0.1}_{-0.1} $ & $ 36.1^{+6.9}_{-6.2} $ & $10^{4*}              $  \\
            \hline
        \end{tabular} 
    \end{center}
    \caption{\gls{mcmc} sampler results for the expanding blob model. Same as Table~\ref{tab:mcmc_fit_par_list}. $^*$: Parameter set to constant for given \gls{eed}. }
    \label{tab:mcmc_fit_par_list_exp_blob}
\end{table*}

\subsection{Emission region size, bulk magnetic field and integrated particle density evolution}
\begin{figure}
    \centering
    \includegraphics[width=0.48\textwidth]{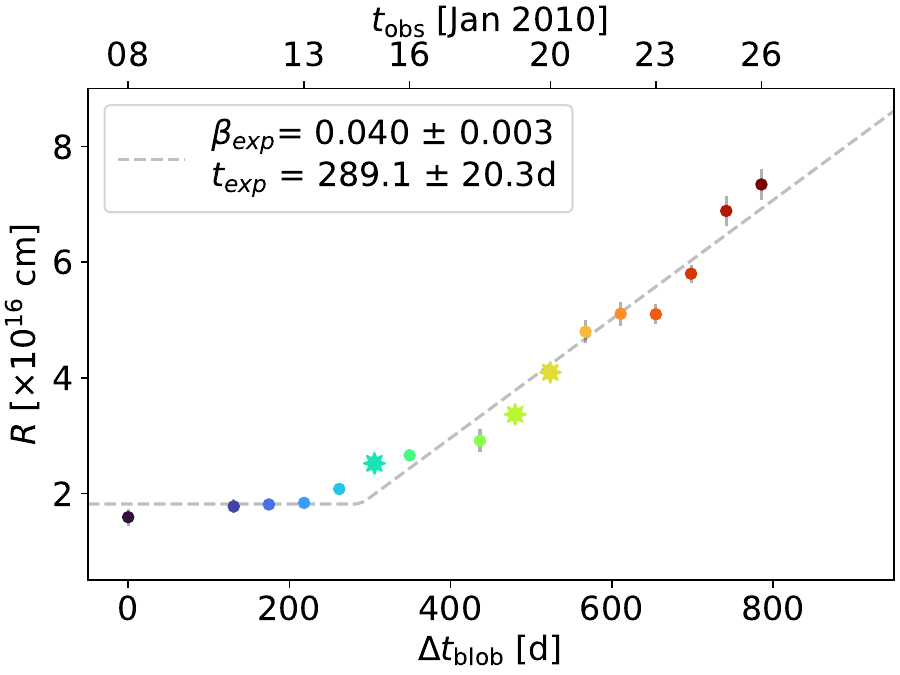}
    \caption{Temporal evolution of the radius of the emitting region ($R$) under the expanding blob model fit with Eq.~\ref{eq:R_conical_expansion} using \texttt{scipy curvefit}. Colour key and markers same as Fig.~\ref{fig:r3p_vs_g3p}.}
    \label{fig:R_exp}
\end{figure}

Conical expansion has been observed for the source BL Lac in radio frequencies by \cite{2021A&A...649A.153C}. 
We fit our emission region radius versus time to test for such a geometry.
Fig.~\ref{fig:R_exp} shows the expansion of the emitting region size for the expanding blob model which agrees well with the conical jet expansion geometry proposed by \cite{tramacere2022} studying the radio-$\gamma$ delay for \gls{421}. Converting between the observer and blob frame timescales, $t_\text{obs}
= t_\text{blob} (1 + z)/\delta$, we fit the temporal evolution of $R$ with the following equation:
\begin{equation}
    R(t) = R_0 + \bexp \, c \, (t - \texp) \cdot \mathcal{H} (t - \texp)
    \label{eq:R_conical_expansion}
\end{equation}
where $\mathcal{H}$ is the Heaviside function and time is in the blob frame.  
The fit suggests a cylindrical profile up to 15 January and a subsequent conical expansion starting $\texp \sim 290$ days from the reference in the blob frame.  
The blob expansion velocity thus obtained, $\bexp=0.040$ ($\log(\bexp) \sim -1.39$) is compatible with $\log(\bexp)=-1.89 \pm 0.59$ reported in \cite{tramacere2022}. 
Additionally, our value for $R_0=1.81 \times 10^{16} cm$ is compatible with the $\log_{10}(R_0^\text{obs})\geq 15.67 \pm 0.59$ reported by the authors, with their flat posterior distribution of the \gls{mcmc} indicative of a lower limit on the parameter. 
In Fig.~\ref{fig:R_exp} we notice a hint of acceleration of blob expansion before the pile-up develops and a deceleration after the pile-up ends, the implications and consequences of which will be explored in a future work. 

The evolution of the bulk magnetic field strength ($B$) versus the radius of the emission region ($R$) in a jet follows a power-law index $m=-1$ for toroidal and $m=-2$ for poloidal bulk magnetic field configuration along the jet assuming magnetic flux pinning and adiabatic conditions \citep{begelman1984}:
\begin{equation} 
    B = B_0 \left( \frac{R}{R'} \right) ^m
    \label{eq:b_vs_r_index}
\end{equation}

\begin{figure}
    \centering
    \includegraphics[width=0.48\textwidth]{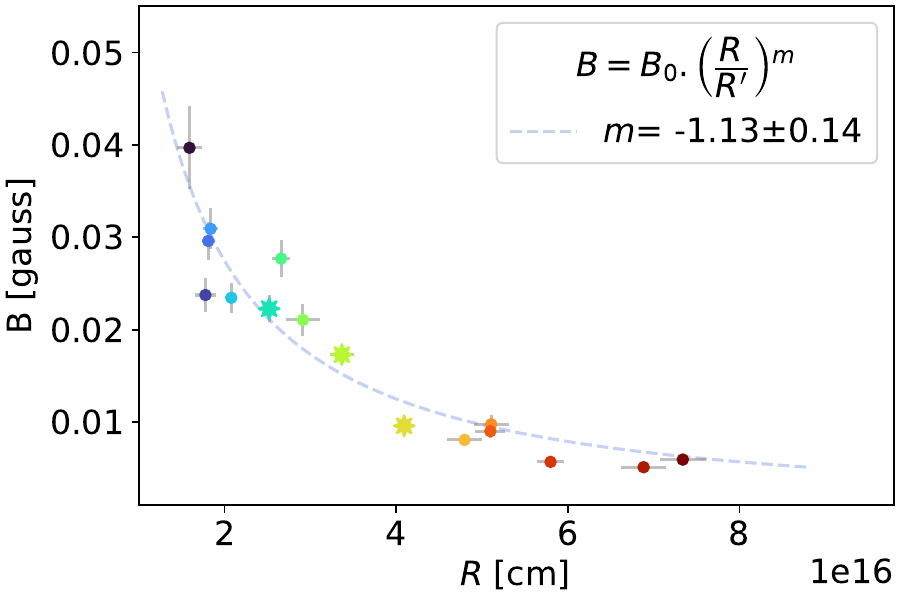}
    \caption{$B$ vs $R$ index fit for the expanding blob model done using \texttt{scipy curvefit}. $m=-1.13$, $B_0=8.0 mG$, $R'=5.9\times 10^{16} cm$. Colour key and markers same as Fig.~\ref{fig:r3p_vs_g3p}.}
    \label{fig:b_vs_r}

    \includegraphics[width=0.48\textwidth]{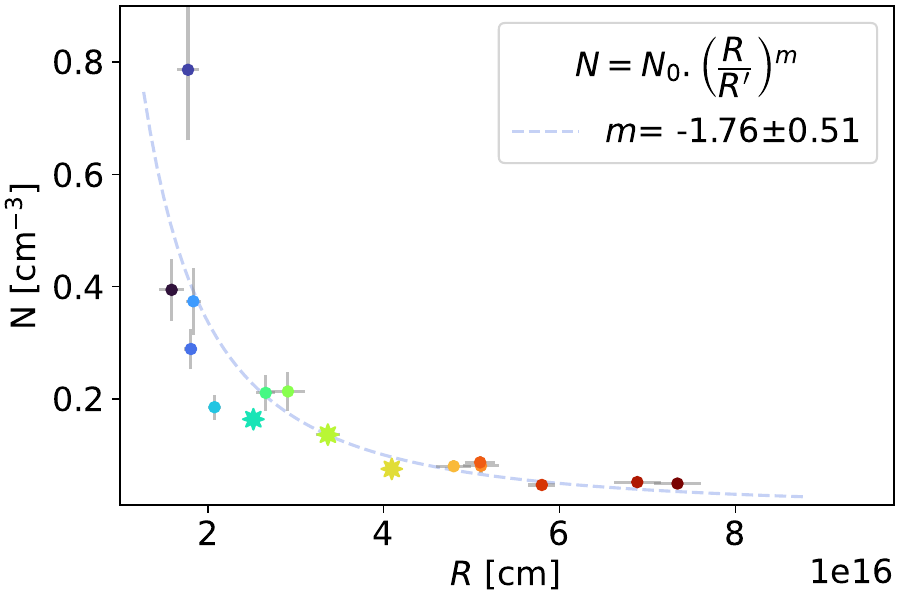}
    \caption{$N$ vs $R$ index fit for the expanding blob model. $m= -1.75$, $N_0 = 0.067 cm^{-3}$,  $R'=5.0\times 10^{16} cm$. Colour key and markers same as Fig.~\ref{fig:r3p_vs_g3p}.}
    \label{fig:n_vs_r}
\end{figure}

\cite{tramacere2022} found the index $m$ to be $-1.39^{+0.38}_{-0.29}$ for adiabatic blob expansion for \gls{421}. 
Our expanding blob model recovers a compatible index $m \sim -1.1$ for the January flare (Fig.~\ref{fig:b_vs_r}), also consistent with the striking parsec scale toroidal magnetic field structure reported by \cite{vlba_toroidal_jet_B} for the blazar PKS~1424+240. 
However, one must note that in \cite{tramacere2022}, the authors mathematically linked $B$ and $R$ via Eq.~\ref{eq:b_vs_r_index} such that $R$ and the index $m$ were the free parameters of their model, while in our work, $R$ and $B$ are independent parameters. 
In this work, the $B$ vs $R$ index thus arises only as a consequence of the time evolution of the best fit to the \glspl{sed} in the expanding blob model.

Similarly, we computed the index for $N$ vs $R$ in Fig.~\ref{fig:n_vs_r} with the same functional form as Eq.~\ref{eq:b_vs_r_index} with the index being $\sim -1.7$ which indicates that there is significant particle injection, otherwise one would expect and index of $-3$ for pure adiabatic expansion with no particle injection or escape. 
In context of the adiabatic expansion model presented in \cite{tramacere2022}, our average spectral slope $s \sim 1.65$ is also in agreement with the \gls{eed} index of $1.97^{+1.26}_{-0.72}$ reported by the authors. 

\subsection{Particle cooling timescales}
We investigated the adiabatic cooling timescales for the expanding blob model and compared them with the synchrotron cooling timescales using the methodology reported in \cite{tramacere2022}.
The adiabatic cooling timescale $t_\text{adiabatic} \sim R(t)/\bexp c$ \citep{2011hea..book.....L} is independent of particle Lorentz factor and was computed using the expansion velocity $\bexp=0.038$c obtained in the fit to Eq.~\ref{eq:R_conical_expansion}. 
On the other hand, the synchrotron cooling rate increases with the particle Lorentz factor. 

\begin{figure}[t]
    \centering
    \includegraphics[width=0.48\textwidth]{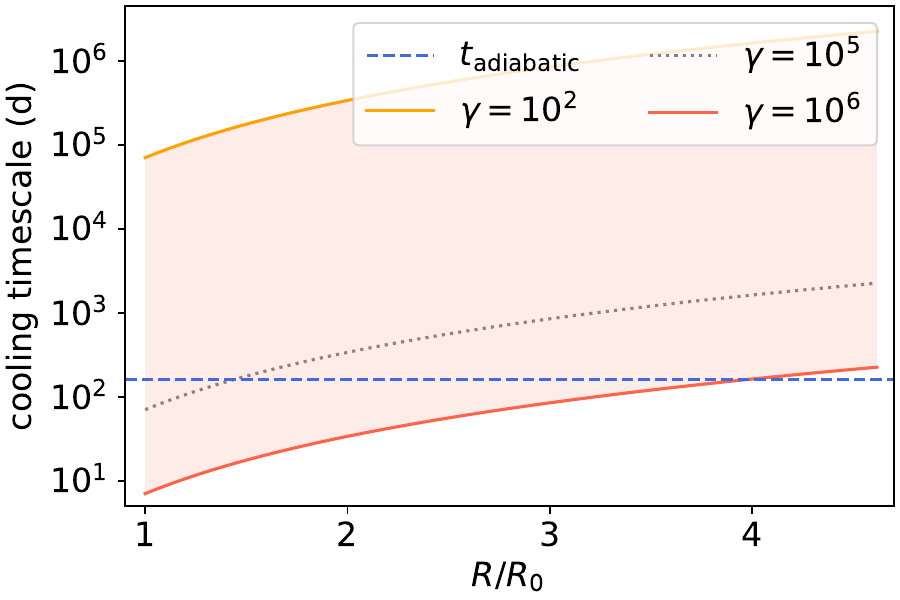}
    \caption{Comparison of the particle cooling time scales (in blob frame) up to an expansion factor of $\sim 4.5$ found in our expanding blob model, for a range of Lorentz factors. The shortest timescale for a given Lorentz factor determines the dominant mechanism for cooling. }
    \label{fig:adiabatic_timescale}
\end{figure}

Once the emission region starts to expand, the magnetic field strength drops and the synchrotron cooling timescales get longer and longer. 
In our model, electrons less energetic than Lorentz factor $\sim 10^5$ have much longer synchrotron cooling timescales compared to the adiabatic cooling timescale (see Fig.~\ref{fig:adiabatic_timescale}) which is of the order of $\sim 150$ days in the blob frame.
$\gamma \sim 10^5$ is also of the same order as $\gammacut$ and $\gtp$ which implies that curvature in the \gls{eed} might come from a combination of stochastic acceleration and adiabatic cooling once the expansion starts after $\texp \sim 290$ days in the blob frame (14-15 January in observer frame).

Given the phenomenology reported above, we conclude that the pile-up might be triggered by the sudden expansion of the blob as it travels along the jet.
However, a more robust analysis using self-consistent temporal evolution is necessary to make stronger statements regarding the origin of the pile-up and the transitions between the acceleration and cooling dominated states during the multiple flaring events.

\subsection{VLBI knots as expanding blobs propagating along the Jet}

There were two radio knots $K1$ and $K2$ reported during 2010-2011 with potential ejection time window in agreement with the giant flare in February. 
We computed the observed apparent velocity of the emission region based on the Doppler factor from our model and compared the values to the \gls{vlbi} angular distances and the apparent radio knot velocities described in \cite{jorstad2017,felix_mrk421}.

Unlike $K2$, the existence of $K1$ cannot be robustly confirmed \citep{jorstad2017, felix_mrk421}. 
The apparent velocity of knot $K2$ was determined to be $\beta_\text{app} \sim 0.3$. 
Given the large uncertainty of the knot emission time ($\sim 300$ days for $K2$), it is possible that the \gls{vhe} emission region and the radio knots are completely unrelated. 
Nevertheless, under the assumption that $K2$ is associated to the high-energy flare, the jet viewing angle ($\thetaview$) must be very close to zero, $\thetaview \leq 0.02^\circ$ to replicate the high Doppler beaming factor of $45$ that we found necessary to capture the broadband \gls{sed}. 
\acrshort{bllac} type objects typically have a viewing angle of $\sim 0^\circ-5^\circ$ and an extremely small value is statistically unlikely for \gls{421}, the closest \acrshort{bllac} \citep{2013A&A...559A..75B}.

Despite these caveats, we estimated the distance travelled by the blob over the $\sim 20$ day flare (in the observer frame), amounting to $\sim15$ parsec for $\thetaview\sim0^\circ$ (see Appendix \ref{appendix:exp_blob}, Fig.~\ref{fig:d_vs_R_exp_blob}). 
This gives an approximate distance scale along the jet over which the physical properties of the emitting region must vary (radius, magnetic field, etc. ).
Finally, we note that it is possible that we observe combined emission from multiple different emission regions travelling at high bulk factor close to the base of the jet and then decelerating as they travel farther, or via a stationary shock much closer to the \gls{smbh}.

\section{Summary and conclusions}\label{sec:conclusion}
We processed the January 2010 \gls{mwl} data of \gls{421} to produce a sequence of \glspl{sed}.
This dataset was then modelled and tested for different underlying \gls{eed}s in a single emission zone scenario.
We found the phenomenology to be consistent with stochastic acceleration, independent of the \gls{eed} used in our model through the trend between synchrotron peak frequency and the peak curvature.

The standard \gls{lppl} distribution ($n_\text{LPPL}$) worked well as a first approximation. 
The peak position and curvature parameters were consistent with an acceleration-dominated phase of a flare and the trend of anti-correlation seen in the \gls{uv}--X-ray data. 
However, the model failed to capture the nuanced shape of the synchrotron peak on 15, 19 and 20 January 2010, with these dates aligning with a decrease in the X-ray flux which probes the leptons with the strongest cooling. 
Since the \gls{lppl} distribution appears in an acceleration dominated phase of the evolution of the particle population, this hinted at a transition to a cooling dominated phase of evolution which we tested with a pile-up distribution ($n_\text{pile-up}$). 
We found that the pile-up model improved the fits on these 3 days at a \gls{ts}>4 level. 
The synchrotron peak of these three \glspl{sed} also had increased curvature in agreement with theoretical work and numerical simulations of \gls{eed} evolution under stochastic acceleration in the literature. 


Based on the results of these fits, we found hints of expanding emission region in the evolution of the emission region parameters irrespective of the choice of the \gls{eed} and tested it using our modified sequential fitting method in the expanding blob model. 
The temporal evolution of the emitting region parameters $R$ and $B$ in the expanding blob model agrees well with the evolution predicted by the long term radio-$\gamma$ delay study of \gls{421} and the helical magnetic field structure of the blazar jet. 

The snapshot analysis presented here is a step towards a fully self-consistent non-equilibrium model using the diffusion kinetic equations which will be a natural follow-up study to the models presented here as seen in \cite{2021A&A...654A..96Z} and \cite{2021MNRAS.505.2712D}. 
The \glspl{eed} utilised in this work are analytical approximations of steady-state solutions to the diffusion equations including contributions from both first and second order Fermi-acceleration processes. 
The use of parametrised lepton distributions abstracts out the microphysical parameters while still capturing the evolution of the particle population which is the main driver of the flaring activity of \gls{421} in our model. 


Since our goal was to study acceleration-cooling transitions, we used a homogenous single zone model {(see \cite{2019MNRAS.487..845B} for an inhomogenous emission region model of \gls{421})}. 
More complex scenarios like single-zone \gls{ssc} with two population \gls{eed}, multi-zone \gls{ssc} and pair-production cascade emission reported in \cite{mrk501_pileup} can also replicate some of the features of our dataset at the cost of additional degeneracy of parameters.
The increased curvature in the synchrotron peak is consistent with multi-zone or two population models sometimes used in the literature, including replicating the hint of narrow feature hinted in the \gls{vhe} spectrum, albeit with more assumptions in the model.
Some of these scenarios are also discussed in \cite{felix_mrk421} in context of the results for the full 2009-2010 observation campaign. 

Our results also highlight that monitoring and deep observations of the brightest blazars is still a relevant science goal in the \gls{ctao} era, with higher time resolution and consistent coverage allowing for detailed analysis of the temporal evolution of the jet. 
Obtaining spectra at the light crossing timescale for a $\sim 10^{16}$ cm \gls{vhe} emission region is still outside the capabilities of the current generation of \glspl{iact} except for the rare cases of exceptionally strong flares, which restricts the amount of information one can possibly extract from the \gls{sed} models without major assumptions. 
Matching the temporal resolution of the X-ray bands provided by space telescopes like \textit{XMM}-Newton and \gls{ixpe} with the Large Size Telescopes of \gls{ctao} will open new windows into the theoretical understanding of blazar jets.
The X-ray polarisation observations made possible by \gls{ixpe} also provide new insights into the magnetic field structure of \gls{agn} jet, challenging existing theoretical understanding based on polarisation in optical and radio bands. 
The successor to the \textit{Fermi} telescope is also long overdue and highly relevant for constraining the \gls{ic} peak of the \gls{sed}.
While the temporal resolution offered by \gls{lat} was barely enough given the comparatively low fractional variability of \gls{421} in the \gls{he} band and the \gls{ic} peak being in the optimal range for \gls{vhe} observations by \gls{magic} and VERITAS telescopes in this campaign, plenty of other sources will benefit from higher sensitivity in the MeV-GeV bands. 

\section*{Author contributions}
List of the main authors in alphabetical order -- 
J. Abhir: project management, data analysis and modelling, paper drafting; 
A. Arbet-Engels: theoretical interpretation, modelling, paper drafting; 
D. Paneque: coordination of \gls{mwl} data collection and processing; 
F. Schmuckermaier: MAGIC analysis cross-check; 
A. Tramacere: theoretical interpretation, modelling, paper drafting. 
The rest of the authors have contributed in one or several of the following ways: design, construction, maintenance and operation of the instrument(s); preparation and/or evaluation of the observation proposals; data acquisition, processing, calibration and/or reduction; production of analysis tools and/or related Monte Carlo simulations; discussion and approval of the contents of the draft.

\begin{acknowledgements}
We are grateful for the feedback provided by the anonymous referee which helped us improve the quality and readability of the paper. 
We would like to thank the Instituto de Astrof\'{\i}sica de Canarias for the excellent working conditions at the Observatorio del Roque de los Muchachos in La Palma. The financial support of the German BMFTR, MPG and HGF; the Italian INFN and INAF; the Swiss National Fund SNF; the grants PID2022-136828NB-C41, PID2022-137810NB-C22, PID2022-138172NB-C41, PID2022-138172NB-C42, PID2022-138172NB-C43, PID2022-139117NB-C41, PID2022-139117NB-C42, PID2022-139117NB-C43, PID2022-139117NB-C44, CNS2023-144504 funded by the Spanish MCIN/AEI/ 10.13039/501100011033 and "ERDF A way of making Europe"; the Indian Department of Atomic Energy; the Japanese ICRR, the University of Tokyo, JSPS, and MEXT; the Bulgarian Ministry of Education and Science, National RI Roadmap Project DO1-400/18.12.2020 and the Academy of Finland grant nr. 320045 is gratefully acknowledged. This work has also been supported by Centros de Excelencia ``Severo Ochoa'' y Unidades ``Mar\'{\i}a de Maeztu'' program of the Spanish MCIN/AEI/ 10.13039/501100011033 (CEX2019-000918-M, CEX2021-001131-S, CEX2024001442-S), by AST22\_00001\_9 with funding from NextGenerationEU funds and by the CERCA institution and grants 2021SGR00426, 2021SGR00607 and 2021SGR00773 of the Generalitat de Catalunya; by the Croatian Science Foundation (HrZZ) Project IP-2022-10-4595 and the University of Rijeka Project uniri-prirod-18-48; by the Deutsche Forschungsgemeinschaft (SFB1491) and by the Lamarr-Institute for Machine Learning and Artificial Intelligence; by the Polish Ministry of Science and Higher Education grant No. 2025/WK/04; by the European Union (ERC, MicroStars, 101076533); and by the Brazilian MCTIC, the CNPq Productivity Grant 309053/2022-6 and FAPERJ Grants E-26/200.532/2023 and E-26/211.342/2021.
J.A. acknowledges support from the Swiss National Science Foundation (SNSF) Grant \text{200020\_197007}.
A.A.-E. acknowledges support from the Deutsche Forschungs gemeinschaft (DFG, German Research Foundation) under Germany’s Excellence Strategy – EXC-2094 – 390783311. 
\end{acknowledgements}

\begin{figure*}
    \resizebox{\textwidth}{!}{
        \begin{tabular}{ccc}
            \subfloat[\textcolor{c08}{$\blacksquare$} 08 January 2010]{\includegraphics[width=0.32\textwidth]{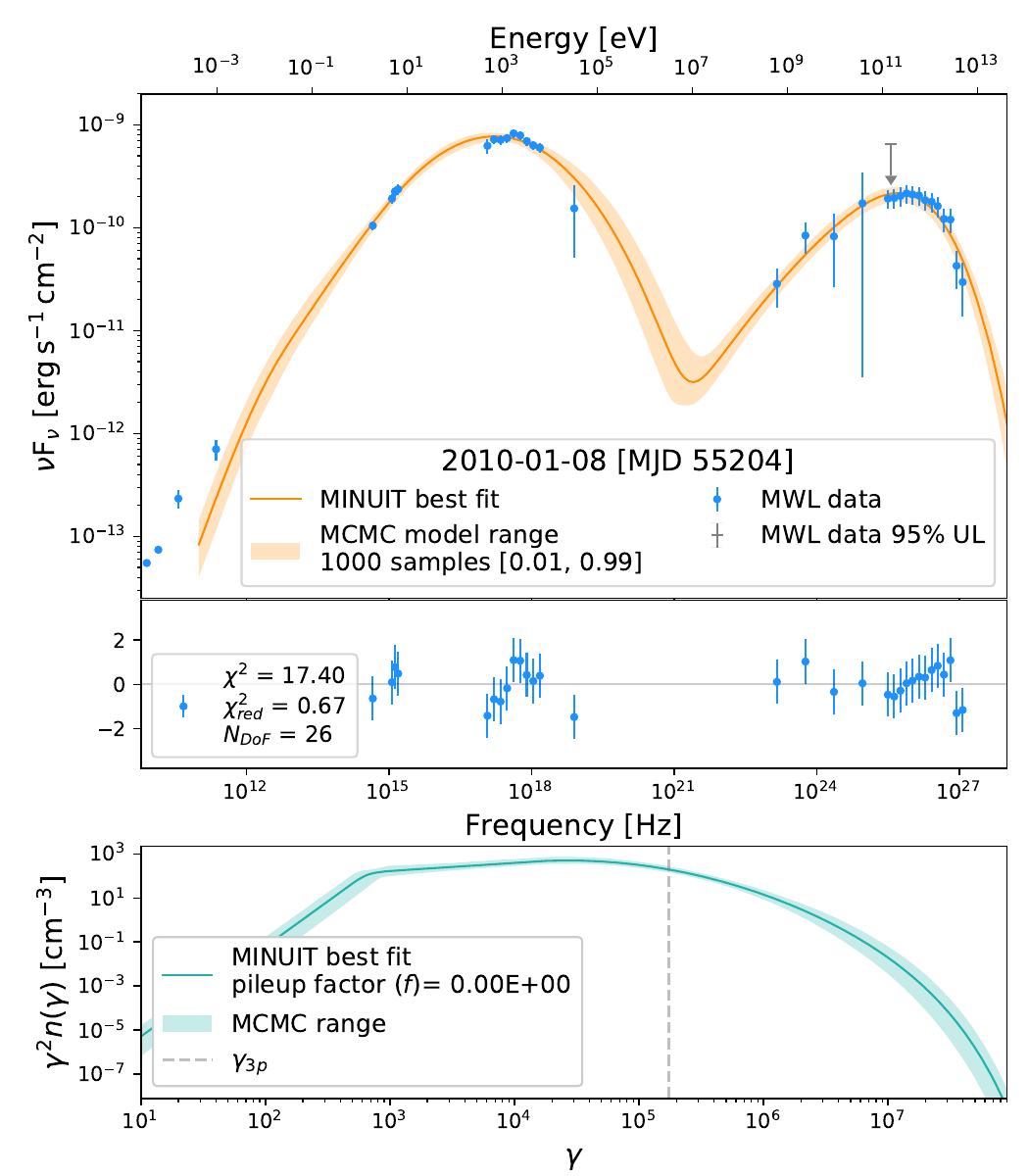}} &
            \subfloat[\textcolor{c11}{$\blacksquare$} 11 January 2010]{\includegraphics[width=0.32\textwidth]{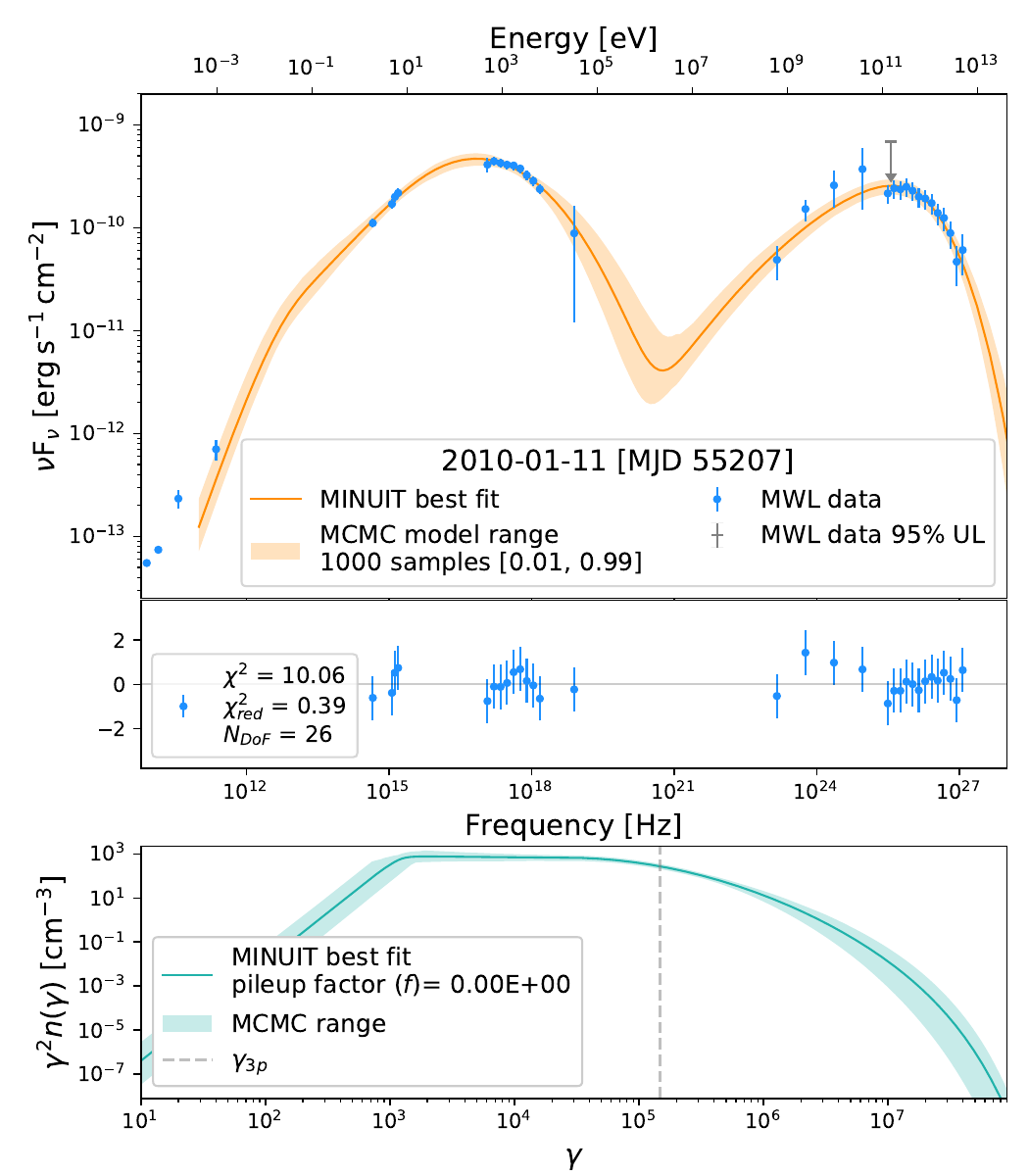}} &
            \subfloat[\textcolor{c12}{$\blacksquare$} 12 January 2010]{\includegraphics[width=0.32\textwidth]{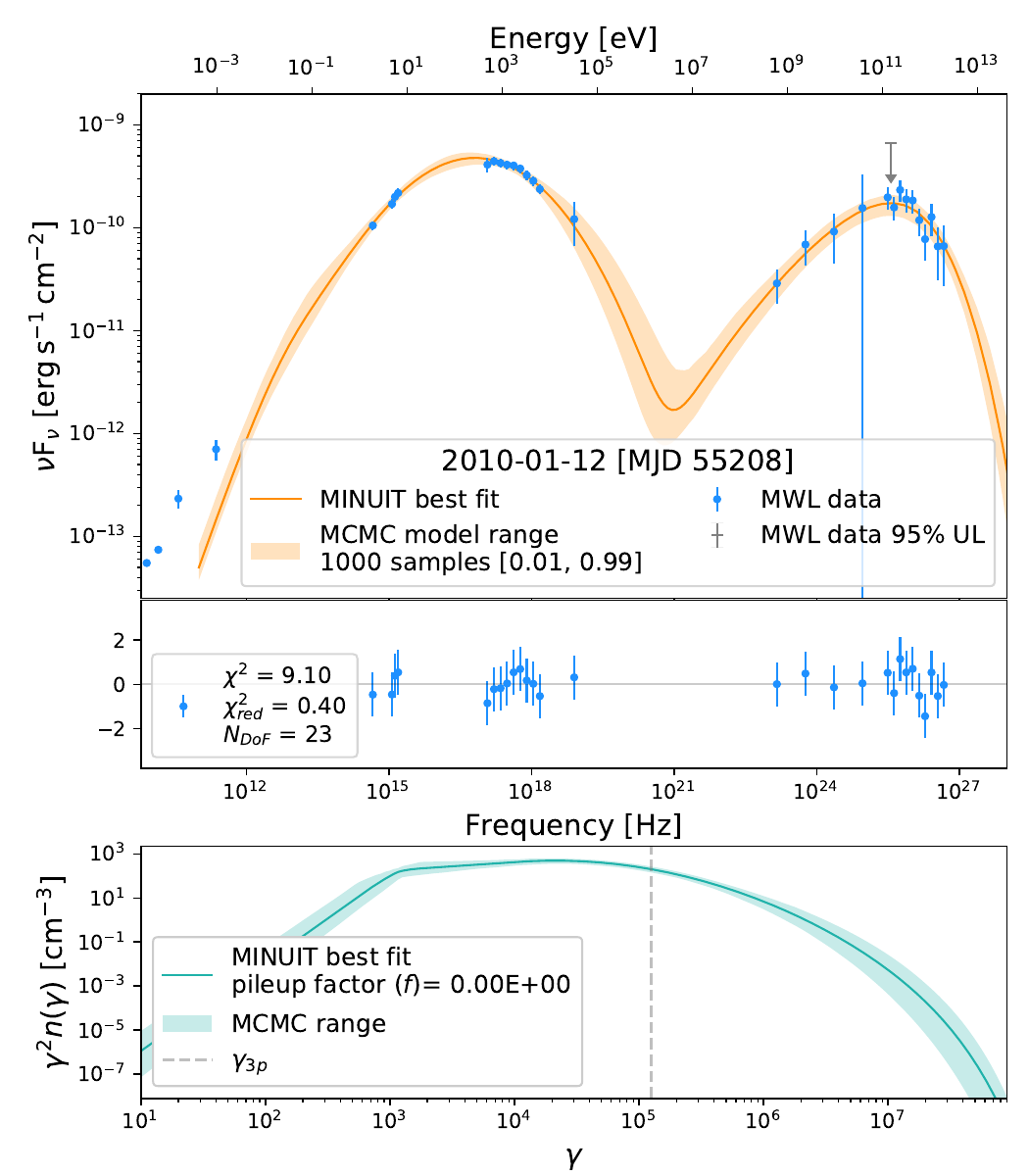}} \\
            \subfloat[\textcolor{c13}{$\blacksquare$} 13 January 2010]{\includegraphics[width=0.32\textwidth]{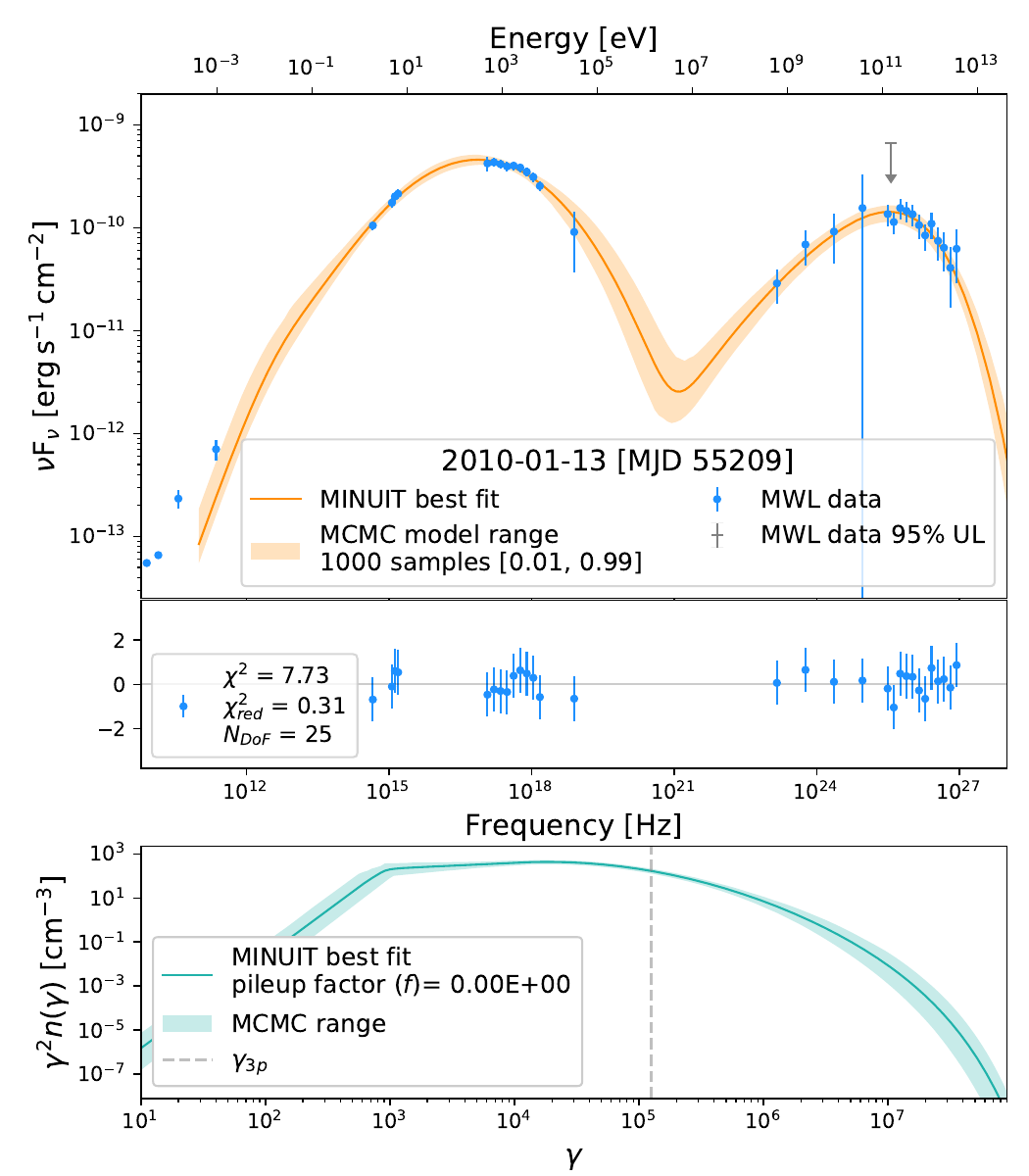}} &
            \subfloat[\textcolor{c14}{$\blacksquare$} 14 January 2010]{\includegraphics[width=0.32\textwidth]{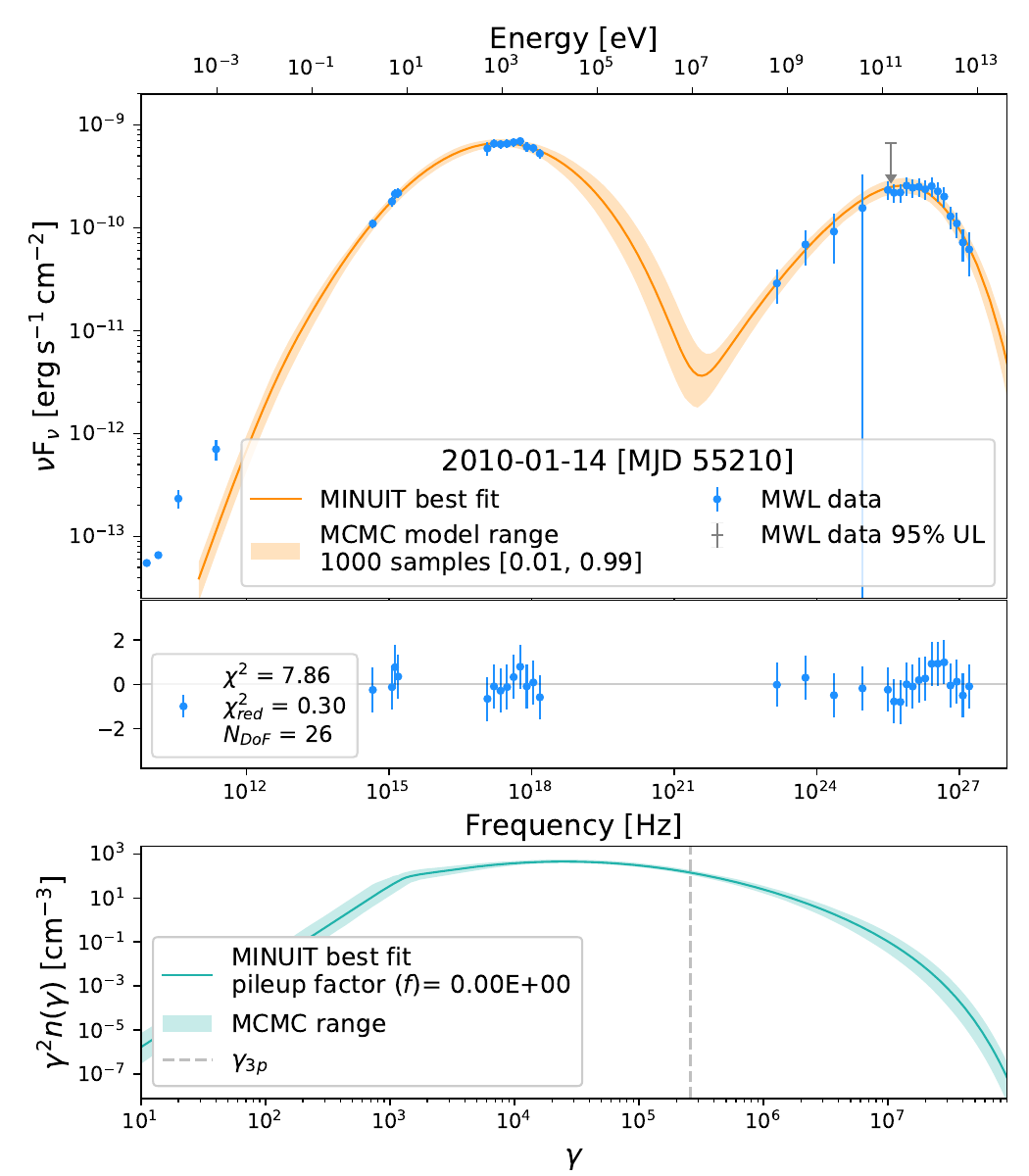}} &
            \subfloat[\textcolor{c15}{$\blacksquare$} 15 January 2010]{\includegraphics[width=0.32\textwidth]{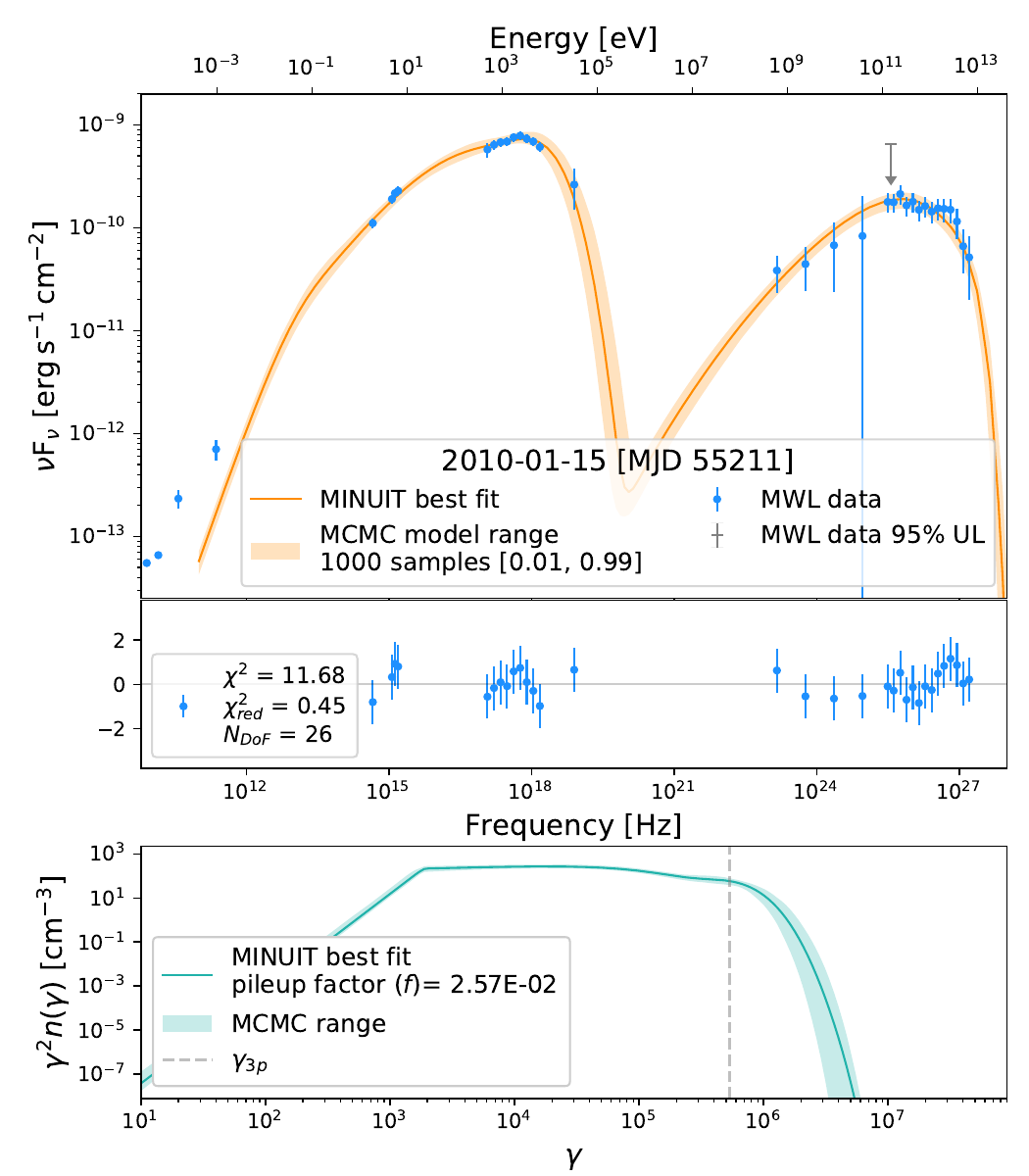}} \\ 
            \subfloat[\textcolor{c16}{$\blacksquare$} 16 January 2010]{\includegraphics[width=0.32\textwidth]{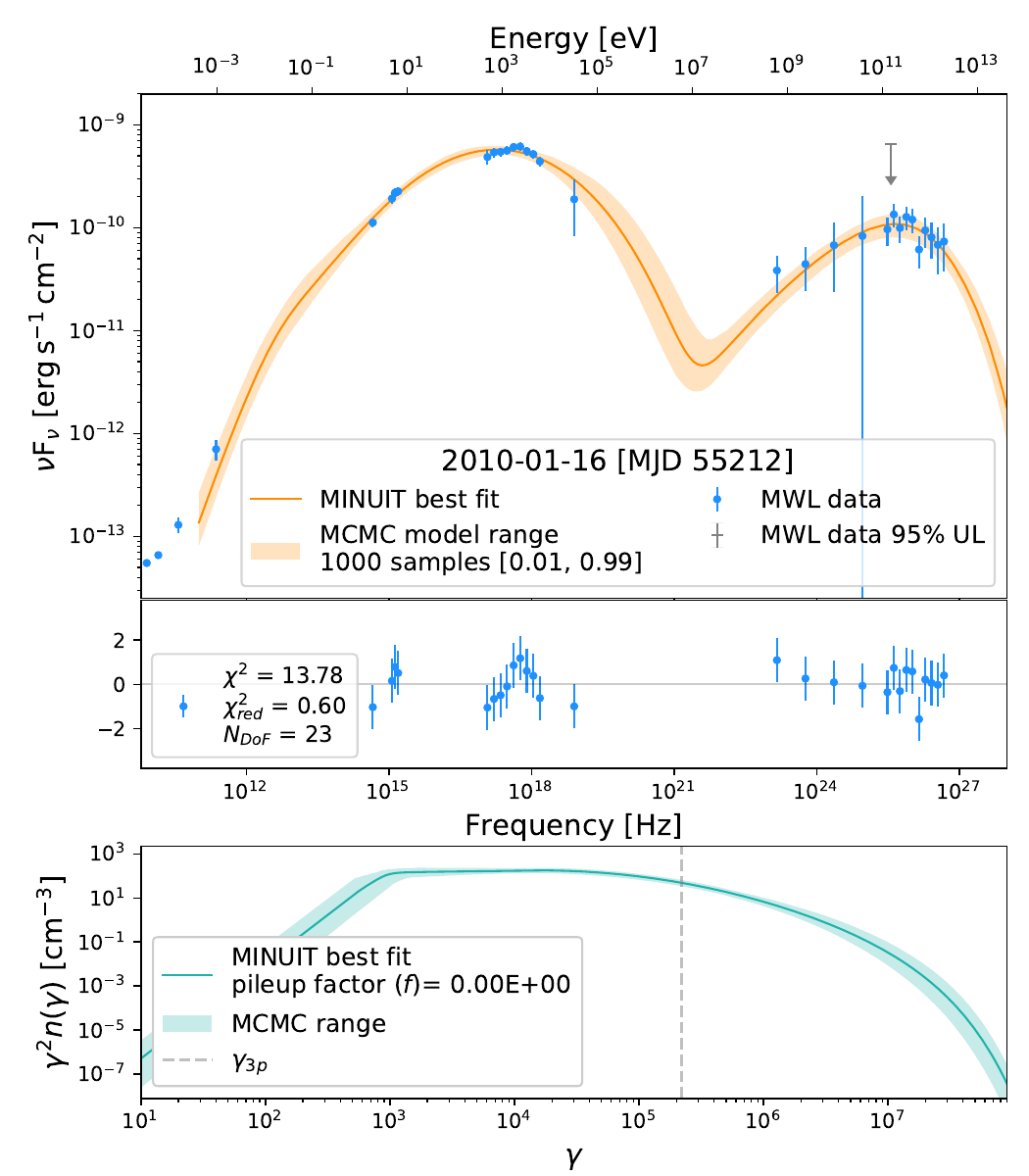}} &
            \subfloat[\textcolor{c18}{$\blacksquare$} 18 January 2010]{\includegraphics[width=0.32\textwidth]{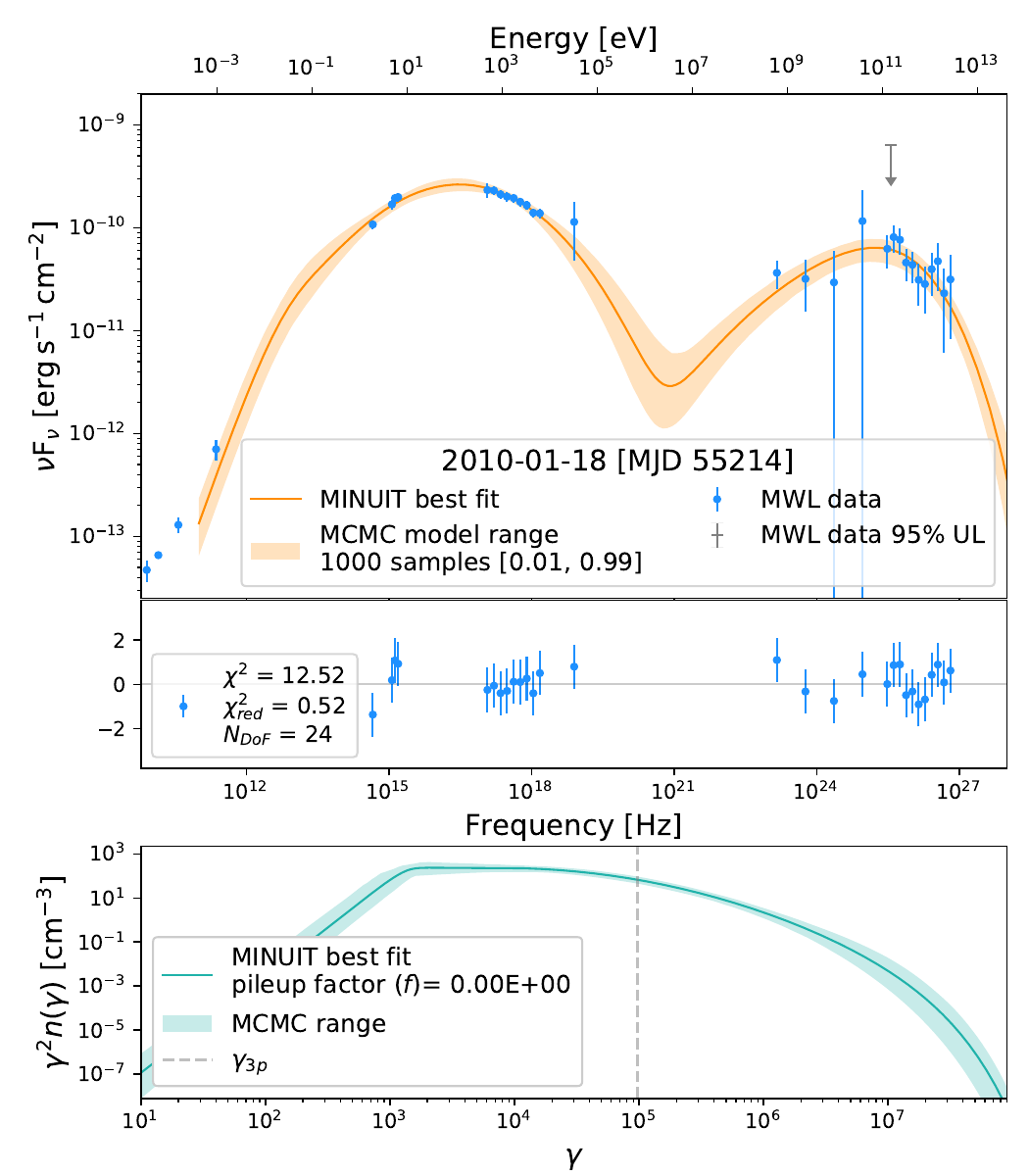}} &
            \subfloat[\textcolor{c19}{$\blacksquare$} 19 January 2010]{\includegraphics[width=0.32\textwidth]{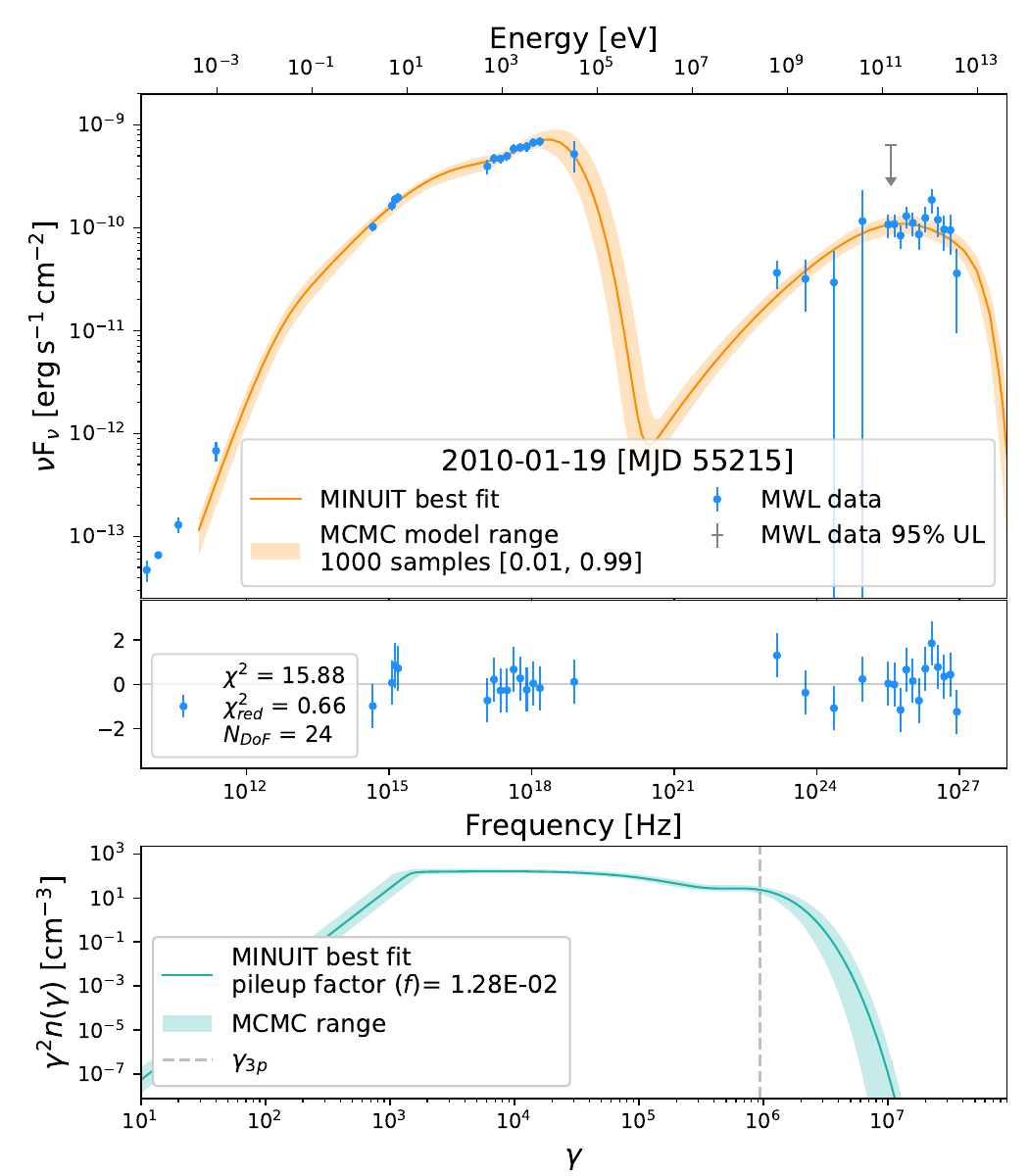}} \\
        \end{tabular}}
    \caption{\gls{sed} fits: expanding blob model. Same as Fig.~\ref{fig:sed_combined_1} for the expanding blob model. The best fit and \gls{mcmc} range for the \gls{sed} (top subplot) and the \gls{eed} (bottom subplot)  for each observation date in January 2010 are shown. The residuals of the model and the $\chisq$, reduced $\chisq$ ($\chisq_\text{red}$) and degrees-of-freedom (N$_\text{DoF}$) of the fit are mentioned in the subplot underneath the \gls{sed}.}
    \label{fig:sed_exp_blob_1}
\end{figure*}

\begin{figure*}
    \resizebox{\textwidth}{!}{
        \begin{tabular}{ccc}
            \subfloat[\textcolor{c20}{$\blacksquare$} 20 January 2010]{\includegraphics[width=0.32\textwidth]{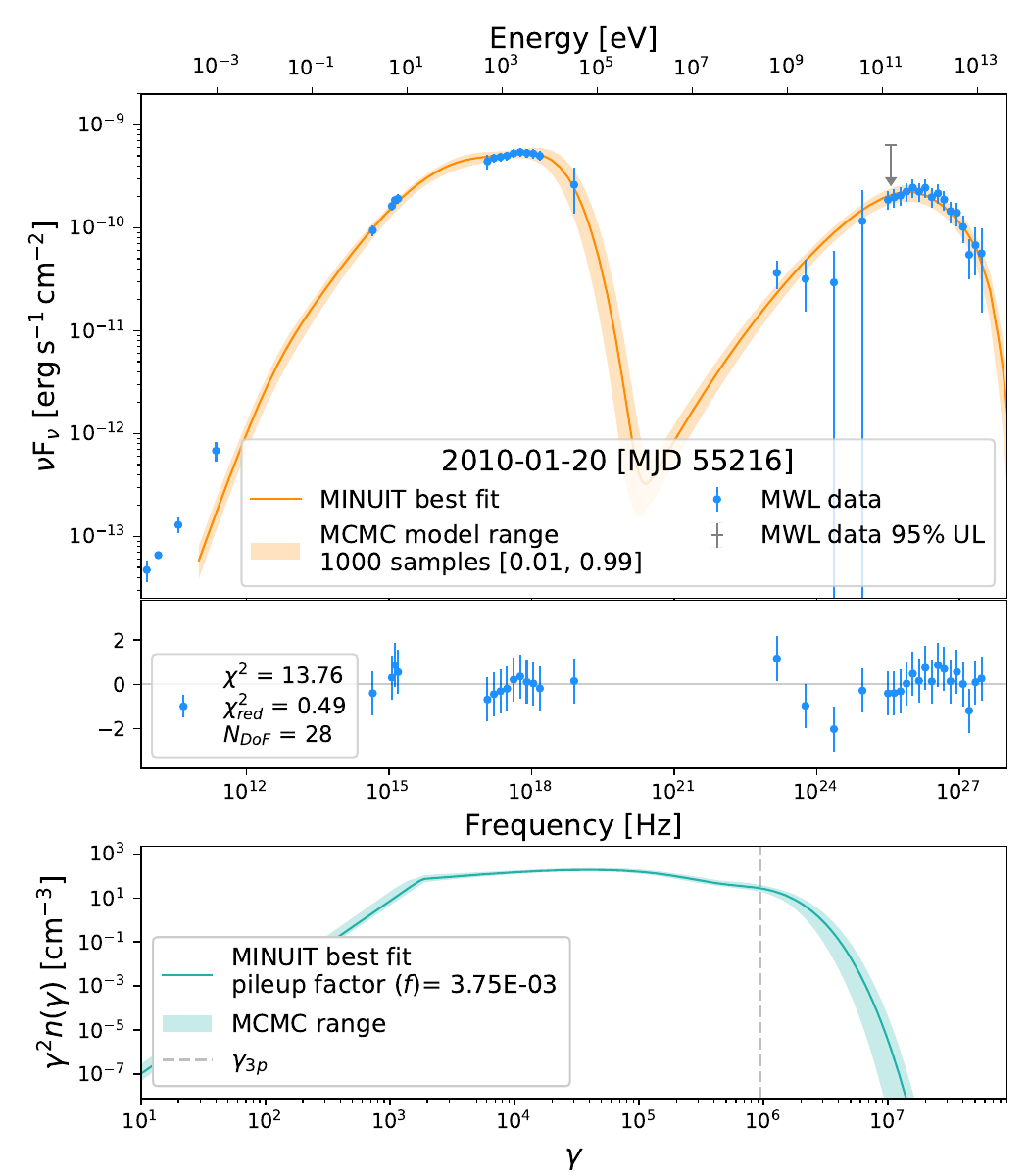}} &
            \subfloat[\textcolor{c21}{$\blacksquare$} 21 January 2010]{\includegraphics[width=0.32\textwidth]{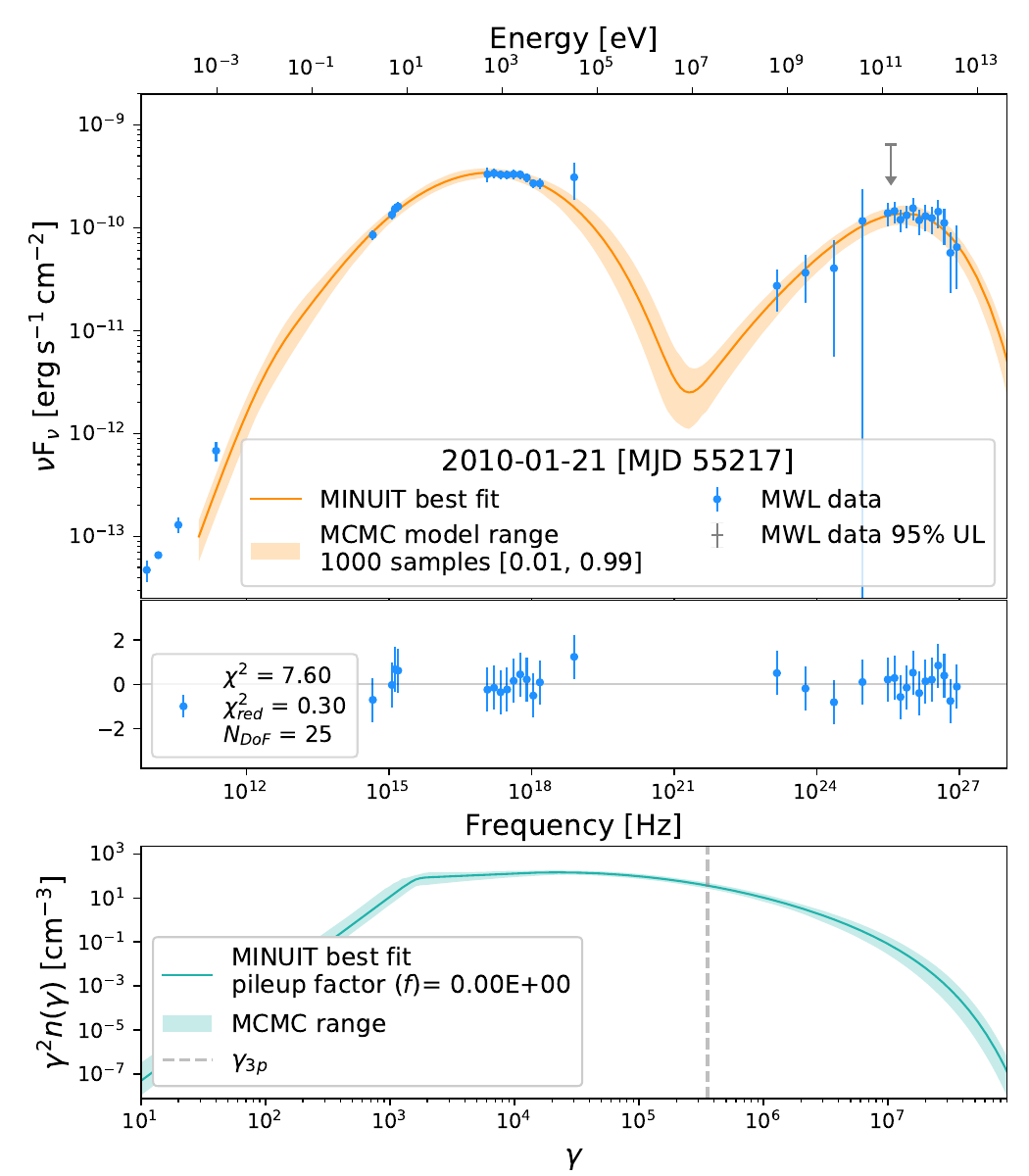}} &
            \subfloat[\textcolor{c22}{$\blacksquare$} 22 January 2010]{\includegraphics[width=0.32\textwidth]{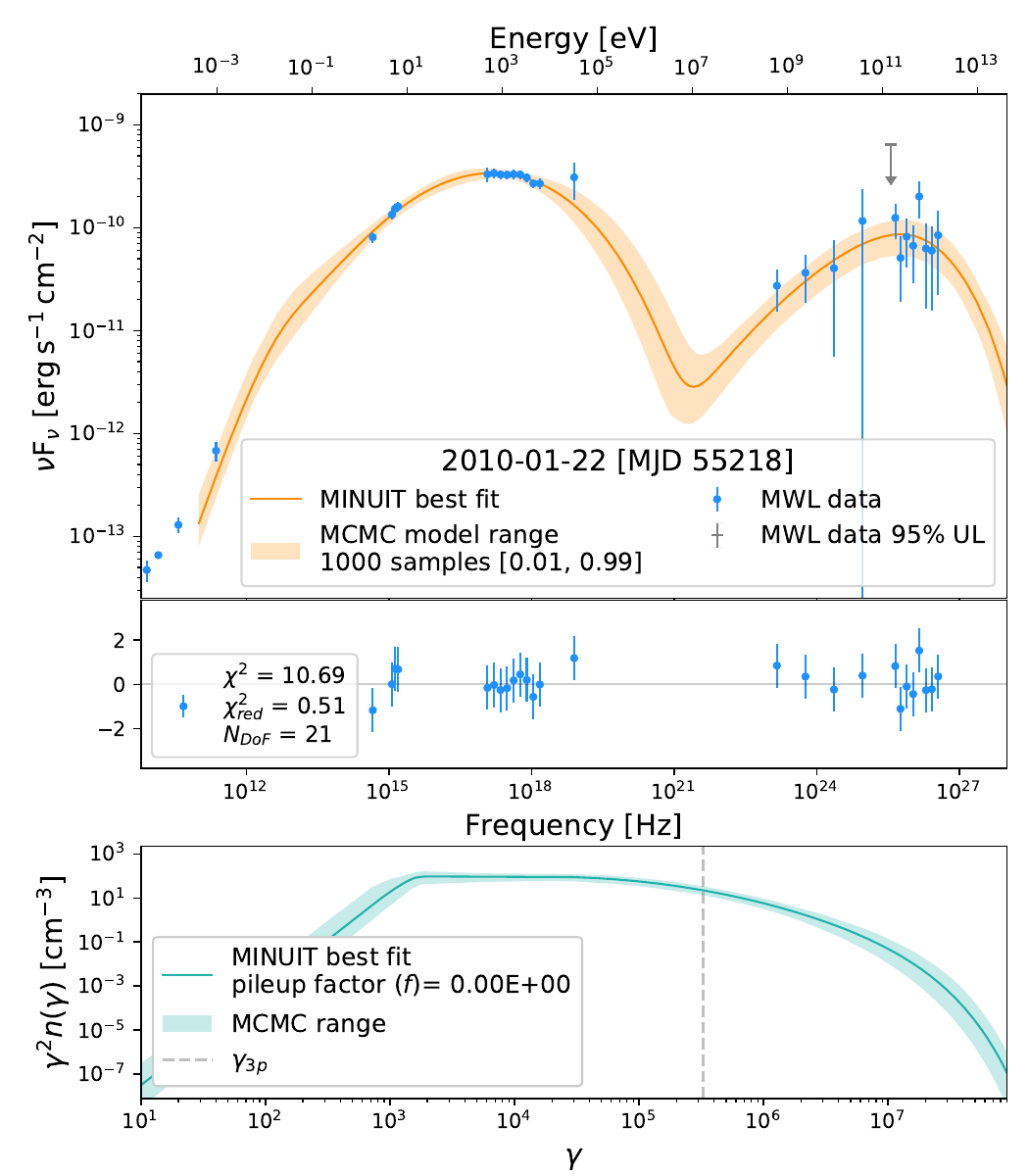}} \\
            \subfloat[\textcolor{c23}{$\blacksquare$} 23 January 2010]{\includegraphics[width=0.32\textwidth]{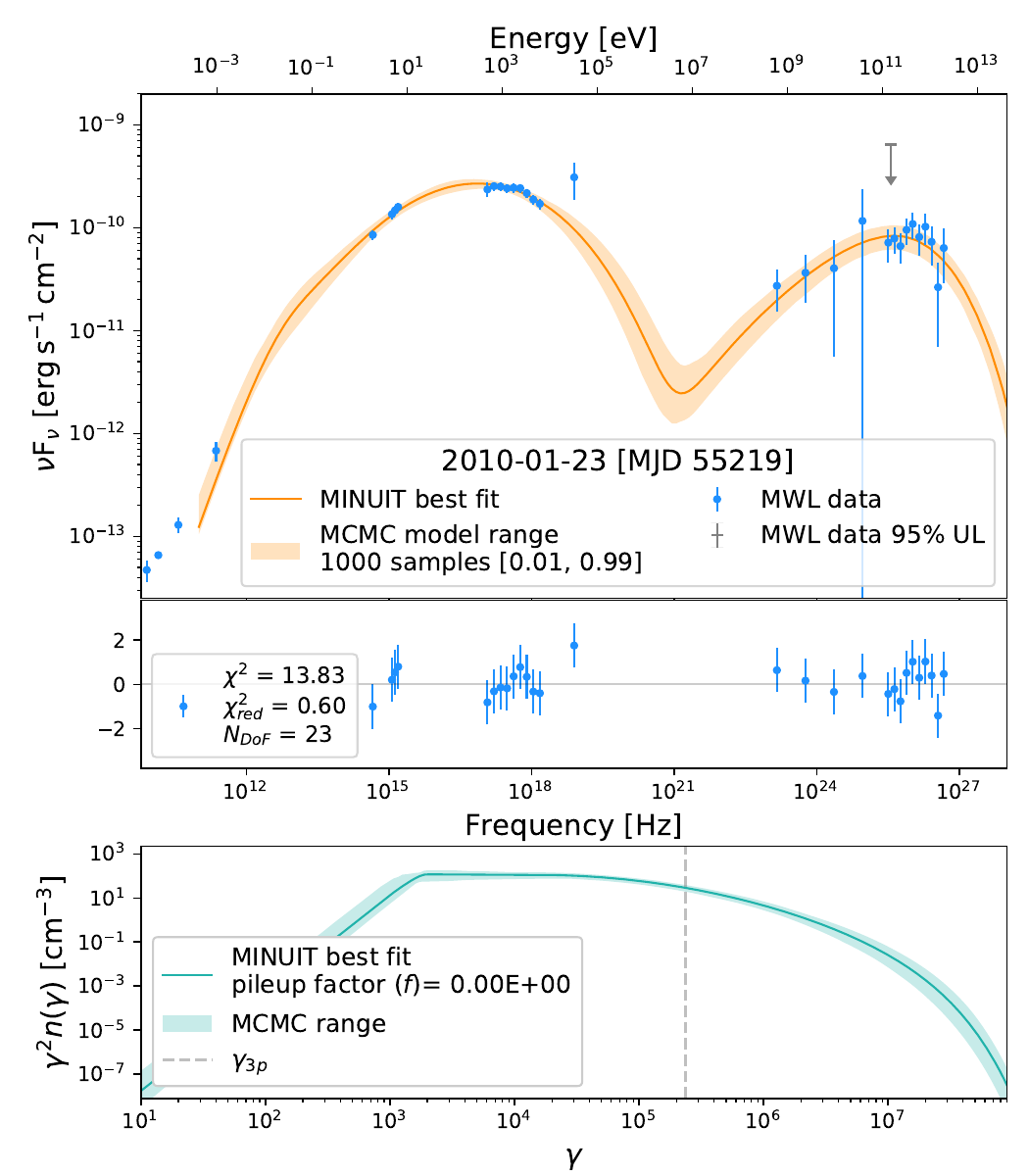}} &
            \subfloat[\textcolor{c24}{$\blacksquare$} 24 January 2010]{\includegraphics[width=0.32\textwidth]{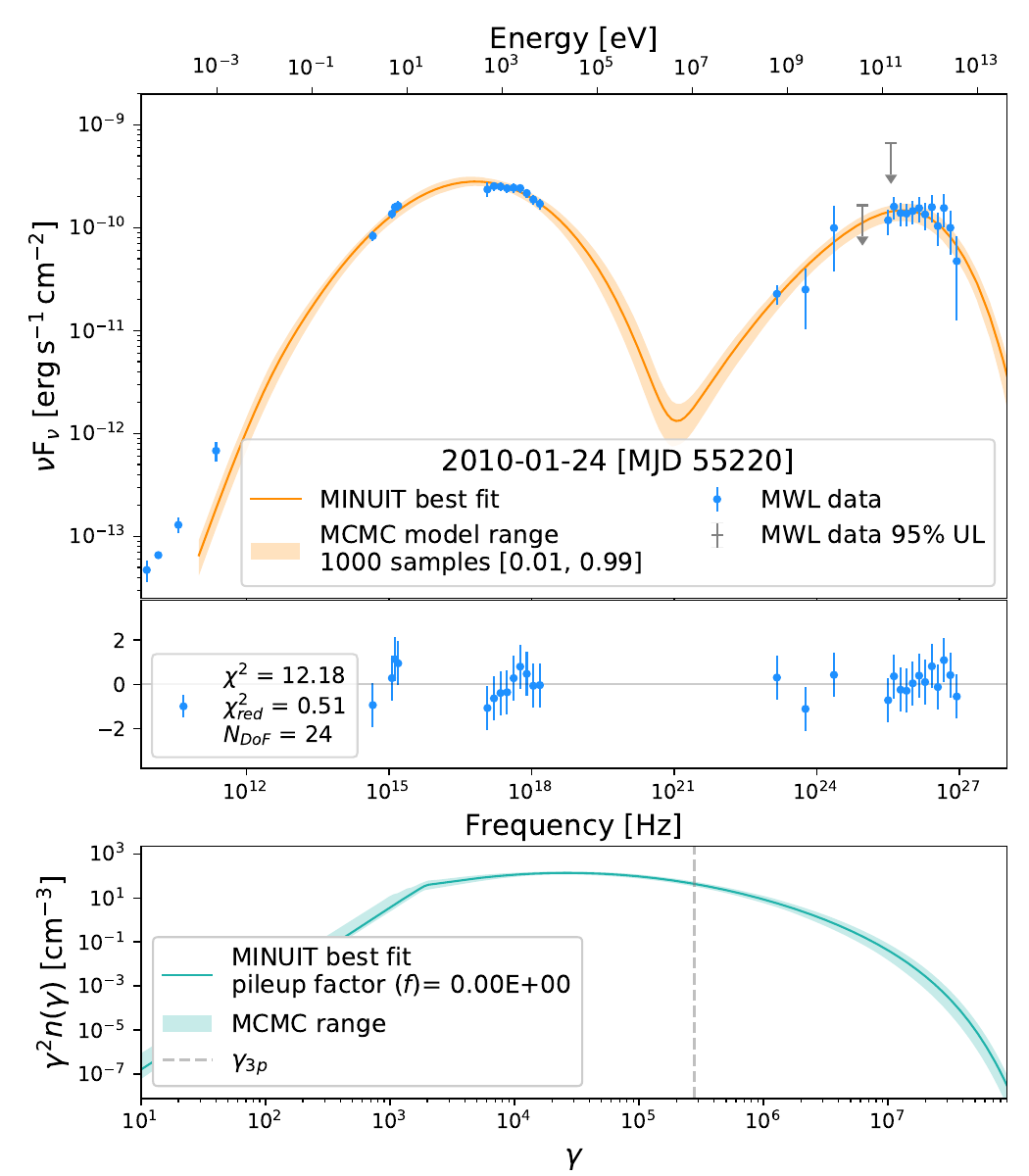}} &
            \subfloat[\textcolor{c25}{$\blacksquare$} 25 January 2010]{\includegraphics[width=0.32\textwidth]{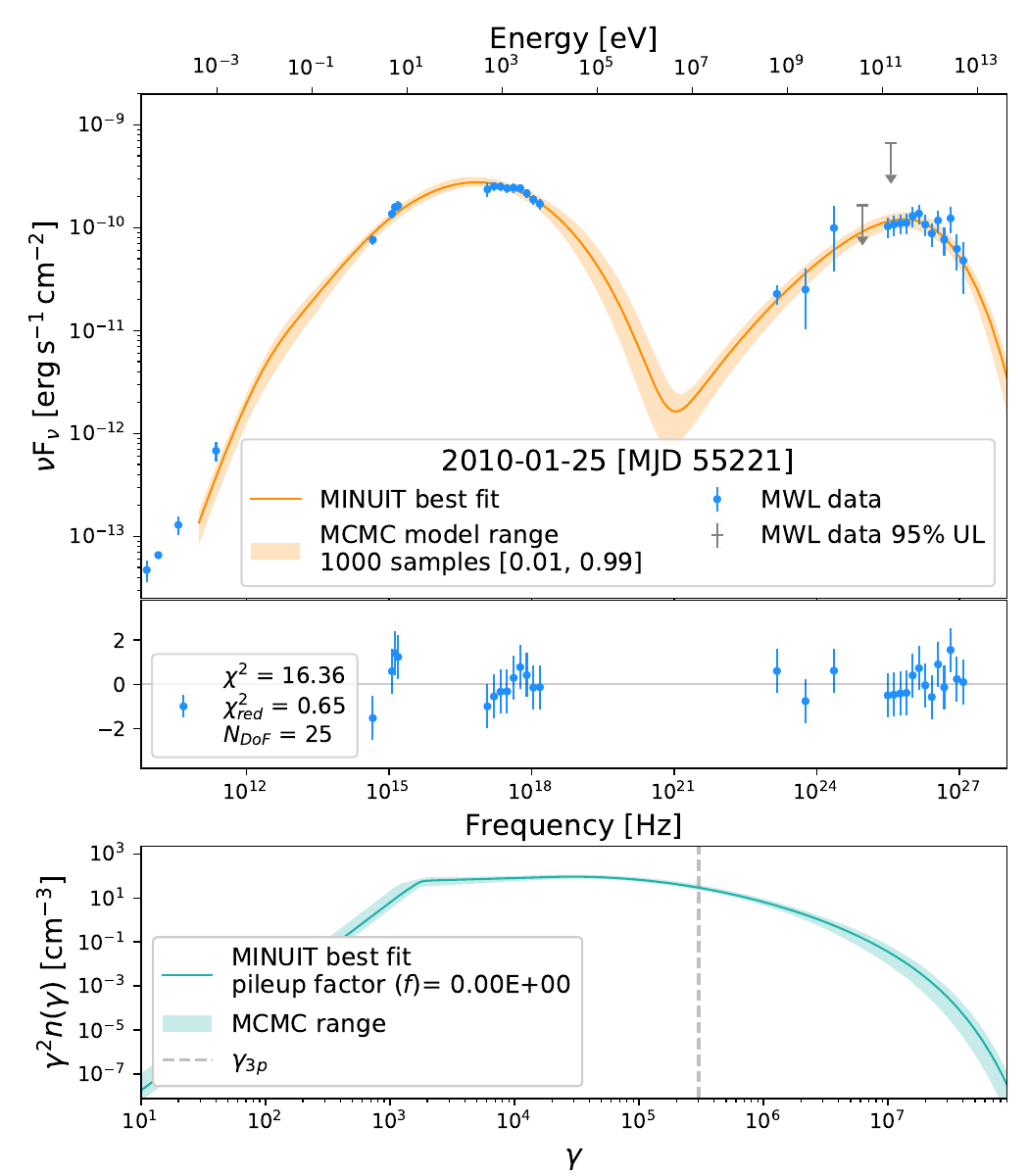}} \\
            \subfloat[\textcolor{c26}{$\blacksquare$} 26 January 2010]{\includegraphics[width=0.32\textwidth]{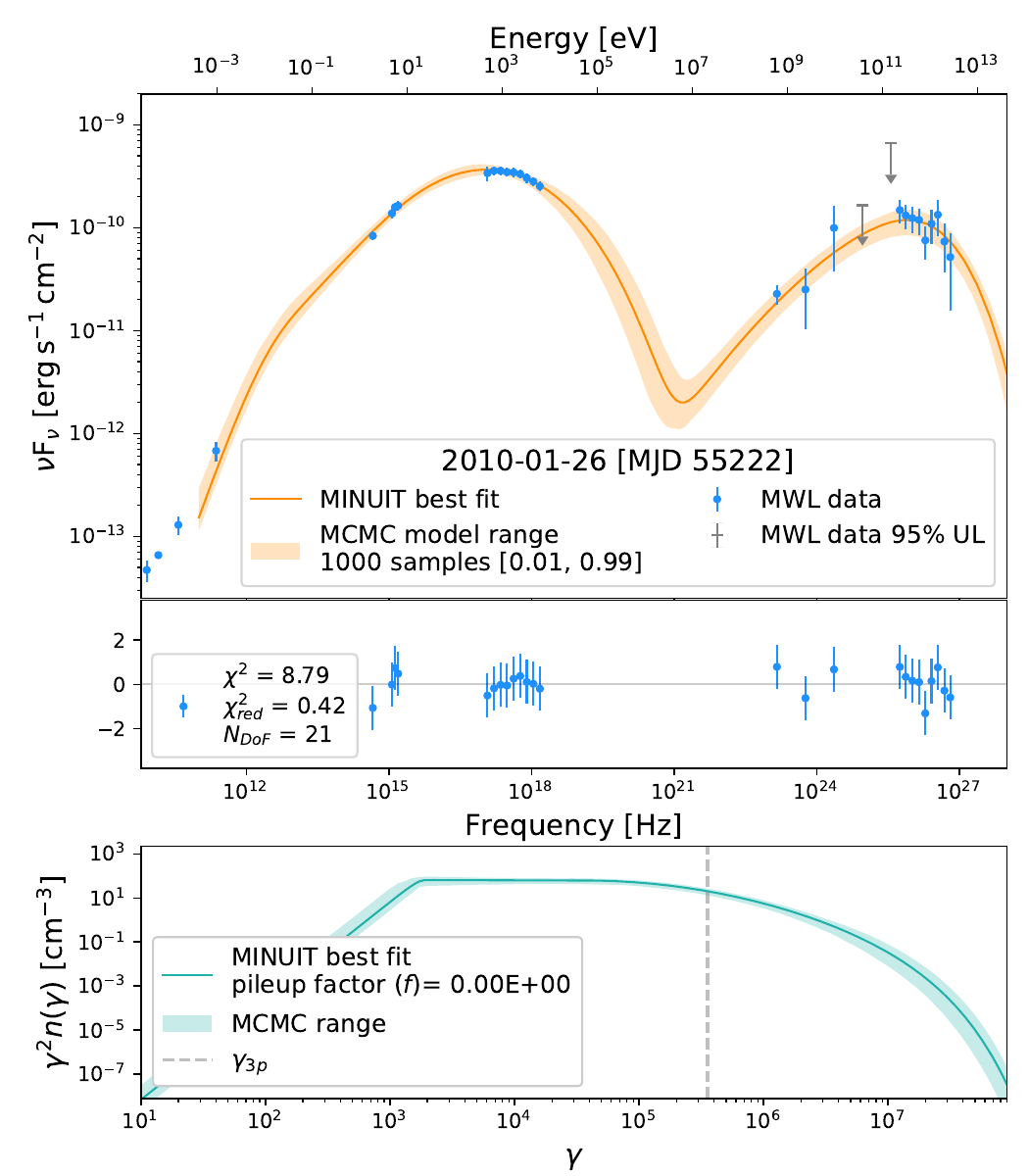}} \\
        \end{tabular}}
    \caption{\gls{sed} fits: expanding blob model. Continued from Fig.~\ref{fig:sed_exp_blob_1}.}
    \label{fig:sed_exp_blob_2}
\end{figure*}

%
%

\bibliographystyle{aa}
\bibliography{bibliography}{}



\appendix{
\section{\texorpdfstring{\gls{mwl}}{MWL} data} \label{appendix:mwl_data}

\subsection{MAGIC VHE data}
The \gls{mwl} \gls{lc} produced in \cite{felix_mrk421} was used to identify interesting flaring periods and the data from January 2010 was analysed in more detail to extract spectral points for the \glspl{sed} by integrating data into daily time bins using the standard \gls{magic} data reduction software \texttt{MARS} \citep{zanin2013mars}. 
The daily binned spectral data points for \gls{magic} were processed to correct for \gls{ebl} absorption using the \cite{ebl_model_D11} model to get the intrinsic spectrum required for modelling the emission at the source.

\subsection{\textit{Fermi}-LAT HE data}\label{sec:mwl_fermi_lat}
The \gls{lat} data were processed using \texttt{fermipy v1.2} and \texttt{ScienceTools v2.2.0}. 
The low energy threshold was set to $300$ MeV for the analysis as described in \cite{felix_mrk421} to improve the signal-to-noise ratio and angular resolution. 
For \gls{lat} analysis, the \gls{ts} is defined as $\gls{ts} = -2 \, ln (\mathcal{L}_\text{1} / \mathcal{L}_\text{0})$ where $\mathcal{L}_\text{1}$ is the likelihood of the model with the source and $\mathcal{L}_\text{0}$ for the null hypothesis, as implemented in \texttt{fermipy}. 
We used a \gls{ts} > 5 cut-off when deciding for the significance of the individual spectral points in the \gls{sed}, choosing to show the $95\%$ confidence upper-limits if the condition is not met. 
Despite these data quality cuts, we still have large errorbars in the data given the relatively low flux of \gls{421} compared to \gls{lat} sensitivity. 

\subsection{\textit{Swift}-XRT X-ray data}\label{sec:mwl_xrt}
We analysed the curvature in the spectral points of \gls{xrt} since conclusions from our stochastic model rely on accurate estimation and modelling of curvature in the synchrotron peak. 
Given that \gls{421} is an extra-galactic source and outside the galactic plane of the Milky-way (J2000 Galactic coordinates ${179.83}\; {+65.03}$), the correction due to X-ray interactions with the atomic \gls{nH} is small and in the preliminary analysis, this effect was ignored. 
However, the fact that the uncertainty in \gls{nH} column density estimation can mimic additional curvature in the spectrum via energy dependent absorption of X-rays, further analysis was carried out to estimate this energy dependent systematic uncertainty. 
It was found that the bins close to $\sim 0.5$ keV were affected by higher uncertainties in the column density of galactic \gls{nH} along the line of sight to the source.
For getting an estimate of the increased uncertainty, \gls{nH}$\, = 1.34 \times 10^{20} \, cm^{-2}$ from the 2D HI4PI map \citep{nh_density_2016A&A...594A.116H} and \gls{nH}$\, = 1.94 \times 10^{20} \, cm^{-2}$ \citep{nh_denisty_old_2005A&A...440..775K} were used as reference and this resulted in a maximum systematic uncertainty of $\sim16 \%$ in the first two bins of the \gls{xrt} spectrum. 
Spectral bins at energies higher than $\sim 0.5$ keV are largely unaffected by this and have the standard instrument uncertainty of $10 \%$ \citep{2009AA...494..775G}.

\subsection{\texorpdfstring{\acrshort{uvot} \gls{uv} data}{\textit{Swift}-UVOT UV data}}
The \gls{uvot} provides simultaneous measurements with \gls{xrt} in the \gls{uv} and optical bands \citep{uvot_2005SSRv..120...95R}.
Observations in the \gls{uv}W1, \gls{uv}M2 and \gls{uv}W2 bands ($2600$ \AA, $2246$ \AA, $1928$ \AA \; central filter wavelengths respectively) were conducted and the data were processed using standard photometry on the total exposure for each filter band in \texttt{HEAsoft v6.23}. 
Data affected by unstable satellite attitude and from starlight of UMa 51 was filtered out. 
Apertures of $5$ arcsec were used for the source region in each filter and $\sim 3$ regions of $16$ arcsec radius were used for the background estimation.  
Official calibrations from the CALDB release \texttt{20201026} were applied to the dataset \citep{uvot_calibration_2011AIPC.1358..373B} and subsequently the data were de-reddened to account for extinction \citep{uvot_dereddening_1999PASP..111...63F, 2011ApJ...737..103S, 1998ApJ...500..525S} and converted to spectral points.

\subsection{Optical R-band data}
The optical band data were captured by the \gls{webt} under the \gls{gasp} which observes \gls{lat} monitored blazars such that contemporaneous \gls{mwl} observations are available for X-ray and \gls{he} observations. 
\cite{carnerero_rband} processed and corrected the data for host galaxy contamination. 
The flux data points thus obtained were converted to spectral points for the R-band. 

\subsection{Radio data}
Observations at $8$ GHz and $14.5$ GHz (University of Michigan Radio Astronomy Observatory), $15$ GHz (Owens Valley Radio Observatory), $37$ GHz (Mets{\"a}hovi Radio Observatory) and $230$ GHz (Sub-Millimeter Array) were used as described in \cite{magic_veritas_sed_2015A&A...578A..22A}. 
The simultaneity of these observations wasn't as good as the other bands but the variability in these bands was also small \citep{felix_mrk421}. 
The fluxes were converted to spectral points using the central band frequency and the closest available observation for each radio telescope was used in the daily binned \glspl{sed}.

\section{SED fits and parameter evolution}
A comparison of the fits at the synchrotron peak on 15 and 20 January can be seen in Fig.~\ref{fig:compare_peak_15_20_jan}, showing a similar short-coming of the \gls{lppl} model in replicating the observed curvature as seen in Fig.~\ref{fig:compare_peak_19_jan}. 

\begin{figure}
    \includegraphics[width=0.47\textwidth]{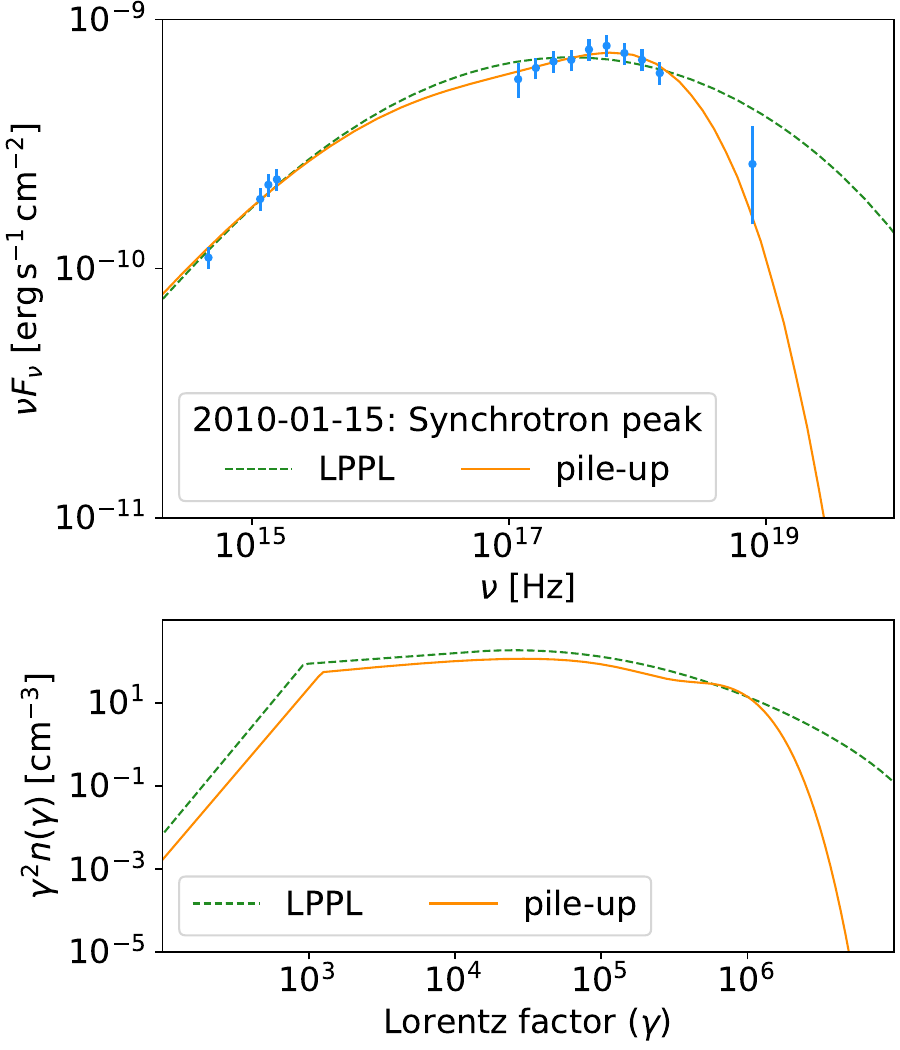}
    \includegraphics[width=0.47\textwidth]{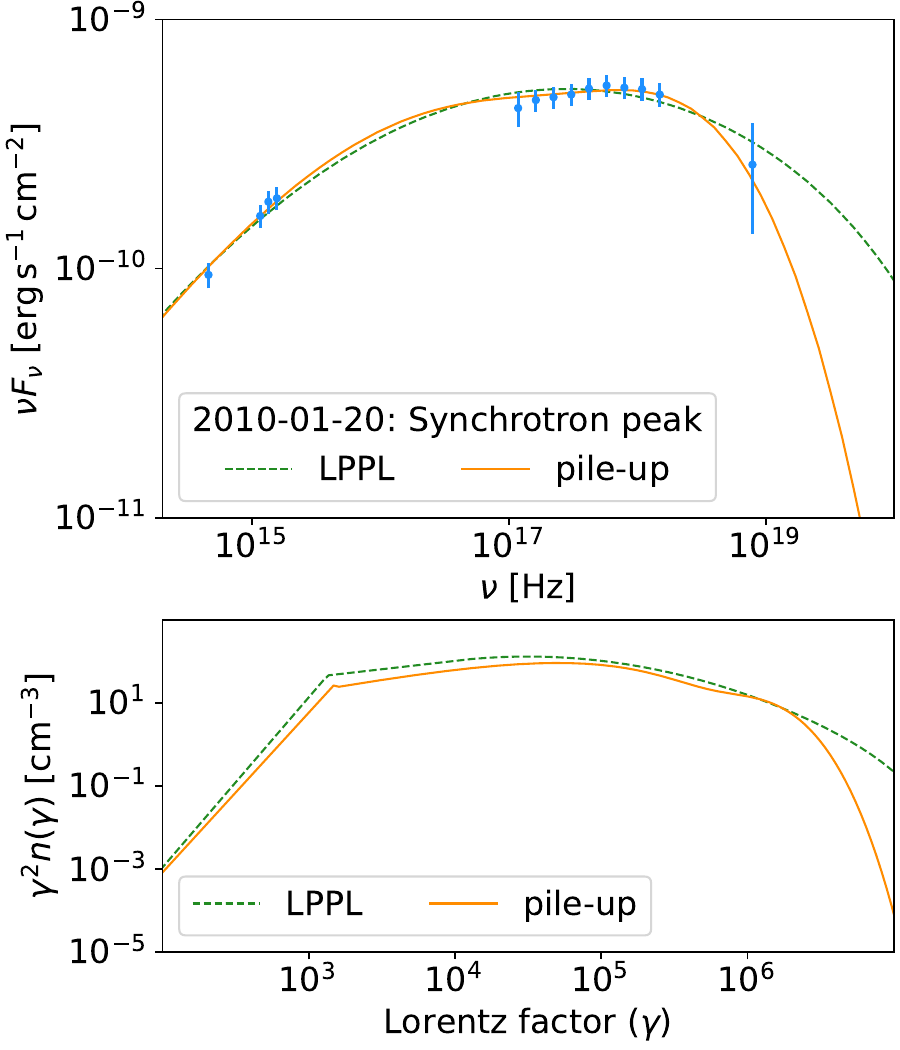}
    \caption{A comparison of the pile-up and \gls{lppl} models with a zoomed inset of the \gls{sed} focused at the synchrotron peak.}    
    \label{fig:compare_peak_15_20_jan}
\end{figure}

As opposed to the comparison of the \gls{lppl} and pile-up models for the same day shown previously, the evolution of the best fit of the \gls{sed} and the corresponding \gls{eed} around the pile-up states can be seen in Fig.~\ref{fig:15jan_evo} and \ref{fig:19jan_evo}.
\begin{figure}
    \resizebox{0.5\textwidth}{!}{
    \includegraphics[width=\textwidth]{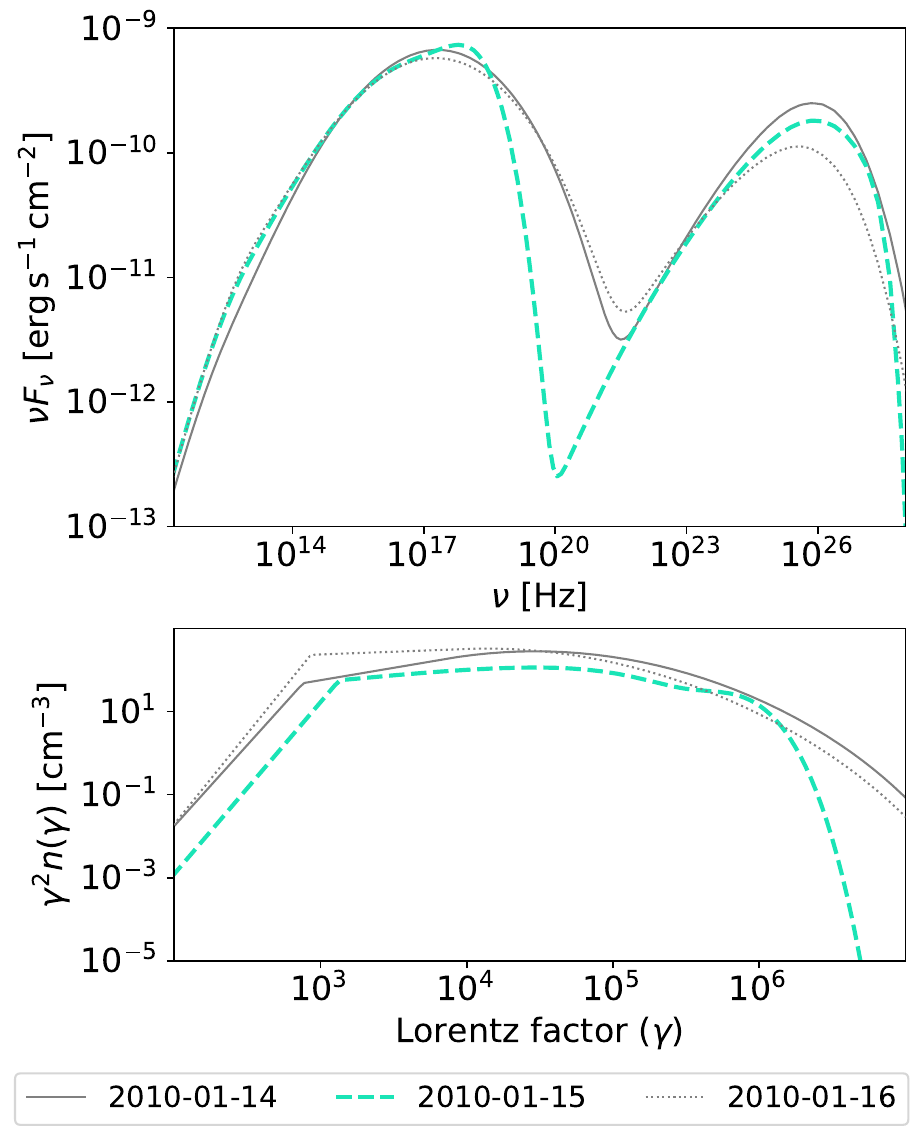}}
    \caption{The evolution of spectrum and the particle population around the pile-up on 15 January. The \gls{lppl} states are plotted in black colour, and the pile-up colour key is the same as the rest of the paper. }
    \label{fig:15jan_evo}
\end{figure}
\begin{figure}
    \resizebox{0.5\textwidth}{!}{
    \includegraphics[width=\textwidth]{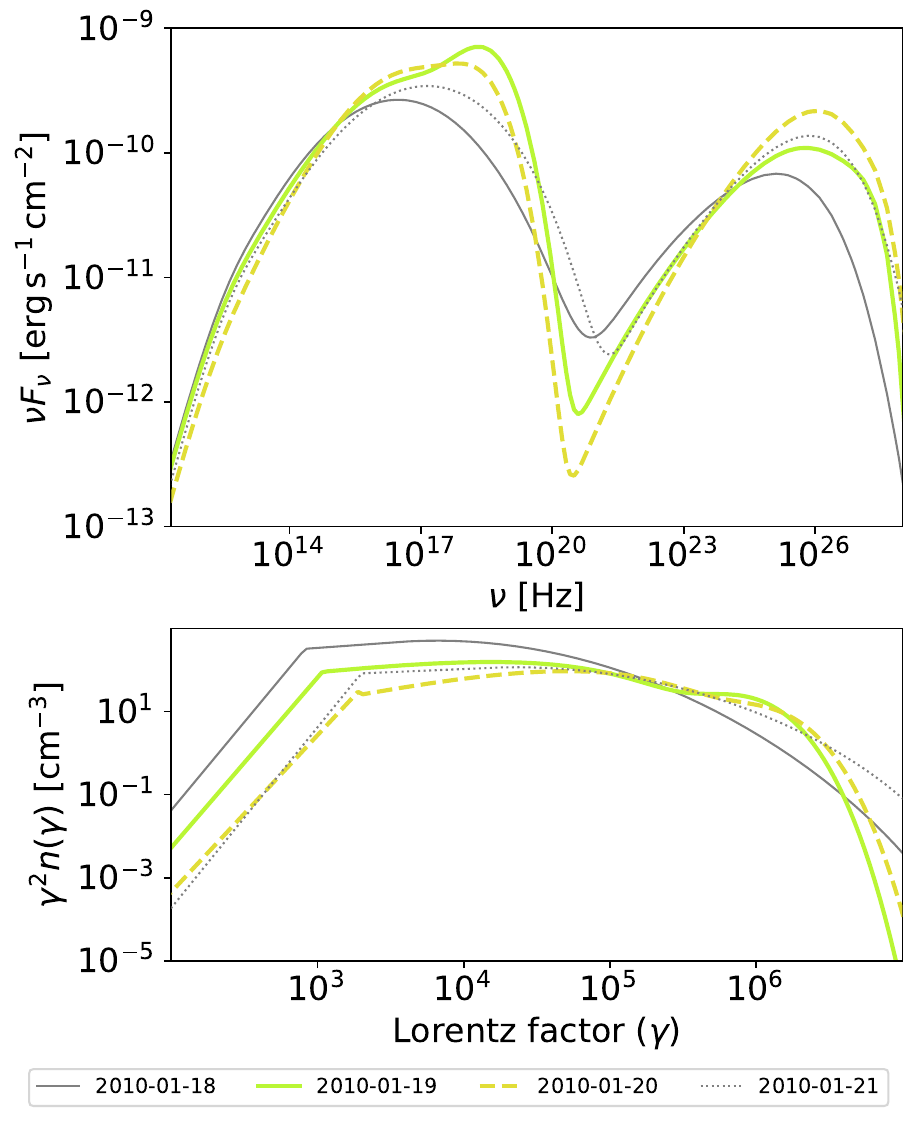}}
    \caption{Same as Fig.~\ref{fig:15jan_evo} for the 19 and 20 January pile-up states.}
    \label{fig:19jan_evo}
\end{figure}

The pile-up model parameter evolution can be seen in Fig.~\ref{fig:mcmc_par_evolution_pileup}. 
The \gls{mcmc} \gls{sed} fits for the expanding blob model can be seen in Fig.~\ref{fig:sed_exp_blob_1}, \ref{fig:sed_exp_blob_2}. 

\begin{figure}
    \resizebox{0.53\textwidth}{!}{
    \includegraphics[width=\textwidth]{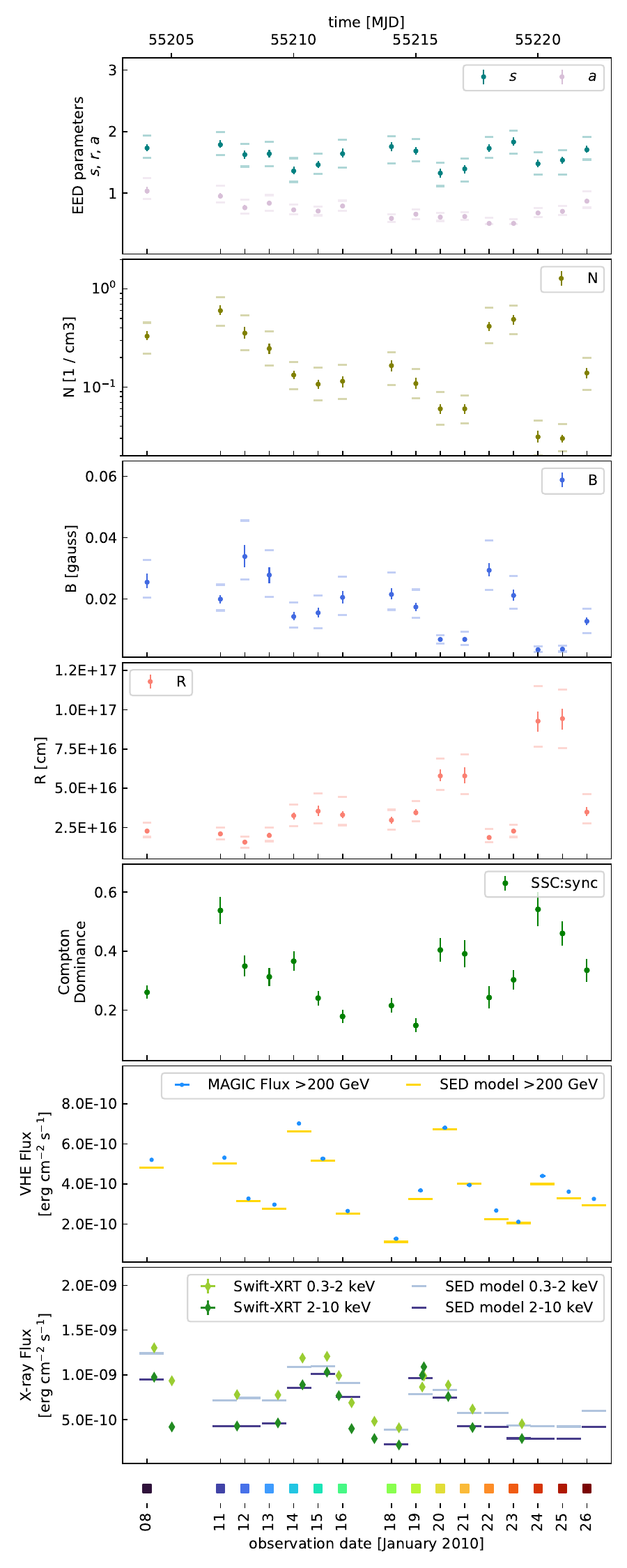}}
    \caption{Same as Fig.~\ref{fig:mcmc_par_evolution_lppl} for the pile-up model.}
    \label{fig:mcmc_par_evolution_pileup}
\end{figure}

\section{Phenomenology - additional context and plots}\label{appendix:phenomenology}

\subsection{\texorpdfstring{Dependence of $\nusync$ on $\gtp$}{Dependence of peak synchrotron frequency}}
As described in Section~\ref{sec:Ep_g3p_r3p}, $\nusync$ has a quadratic dependence on $\gtp$ as long as $B$ and $\delta$ are constants. 
Since our models have the bulk magnetic field as a free parameter while the beaming factor is fixed ($\delta = 45$), the scatter in the trend in Fig.~\ref{fig:gamma3p_vs_nu_sync} is entirely due to variations in $B$, resulting in a index of $m=1.23 \pm 0.24$ instead of $m=2$. 

\begin{figure}
    \centering
    \includegraphics[width=0.48\textwidth]{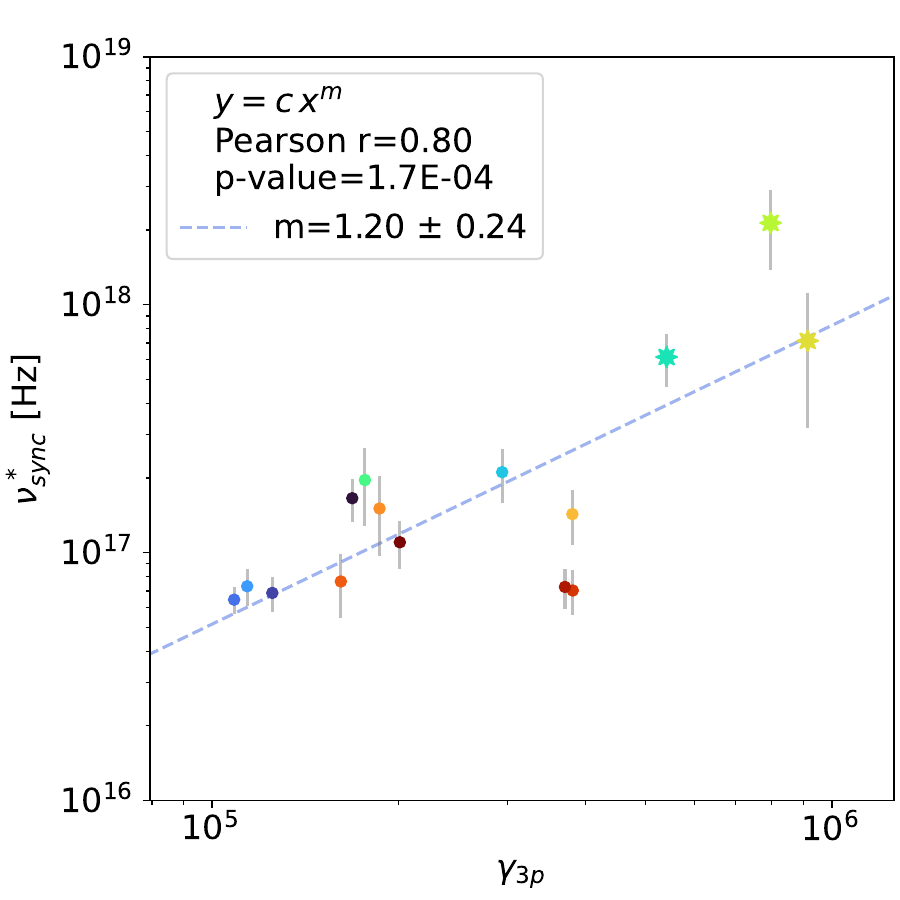}
    \caption{Combined model: Synchrotron peak position ($\nusync$) of the \gls{sed} is directly correlated with the peak of the $\ngt$ distribution ($\gtp$)}    
    \label{fig:gamma3p_vs_nu_sync}
\end{figure}

\subsection{\texorpdfstring{\gls{ic} scattering regime}{IC scattering regime}}
The ratio of the \gls{ic}/\gls{ssc} and synchrotron peak frequencies ($\nussc$/$\nusync$) can be used to identify whether the \gls{ic} scattering is in the elastic Thomson regime which works well as an approximation at lower energies or the energy dependent cross-section of the \gls{kn} regime which applies at higher energy of the scattering electron. The synchrotron (under the $\delta$-approximation of synchrotron emission) and \gls{ic} peak frequencies (in Hz) can be approximated as follows \citep{rybicki_lightman, 1970RvMP...42..237B}: 
\begin{align*}
    \nusync &\approx 10^6 \cdot \gtp^2 \cdot B \cdot \delta \\
    \nussc &\approx 
    \begin{cases}
        \frac{4}{3} \gtp^2 \nusync &\text{Thomson regime}\\
        \frac{m_e c^2}{h} \gtp &\text{\gls{kn} regime}
    \end{cases}
\end{align*}
where $B$ is in Gauss, $m_e c^2$ is the rest mass of the electron and $h$ is the Planck constant. The results from the combined model are shown in \ref{fig:TH_KN_transition_combined}, with $\delta=45$ and the minimum and maximum values of $B$ obtained in the modelling used for the dashed and dotted blue lines. During January 2010, \gls{421} has $\gtp \sim 10^5 - 10^6$  such that the \gls{ic} scattering is in the \gls{kn} regime. 

\begin{figure}
    \centering
    \includegraphics[width=0.48\textwidth]{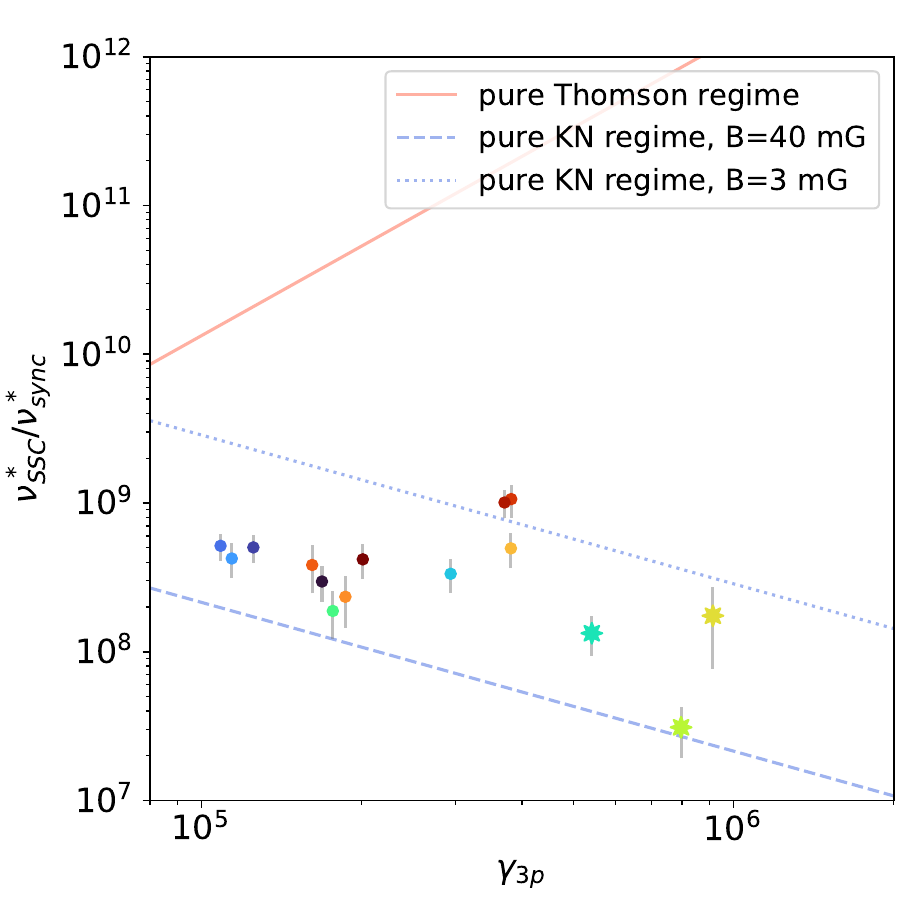}
    \caption{Combined model: Ratio of $\nussc$ and $\nusync$, indicating the \gls{ic} scattering in the \gls{kn} regime.}    
    \label{fig:TH_KN_transition_combined}
\end{figure}

\subsection{Expanding blob model}\label{appendix:exp_blob}
The expanding blob model shows very similar trends and phenomenology as the combined model. The anti-correlation of curvature as a signature of stochastic acceleration is shown in Fig.~\ref{fig:r3p_vs_g3p_exp_blob}. 
A plot of the emission region radius $R$ versus distance from the \gls{smbh} can be seen in Fig.~\ref{fig:d_vs_R_exp_blob}. 

\begin{figure}
    \centering
    \includegraphics[width=0.48\textwidth]{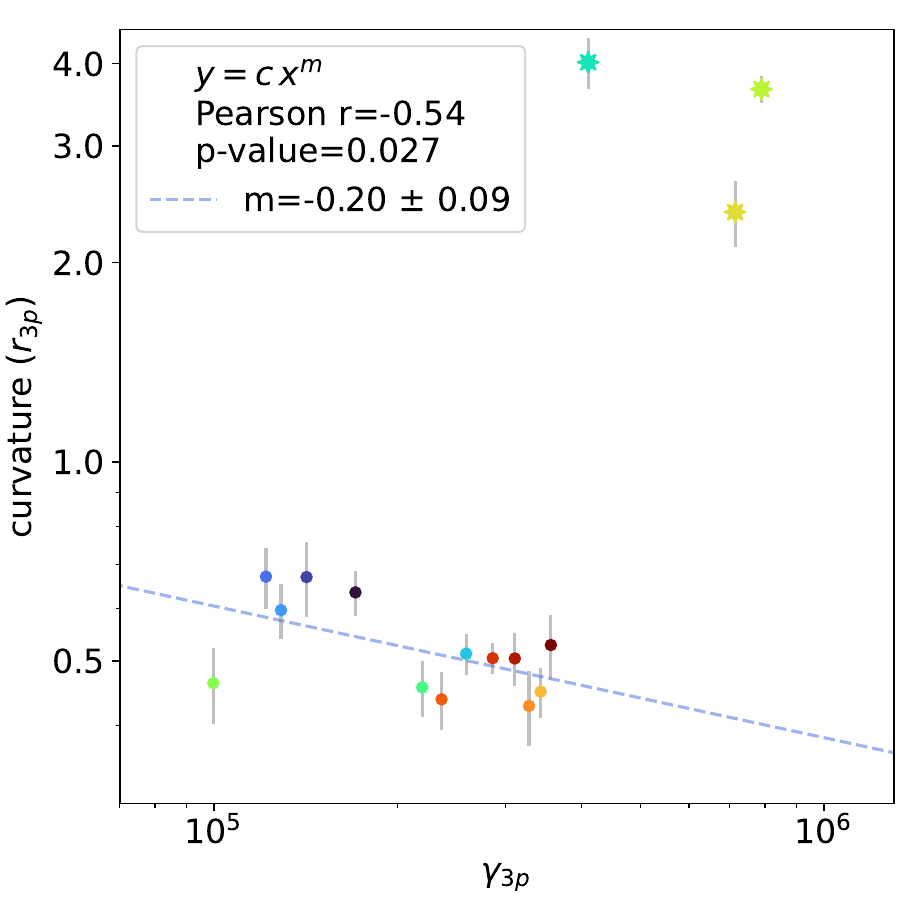}
    \caption{Expanding blob model: Anti-correlation between $\rtp$ and $\gtp$ similar to Fig.~\ref{fig:r3p_vs_g3p}.}
    \label{fig:r3p_vs_g3p_exp_blob}
\end{figure}

\begin{figure}
    \centering
    \includegraphics[width=0.48\textwidth]{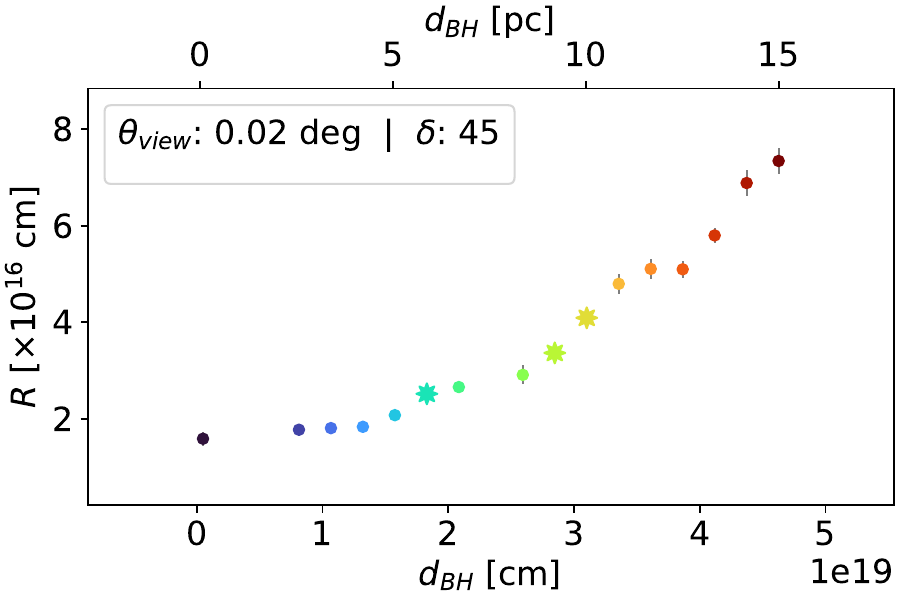}
    \caption{Expanding blob model: The expansion of the blob vs the distance from the \gls{smbh} ($d_{BH}$) assuming that it is a moving emission region with a constant Doppler beaming factor $\delta=45$ and a viewing angle $\thetaview = 0.02^\circ$ as described in Section~\ref{sec:expanding_blob}. The blob travels about 15 parsec under these assumptions over $\sim20$ days.}
    \label{fig:d_vs_R_exp_blob}
\end{figure}

}

\end{document}